\documentclass[aps,prc,showpacs,twocolumn,superscriptaddress,nofootinbib,preprintnumbers,floatfix]{revtex4-1}%preprint
\usepackage{tikz}
\usepackage{longtable}
\usepackage{epstopdf}
\usepackage{graphicx}
\usepackage{slashed}
\usepackage{dcolumn}
\usepackage{bm}
\usepackage{amssymb,amsthm,amsmath,amstext,amsbsy,amsopn}
\usepackage{bbm}
\usepackage{nicefrac}
\usepackage{mathrsfs}
\usepackage{pstricks}
\usepackage{hyperref}
\usepackage{leftidx}
\usepackage{environ}
\usepackage{mathtools}
\usepackage{xspace}
\usepackage{overpic}
\usepackage{subfigure}
\usepackage{tensor}
\usepackage{gensymb} 
\usepackage{cancel}
\usepackage{float}
\usepackage{color}
\usepackage{calligra}
\usepackage{soul}
\usepackage{changes}
\usepackage{appendix}

\newcommand{\bea}{\begin{eqnarray}}
\newcommand{\eea}{\end{eqnarray}}
\newcommand{\beq}{\begin{equation}}
\newcommand{\eeq}{\end{equation}}

\usepackage[normalem]{ulem}  % \sout{old text} for strikeout

\renewcommand\sout{\bgroup \color{red} \ULdepth=-.5ex \ULset}
\begin{document}

%\preprint{}

% Use the \preprint command to place your local institutional report
% number in the upper right-hand corner of the title page in preprint mode.
% Multiple \preprint commands are allowed.
% Use the 'preprint numbers' class option to override journal defaults
% to display numbers if necessary

%%%%%%%%%%%%%%%%%%%%%%%%%%%%%%%%%%%%%%%%%%%%%%%%%%%%%%%%%%%%%%%%%%%%%%%%%%%%%%%%%%%%%%%%%%%%%%%%%%%%%%%
\title{Lepton-proton two-photon exchange with improved soft-photon approximation including proton's 
structure effects in HB$\chi$PT }
%%%%%%%%%%%%%%%%%%%%%%%%%%%%%%%%%%%%%%%%%%%%%%%%%%%%%%%%%%%%%%%%%%%%%%%%%%%%%%%%%%%%%%%%%%%%%%%%%%%%%%%
\author{Rakshanda Goswami}
     \email[]{r.goswami@iitg.ac.in}
     \affiliation{Department of Physics, Indian Institute of Technology Guwahati, 
                 Guwahati - 781039, Assam, India.}%
\author{Pulak Talukdar}
     \email[]{t.pulak@iitg.ac.in}
     \affiliation{Department of Physics, S. B. Deorah College, Guwahati - 781007, Assam, India.}%  
\author{Bhoomika Das}
     \email[]{bhoomika.das@iitg.ac.in}
     \affiliation{Department of Physics, Indian Institute of Technology Guwahati, 
                 Guwahati - 781039, Assam, India.}%     
\author{Udit Raha}
     \email[]{udit.raha@iitg.ac.in}
     \affiliation{Department of Physics, Indian Institute of Technology Guwahati, 
                 Guwahati - 781039, Assam, India.}%
\author{Fred Myhrer}
     \email[]{myhrer@mailbox.sc.edu}
     \affiliation{Department of Physics and Astronomy, 
                 University of South Carolina, Columbia, SC 29208, USA.}%
%%%%%%%%%%%%%%%%%%%%%%%%%%%%%%%%%%%%%%%%%%%%%%%%%%%%%%%%%%%%%%%%%%%%%%%%%%%%%%%%%%%%%%%%%%%%%%%%%%%%%%%
%\date{}

%%%%%%%%%%%%%%%%%%%%%%%%%%%%%%%%%%%%%%%%%%%%%%%ABSTRACT%%%%%%%%%%%%%%%%%%%%%%%%%%%%%%%%%%%%%%%%%%%%%%%%
\begin{abstract}
\noindent
We present an improved evaluation of the two-photon exchange correction to the unpolarized lepton-proton
elastic scattering process at very low-energies relevant to the MUSE experiment, where only the dominant 
intermediate elastic proton is considered. We employ the framework of heavy baryon chiral perturbation 
theory and invoke the soft-photon approximation in order to reduce the intricate 4-point loop-integrals 
into simpler 3-point loop-integrals. In the present work, we adopt a more robust methodology compared 
to an earlier work along the same lines for analytically evaluating the loop-integrations, and 
incorporate important corrections at next-to-next-to-leading order. These include the proton's structure
effects which renormalize the proton-photon interaction vertices and the proton's propagator. Finally, 
our results allow a model-independent estimation of the charge asymmetry for the scattering of 
unpolarized massive leptons and anti-leptons. 
\end{abstract}

%\pacs{12.38.Bx, 12.38.Cy, 12.39.St, 13.40.Gp}
\maketitle
%%%%%%%%%%%%%%%%%%%%%%%%%%%%%%%%%%%%%%%%%%%%%%%%%%%%%%%%%%%%%%%%%%%%%%%%%%%%%%%%%%%%%%%%%%%%%%%%%%%%%%%

%%%%%%%%%%%%%%%%%%%%%%%%%%%%%%%%%%%%%%%%%%%%%%%%%%%%%%%%%%%%%%%%%%%%%%%%%%%%%%%%%%%%%%%%%%%%%%%%%%%%%%%
\section{Introduction}
\label{sec:intro}
%%%%%%%%%%%%%%%%%%%%%%%%%%%%%%%%%%%%%%%%%%%%%%%%%%%%%%%%%%%%%%%%%%%%%%%%%%%%%%%%%%%%%%%%%%%%%%%%%%%%%%%
The two-photon exchange (TPE) contribution in the context of unpolarized lepton-proton ($\ell$-p, where 
$\ell\equiv e^\pm, \mu^\pm $) elastic scattering refers to the quantum mechanical effect in which two virtual photons 
are exchanged between a lepton and a target proton. This process constitutes an important higher-order radiative 
correction to the leading one-photon exchange (OPE) mechanism, which dominates the low-energy scattering process. The
advent of the form factor discrepancy~\cite{Jones:1999rz,Perdrisat:2006hj,Punjabi:2015bba,Puckett:2010} and the proton's
radius puzzle~\cite{Pohl:2010zza,Pohl:2013,Antognini:1900ns,Bernauer:2014,Carlson:2015,Bernauer:2020ont} has generated 
renewed interest in the scientific community concerning the TPE mechanism, due to its potential role in explaining both 
discrepancies, simultaneously. Significant efforts have been invested in the last two decades to estimate the TPE 
contributions in elastic lepton- and anti-lepton-proton scattering both experimentally and theoretically. Earlier 
analyses of radiative corrections to $\ell$-p scattering included the TPE merely to cancel the infrared (IR) divergence
arising from the so-called charge-odd  soft-photon bremsstrahlung counterparts (namely, from the interference process 
between a lepton and a proton each emitting a single soft-photon), focusing mainly on high-energy 
kinematics~\cite{Arrington:2003,Guichon:2003,Blunden:2003sp,Rekalo:2004wa,Blunden:2005ew,Carlson:2007sp,Arrington:2011}). 
However, persuaded by a host of ongoing/planned high-precision experiments targeted at low-energy kinematics (e.g., see 
Ref.~\cite{Gao:2021sml} for a recent review), the current demand, therefore, requires improved reliability of modeling 
and precision evaluation of the TPE intermediate hadronic states and their interaction. This motivated an entire gamut of 
recent rigorous TPE analyses pertinent to low-momentum transfer kinematics employing different perturbative and 
non-perturbative 
techniques~\cite{Kivel:2012vs,Lorenz:2014yda,Tomalak:2014sva,Tomalak:2014dja,Tomalak:2015aoa,Tomalak:2015hva,Tomalak:2016vbf,Tomalak:2017npu,Koshchii:2017dzr,Tomalak:2018jak,Bucoveanu:2018soy,Talukdar:2019dko,Peset:2021iul,Talukdar:2020aui,Kaiser:2022pso,Guo:2022kfo,Engel:2023arz,Cao:2021nhm,Choudhary:2023rsz}.

One such approach is to systematically analyze the non-perturbative dynamics of hadrons in the model-independent framework 
of a chiral effective field theory (EFT). 
%
% FM: EFT = Effective Field Theory
%
In EFT the hadrons are effectively treated as elementary degrees of freedom with a 
low-energy perturbative expansion for typical momenta far below a well-defined breakdown scale. The isospin SU(2) version 
of the heavy baryon chiral perturbation theory (HBChPT) is one such EFT. It provides a systematic computational framework to 
investigate the low-energy structure and dynamics of mesons and nucleons, and it also allows a study of their responses to 
external weak and electromagnetic currents~\cite{Gasser:1982ap,Jenkins:1990jv,Bernard:1992qa,Ecker:1994pi,Bernard:1995dp}. 
However, the applicability is limited to typical momentum scales not much larger than the pion mass and far below the chiral
symmetric breaking scale $\Lambda_\chi\simeq 1$~GeV/c. In addition, HBChPT entails a chiral expansion scheme in powers of the
inverse nucleon mass $M$, i.e., ${\mathcal Q}/M$ where ${\mathcal Q}$ is the generic momentum scale. In our analysis, 
$\mathcal Q$ will correspond to the typical magnitude of the 4-momentum transfer $|Q|=\sqrt{-Q^2}$ for the elastic 
$t$-channel scattering process. The framework is ideally suited for a gauge-invariant estimation of the non-perturbative 
{\it recoil-radiative} corrections relevant to low-energy $\ell$-p scattering. Recently, the corresponding TPE radiative 
corrections were estimated up-to-and-including next-to-leading order (NLO), both in the context of the soft photon 
approximation (SPA) by Talukdar {\it et al.}~\cite{Talukdar:2019dko,Talukdar:2020aui}, as well as the {\it exact} TPE 
evaluation in Choudhary {\it et al.}~\cite{Choudhary:2023rsz}, all at low-$Q^2$ kinematics relevant to the ongoing MUSE 
experiment at PSI~\cite{Gilman:2013eiv}. In particular, in the SPA-based approach~\cite{Talukdar:2019dko}, one of the two 
photon propagator momenta is set to zero, following the method outlined in the work of Mo and Tsai~\cite{Mo:1968cg}. This is
a well-known technique widely adopted in works on radiative corrections for simplifying the evaluation of complicated 4-point 
TPE loop-integrals (see e.g., 
Refs.~\cite{Koshchii:2017dzr,Bucoveanu:2018soy,Maximon:1969nw,Maximon:2000hm,Vanderhaeghen:2000ws}). Furthermore, we note the
following two features regarding the SPA analysis of Talukdar {\it et al}~\cite{Talukdar:2019dko} that deserve some 
introspection: 
\begin{itemize}
\item As only the dominant real (or dispersive) part of the low-energy TPE contribution is relevant for evaluating the 
unpolarized elastic scattering cross-section, certain simplifying approximations were adopted in evaluating the TPE loops, 
such as dropping the complex $\pm i\epsilon$ parts from the propagators within the TPE loops.\footnote{The sub-dominant 
complex or imaginary absorptive parts of the low-energy TPE loops are generally attributed to the polarized and inelastic 
cross-sections and therefore irrelevant to our elastic TPE contributions of interest. However, small changes in the relevant
real parts of the loop-integrals are still possible without using the correct $\pm i\epsilon$ prescription. Thus, either 
neglecting the proper $\pm$ signs or foregoing the $i\epsilon$ factors in the propagators may introduce small systematic 
errors in the evaluation of loop-integrals.} Such an approximation could, however, be a source of unwarranted systematics 
that should be avoided to achieve better precision. 
\item The infrared (IR) divergences which were isolated {\it via} dimensional regularization (DR) led to a contribution
to the cross-section proportional to $$-\frac{2}{D-4} - \gamma_E + \ln\left(\frac{4\pi\mu^2}{-Q^2}\right)\,,$$ 
where $D>4$ is the analytically continued space-time dimension, $\gamma_E=0.577216...$ is the Euler-Mascheroni constant and 
$\mu$ is an arbitrary subtraction scale. Since the subtraction scheme adopted to eliminate such singularities contains 
$Q^2$-dependent terms, certain dynamical contributions from the residual finite part of the TPE could be lost. Therefore, it
is more appropriate to choose a subtraction scheme for the TPE analysis that is $Q^2$-independent, for example, proportional 
to $\ln (m^2_l)$, namely,  $$-\frac{2}{D-4} - \gamma_E + \ln\left(\frac{4\pi\mu^2}{m^2_l}\right)\,,$$ where $m_l$ is the 
lepton mass. 
\end{itemize}

The purpose of the current work is to improve upon the SPA results of Talukdar {\it et al.}~\cite{Talukdar:2019dko} (a) by 
including some of the dominant TPE corrections of next-to-next-to-leading order (NNLO) that also include effects due to the 
proton's radius and anomalous magnetic moment at the level of one-loop accuracy,\footnote{In this work we exclude the subset 
of two-loop TPE diagrams of NNLO accuracy consisting of an additional pion-loop which is expected to be much suppressed (cf. 
Fig.~\ref{fig:pi-loops}).} and (b) by refining some of the approximations used in that study by employing a more rigorous 
treatment of the loop-integrations with an improved IR-subtraction scheme. In addition, we shall present the corresponding 
charge asymmetry results in SPA by including the soft-photon bremsstrahlung contributions at NLO as was considered in the 
work of Ref.~\cite{Talukdar:2020aui}. The paper is organized as follows. In Sec.~\ref{sec:theory}, we present the specifics 
of our {\it modus operandi} in regard to the low-energy EFT framework, along with the kinematical settings used in our 
analytical calculations. The analytical expressions for the various orders of the fractional TPE corrections up to NNLO are
evaluated in this section. We then utilize our SPA-based TPE results up to NLO in this section, along with the rest of the 
NLO radiative corrections results obtained earlier by Talukdar {\it et al.}~\cite{Talukdar:2020aui}, to obtain an estimate
of the charge asymmetry observable (between lepton-proton and anti-lepton-proton scatterings) in Sec.~\ref{sec:asym}. 
Finally, in Sec.~\ref{sec:results}, we discuss our numerical results in the context of low-energy kinematics relevant to the
proposed MUSE experiment kinematical region. We also provide an appendix where we collect the analytical expressions of all 
relevant loop-integrations used to obtain our results.

%%%%%%%%%%%%%%%%%%%%%%%%%%%%%%%%%%%%%%%%%%%%%%%%%%%%%%%%%%%%%%%%%%%%%%%%%%%%%%%%%%%%%%%%%%%%%%%%%%%%%%%
\section{Effective Lagrangian and Kinematics of TPE}
\label{sec:theory}
%%%%%%%%%%%%%%%%%%%%%%%%%%%%%%%%%%%%%%%%%%%%%%%%%%%%%%%%%%%%%%%%%%%%%%%%%%%%%%%%%%%%%%%%%%%%%%%%%%%%%%%
The relevant parts of the effective $\pi N$ chiral Lagrangian up to NNLO accuracy that is needed for our evaluation of the 
TPE amplitudes are displayed as follows~\cite{Talukdar:2020aui,Bernard:1995dp,Fettes:2000fd}:
\begin{eqnarray}
\mathcal{L}_{\pi N}=\mathcal{L}^{(0)}_{\pi N}+\mathcal{L}^{(1)}_{\pi N}+\mathcal{L}^{(2)}_{\pi N}+ ...\,\,,
\label{eq:LpiN}
\end{eqnarray}
where we need only the $\nu=0,1,$ and $2$ chiral order terms, namely,
\begin{eqnarray}
\mathcal{L}_{\pi N}^{(0)} \!\!&=&\!\! {\bar N}_v (i v\cdot {\mathcal D} + g_A S \cdot u) N_v\,,
\end{eqnarray}
\begin{eqnarray}
\mathcal{L}_{\pi N}^{(1)} \!\!&=&\!\! {\bar N}_v\! \left[\frac{1}{2 M}(v \cdot {\mathcal{D}})^2 
- \frac{1}{2 M} \mathcal{D} \cdot \mathcal{D} - \frac{i g_A}{2M}\big\{S\cdot {\mathcal D} , v\cdot u\big\}  \right.
\nonumber\\
&&\hspace{0.4cm} \left. \!+\, c_1 {\rm Tr}(\chi_+) + \left( c_2 - \frac{g_A^2}{8M}\right) ( v \cdot u)^2 
+ c_3 u \cdot u  \right.
\nonumber\\
&&\hspace{0.4cm} \left. \!+\, \left( c_4 +\frac{1}{4M}\right) [S^\mu , S^\nu]u_\mu u_\nu 
+ c_5 {\rm Tr}(\tilde{\chi}_+) \right.
\nonumber\\
&&\hspace{0.4cm} \left. \!-\, \frac{i}{4M}[S^\mu , S^\nu] \left[(1 +c_6)f^+_{\mu \nu} 
+ c_7 {\rm Tr}(f^+_{\mu\nu}) \right] \right]\!\! N_v\,,  
\nonumber\\
\text{and} &&\\
{\mathcal{L}}_{\pi N}^{(2)} \!\!&=&\!\! \bar{N_v} \left[  - \frac{i}{4M^2} (v \cdot \mathcal{D})^3 
+ \frac{i}{8 M^2} \Big \{ \mathcal{D}^2 (v \cdot \mathcal{D}) \right. 
\nonumber\\
&&\hspace{0.4cm} \left.  - (v \cdot \mathcal{D}^\dag) {\mathcal{D}^\dag}^2 \Big \} +\cdots  \right] N_v\,.
\label{eq:LpiN123}
\end{eqnarray}
The ellipsis above denotes terms not explicitly needed in this work. In particular, we have not displayed those parts 
of the Lagrangian dealing with the purely pionic interactions, as well as the standard (relativistic) quantum 
electrodynamical (QED) interactions with the leptons. $N_v=\,$(p${}_v$,\,n${}_v$)${}^T$ denotes the heavy nucleon isospin 
doublet spinor field, $g_A=1.267$ is the nucleon axial-vector coupling constant, $v=(1,{\bf 0})$ is the nucleon velocity
four-vector, and $S=(0,\boldsymbol{\sigma}/2)$, with the Pauli spin vector 
${\boldsymbol{\sigma}}=(\sigma_x,\sigma_y,\sigma_z)$, is the standard choice of the nucleon spin four-vector satisfying 
the condition, $S\cdot v =0$. The chiral covariant derivative $ \mathcal{D}_\mu$ is defined as,
\begin{eqnarray}
\mathcal{D}_\mu \!&=&\! \partial_\mu + \Gamma_\mu - i v_\mu^{(s)} \,, \quad \text{where}
\nonumber\\
\Gamma_\mu \!&=&\! \frac{1}{2} \left[u^\dag (\partial_\mu - ir_\mu)u   + u (\partial_\mu - i  l_\mu) u ^\dag \right]\,,
\end{eqnarray}
is the chiral connection that includes external vector and axial-vector gauge fields and is expressed in terms of the 
non-linearly realized pion fields $u=\sqrt{U}$, namely,
\begin{equation}
U=\sqrt{1-\frac{\boldsymbol{\pi}^2}{f_\pi}}+\frac{i}{f_\pi}\boldsymbol{\tau}\cdot\boldsymbol{\pi}\,,
\end{equation}
with $\boldsymbol{\pi}=(\pi_1,\pi_2,\pi_3)$ being the Cartesian pion fields, $\boldsymbol{\tau}=(\tau_1,\tau_2,\tau_3)$,
the Pauli iso-spin matrices, and $f_\pi$ the pion decay constant. The axial connection or the so-called chiral 
{\it vielbein} is similarly given as
\begin{eqnarray}
u_\mu \!&=&\! i u^\dag \nabla_\mu U u^\dag\,, \quad \text{where}
\nonumber\\
\nabla_\mu U \!&=&\! \partial_\mu U - i r_\mu U + i U l_\mu\,.
\end{eqnarray}
The NLO (i.e., chiral order $\nu=1$) Lagrangian $\mathcal L^{(1)}_{\pi N}$ contains seven phenomenologically determined 
finite low-energy constants (LECs), $c_i$, $i=1,...,7$~\cite{Scherer:2002tk,Fettes:1998ud}. The LECs $c_6$ and $c_7$ can 
be determined {\it via} the relations $c_6 = \kappa_v$,  and $c_7 = (\kappa_s - \kappa_v)/2$, where $\kappa_s = -0.12$ and 
$\kappa_v = 3.71$ are the nucleon iso-scalar and iso-vector anomalous magnetic moments, respectively~\cite{Bernard:1998gv}. 
In addition, the NNLO (i.e., chiral order $\nu=2$) Lagrangian $\mathcal L^{(2)}_{\pi N}$ contains a number of counterterms
LECs for the sake of renormalizing UV divergences arising from pion-loop diagrams. In this work, we shall exclude the 
explicit appearance of pion-loop within the TPE diagrams. Nevertheless, they enter implicitly {\it via} the proton's root 
mean squared (rms) charge radii taken as phenomenological input~\cite{Talukdar:2020aui} to fix these LECs. Hence, such LECs
that are directly not relevant to our calculations have been omitted in the expression for $\mathcal L^{(2)}_{\pi N}$, as 
represented by the ellipses. The other terms used in the chiral Lagrangian include
\begin{eqnarray}
\chi_+ \!\!&=&\!\! u^\dagger \chi u + u \chi^\dagger u\,, \quad
\tilde{\chi}_+ =  \chi_+ - \frac{\mathbb I}{2}{\rm Tr} (\chi_+)\,,
\nonumber\\
f^{+}_{\mu\nu} \!\!&=&\!\! u^\dagger \left(f^R_{\mu\nu}+v^{(s)}_{\mu\nu}\right) u 
+ u \left(f^L_{\mu\nu}+v^{(s)}_{\mu\nu}\right) u^\dagger 
\nonumber\\
&=& eF_{\mu \nu}(u {\mathscr{Q}} u^\dagger + u^\dagger{\mathscr{Q}} u)\,; \quad 
{\mathscr{Q}}=\frac{1}{2}({\mathbb I}+\tau_3)
\nonumber\\
f^{R}_{\mu\nu} \!\!&=&\!\! \partial_\mu r_\nu -\partial_\nu r_\mu - i[r_\mu,r_\nu]
=e\frac{\tau^3}{2} F_{\mu\nu}\,,
\nonumber\\
f^{L}_{\mu\nu} \!\!&=&\!\! \partial_\mu l_\nu -\partial_\nu l_\mu - i[l_\mu,l_\nu]
= e\frac{\tau^3}{2} F_{\mu\nu}\,, \quad{\rm and} 
\nonumber\\
v^{(s)}_{\mu\nu} \!\!&=&\!\! 
e\frac{\mathbb I}{2} \left(\partial_\mu A_\nu -\partial_\nu A_\mu\right)\,, 
\label{eq:L_source}
\end{eqnarray}
where the field tensor $f_{\mu \nu}^+$ represents a combination of external iso-scalar and iso-vector sources, namely, 
with $v_\mu^{(s)} = - \frac{e}{2}\mathbb I A_\mu$ being the iso-scalar source, and 
$l_\mu = r_\mu = - \frac{e}{2}\tau^3 A_\mu$, the left and right-handed iso-vector sources expressed in terms of the 
photon vector field $A_\mu$. Here, we have ignored all the sources that violate isospin invariance. Thus, in this limit 
$m_d \rightarrow m_u$ and $\tilde{\chi}_+ \rightarrow 0$, so the term proportional to $c_5$ does not contribute. 

\begin{figure}[tbp]
\centering
\includegraphics[scale=0.55]{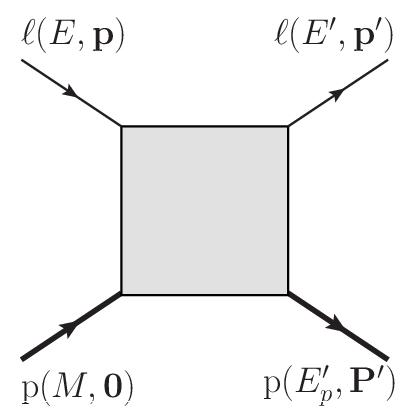}
\caption{\label{kinematics} The kinematics of $\ell$p $\to\ell$p scattering in the laboratory frame. The 
         square-shaded area represents all possible internal graphs contributing to the elastic scattering 
         process.}
        \label{fig:kinematics}
\end{figure}

To facilitate an easy comparison, we have kept the notations and conventions in this work in close analogy with those of 
Ref.~\cite{Talukdar:2019dko}. The choice of a laboratory frame (lab-frame) where the target proton is at rest, greatly 
simplifies our evaluations and kinematics, bearing in mind the MUSE kinematic region of our interest (cf. Table I of 
Ref.~\cite{Talukdar:2019dko}). In this work, we shall consider $Q_{\mu} =(p-p^\prime)_{\mu} = (p^\prime_p-p_p)_\mu$ as 
the four-momentum transfer for the elastic scattering process, with $p(p^\prime)$ and $p_p(p^\prime_p)$ as the incoming 
(outgoing) lepton four-momenta and off-shell or {\it residual} proton four-momentum, respectively (cf. 
Fig.~\ref{fig:kinematics}). Here, we wish to emphasize that terms like $v\cdot Q=E-E^\prime$ and $v\cdot p^\prime_p$ are
of order $\mathcal{O}(1/M)$, namely, with $P^{\mu}_p=Mv^\mu+p^\mu_p$ [where $p_p=(0,{\bf 0})$] and 
$P^{\prime\mu}_p=Mv^\mu+p^{\prime\mu}_p$, defined as the full relativistic four-momenta of the initial and final state 
proton in the lab-frame. Then it follows that $v\cdot Q=v\cdot p^\prime_p=E-E^\prime=-Q^2/(2M)$. Finally, for the sake 
of bookkeeping, we display the form of the proton propagator (with generic off-shell momentum $l$) up to 
${\mathcal O}(1/M^2)$, needed in the TPE loops diagrams:
\begin{eqnarray}
\label{eq:p_prop}
iS^{(1/M^2)}_{\rm full}(l) \!\!&=&\!\!
\frac{i}{v\cdot l+i0} + \frac{i}{2M}\left[1-\frac{l^2}{(v\cdot l +i0)^2}\right] 
\\
&&\!\! +\frac{i}{4M^2}\left[\frac{(v \cdot l)^3-l^2 (v \cdot l)}{(v\cdot l +i0)^2}\right] 
+ {\mathcal O}\left(M^{-3}\right)\,.
\nonumber
\end{eqnarray}

Next, we present the essential steps of our improved analytical evaluation of the various Feynman amplitudes contributing 
to the TPE up to NNLO accuracy invoking the kind of SPA approach as advocated by Maximon and Tjon~\cite{Maximon:2000hm}. 
For specific details regarding the SPA methodology and its adaptation in the context of the HB$\chi$PT framework, we refer 
the reader to the work of Talukdar {\it et al.}~\cite{Talukdar:2019dko}.
%%%%%%%%%%%%%%%%%%%%%%%%%%%%%%%%%%%%%%%%%%%%%%%%%%FIGURE-1%%%%%%%%%%%%%%%%%%%%%%%%%%%%%%%%%%%%%%%%%%%%%%%%%%%%%%
%%%%%%%%%%%%%%%%%%%%%%%%%%%%%%%%%%%%%%%%%%%%%%%%%%%%%%%%%%%%%%%%%%%%%%%%%%%%%%%%%%%%%%%%%%%%%%%%%%%%%%%%%%%%%%%%%

%%%%%%%%%%%%%%%%%%%%%%%%%%%%%%%%%%%%%%%%%%%%%%%%%%%%%%%%%%
\subsection{Leading order TPE contribution}
\label{sec:A}
\begin{figure}[h]
\centering
\includegraphics[scale=0.5]{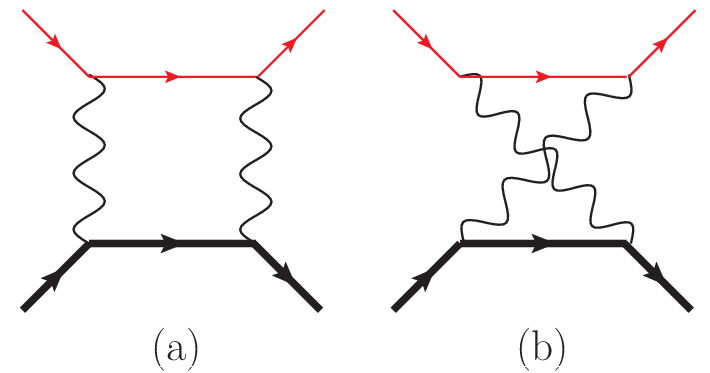}
\caption{The LO TPE box (a), and crossed-box (b) amplitudes of $\mathcal{O}(\alpha^2)$, consisting of 
         interactions from the $\mathcal L^{(0)}_{\pi N}$ chiral Lagrangian, and contributing to the 
         leading radiative corrections to the unpolarized lepton-proton elastic differential cross-section 
         of $\mathcal{O}(\alpha^3)$. The thick, thin, and wiggly lines denote the proton, lepton, and photon 
         propagators. }
\label{fig:LO_TPE}
\end{figure}

%%%%%%%%%%%%%%%%%%%%%%%%%%%%%%%%%%%%%%%%%%%%%%%%%%%%%%%%%%
Figure.~\ref{fig:LO_TPE} displays the so-called ``box" (planar) and ``crossed-box" (non-planar) TPE diagrams, (a) and (b), 
respectively, contributing at the chiral leading order (LO) amplitude in HB$\chi$PT, and can be expressed in the following 
4-point function (namely, with four internal propagators) loop-integral representations:
%%%%%%%%%%%%%%%%%%%%%%%%%%%%%%%%%%%%%%%%%%%%%%%%%%%%%%%%%%
\begin{widetext}
\begin{eqnarray}
{\mathcal{M}^{(a)}_{\rm box}} \!\!&=&\!\! e^4 \int \frac{{\rm d}^4 k}{(2\pi)^4 i}
\frac{\left[\Bar{u}(p^\prime)\gamma^{\mu}(\slashed{p}-\slashed{k}+m_l)\gamma^{\nu}u(p)\right]
\left[\chi^{\dagger}(p^\prime_p)v_{\mu}v_{\nu}\chi(p_p)\right]}{(k^2+i0)\, [(Q-k)^2+i0]\,(k^2-2k\cdot p+i0)\,
(v\cdot k+ v\cdot p_p +i0)} \,,
\\
%%%%%%%%%%%%%%%%%%%%%%%%%%%%%%%%%%%%%%%%%%%%%%%%%%%%%%%%%%
{\mathcal{M}^{(b)}_{\rm xbox}} \!\!&=&\!\! e^4 \int \frac{{\rm d}^4 k}{(2\pi)^4 i}
\frac{\left[\Bar{u}(p^\prime)\gamma^{\mu}(\slashed{p}-\slashed{k}+m_l)\gamma^{\nu}u(p)\right]
\left[\chi^{\dagger}(p^\prime_p)v_{\mu}v_{\nu}\chi(p_p)\right]}{(k^2+i0)\, [(Q-k)^2+i0]\,(k^2-2k\cdot p+i0)\, 
(-v\cdot k+v\cdot p^\prime_p +i0)}\,,
\label{eq:LO_TPE_ab}
\end{eqnarray}
\end{widetext}
%%%%%%%%%%%%%%%%%%%%%%%%%%%%%%%%%%%%%%%%%%%%%%%%%%%%%%%%%%

%%%%%%%%%%%%%%%%%%%%%%%%%%%%%%%%%%%%%%%%%%%%%%%%%%%%FIGURE-2%%%%%%%%%%%%%%%%%%%%%%%%%%%%%%%%%%%%%%%%%%%%%%%%%%%%
\begin{figure*}[tbp]
\centering
\includegraphics[scale=0.5]{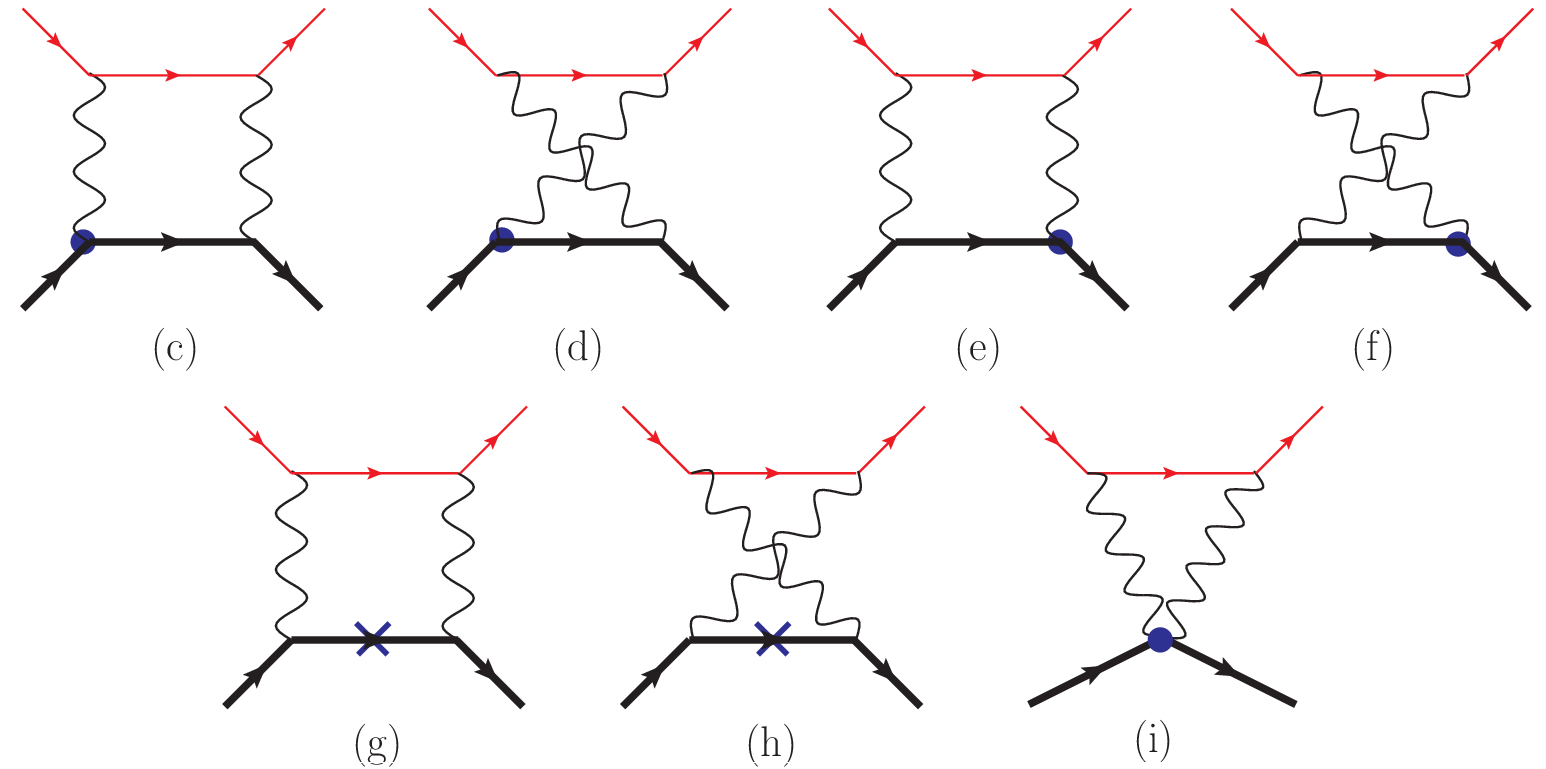}
\caption{The NLO TPE diagrams of $\mathcal{O}(\alpha^2/M)$ contributing to the $\mathcal{O}(\alpha^3/M)$ elastic 
         differential cross-section. The thick, thin, and wiggly lines denote the proton, lepton, and photon 
         propagators. The blobs and crosses denote the insertions of the NLO proton-photon vertices and 
         ${\mathcal O}(1/M)$  proton propagators, respectively. }
\label{fig:NLO_TPE}
\end{figure*}

%%%%%%%%%%%%%%%%%%%%%%%%%%%%%%%%%%%%%%%%%%%%%%%%%%%%%%%%%%%%%%%%%%%%%%%%%%%%%%%%%%%%%%%%%%%%%%%%%%%%%%%%%%%%%%%%%

\noindent where $u(p)$ and $\Bar{u}(p^\prime)$ are Dirac spinors for incoming and outgoing lepton, while $\chi(p_p)$ and
$\chi(p^\prime_p)$ represent the non-relativistic two-component Pauli spinors for the proton. It must be mentioned that the
above two LO TPE diagrams are not {\it strictly} LO in the sense that they also contain ${\mathcal O}(1/M)$ 
{\it kinematical recoil} terms proportional to $v\cdot Q$ or $v\cdot p^\prime_p$ which generate contributions commensurate 
with the {\it dynamical recoil} contributions arising from the ``NLO TPE" diagrams (c)-(i) (cf. Fig.~\ref{fig:NLO_TPE}). 
Likewise, the NLO diagrams generate kinematically suppressed term of ${\mathcal O}(1/M^2)$ commensurate with the dynamical 
contributions from NNLO TPE diagrams (j)-(v) (cf. Fig.~\ref{fig:NNLO_TPE}). Hence, care must be taken to correctly 
distinguish between the kinematically and dynamically suppressed subleading terms, all of which scale as 
${\mathcal O}(1/M^n)$. Thus, to extract the true {\it bona fide} LO TPE cross-section, the resulting ${\mathcal O}(1/M)$ 
terms from diagrams (a) and (b) must be relegated to the NLO part of the TPE 
amplitudes~\cite{Talukdar:2019dko,Choudhary:2023rsz}. Particularly, in Ref.~\cite{Talukdar:2019dko}, for a simplistic 
evaluation of the intricate four-point loop-integration, the authors took recourse to an expansion of the proton propagator
in $\mathcal{M}^{(b)}_{\rm xbox}$ in powers of $1/M$ due to the presence of the $v\cdot p^\prime_p$ term in the denominator 
and subsequently relocating the ${\mathcal O}(1/M)$ terms in the integrand to the NLO level by adding them together with the
genuine NLO TPE diagram (h) in Fig.~\ref{fig:NLO_TPE}. This methodology can certainly be justified for {\it regular} 
integrals leading to finite results. However, in the purview of existing IR-divergences, this simplification potentially 
undermines the accuracy of the results, as emphasized in the recent work on the exact TPE evaluations in HB$\chi$PT, 
Ref.~\cite{Choudhary:2023rsz}. Thus, in this work, we do not expand the proton propagators before evaluating the 
loop-integrals, and instead, follow the approach adopted in Ref.~\cite{Choudhary:2023rsz}. First, we remove the negative
sign in front of $v\cdot k$ by a shift of the integration variable $k\rightarrow -k+Q$ in $\mathcal{M}^{(b)}_{\rm xbox}$ 
leading to
\begin{widetext}
%%%%%%%%%%%%%%%%%%%%%%%%%%%%%%%%%%%%%%%%%%%%%%%%%%%%%%%%%%
\begin{eqnarray}
{\mathcal{M}^{(b)}_{\rm xbox}} = e^4 \int \frac{{\rm d}^4 k}{(2\pi)^4 i}
\frac{\left[\Bar{u}(p^\prime)\gamma^{\mu}(\slashed{p}+\slashed{k}-\slashed{Q}+m_l)\gamma^{\nu}u(p)\right]
\left[\chi^{\dagger}(p^\prime_p)v_{\mu}v_{\nu}\chi(p_p)\right]}{(k^2+i0)\, [(Q-k)^2+i0]\,(k^2+2k\cdot p^\prime+i0)\, 
(v\cdot k+v\cdot p_p +i0)} \,.
\end{eqnarray}
Next,  by invoking SPA (implied henceforth by the symbol $\stackrel{\gamma_{\rm soft}}{\leadsto}$), i.e., letting one of the 
exchange photons as soft (either $k=0$ or $k=Q$) in each of the two LO amplitudes~\cite{Talukdar:2019dko}, we obtain the 
following two 3-point amplitudes: 
\begin{eqnarray}
{\mathcal{M}^{(a)}_{\rm box}} &\stackrel{\gamma_{\rm soft}}{\leadsto}& 
-\,2e^2 (v\cdot p)\mathcal{M}_{\gamma}^{(0)}\int 
\frac{{\rm d}^4 k}{(2\pi)^4 i}\frac{1}{(k^2+i0)\,(k^2-2k\cdot p +i0)\, (v\cdot k +i0)}
\nonumber\\
&&-\,2e^2 (v\cdot p^\prime)\mathcal{M}_{\gamma}^{(0)}\int 
\frac{{\rm d}^4 k}{(2\pi)^4 i}\frac{1}{[(Q-k)^2+i0]\,(k^2-2k\cdot p +i0)\, (v\cdot k +i0)}\,, \quad \text{and}
\nonumber\\
{\mathcal{M}^{(b)}_{\rm xbox}}  &\stackrel{\gamma_{\rm soft}}{\leadsto}& 
-\,2e^2 (v\cdot p)\mathcal{M}_{\gamma}^{(0)} \int 
\frac{{\rm d}^4 k}{(2\pi)^4 i}\frac{1}{ [(Q-k)^2+i0]\,(k^2+2k\cdot p^\prime +i0)\, (v\cdot k +i0)}
\nonumber \\
&&-\,2e^2 (v\cdot p^\prime)\mathcal{M}_{\gamma}^{(0)} \int 
\frac{{\rm d}^4 k}{(2\pi)^4 i}\frac{1}{(k^2+i0)\, (k^2+2k\cdot p^\prime +i0)\, (v\cdot k +i0)}\,.
\end{eqnarray} 
%%%%%%%%%%%%%%%%%%%%%%%%%%%%%%%%%%%%%%%%%%%%%%%%%%%%%%%%%%
\end{widetext} 
Here, we used the fact that $v\cdot p_p=0$ (lab-frame). The resulting sum of the TPE amplitudes in the soft-photon 
approximation gets factorized into the LO Born amplitude $\mathcal{M}_{\gamma}^{(0)}$ times a $Q^2$-dependent function 
$f^{(ab)}_{\gamma\gamma}(Q^2)$, namely,
\begin{eqnarray}
\mathcal{M}_{\gamma \gamma}^{(ab)} \!\!&=&\!\! \mathcal{M}^{(a)}_{\rm box}+\mathcal{M}^{(b)}_{\rm xbox} 
\stackrel{\gamma_{\rm soft}}{\leadsto}  \mathcal{M}_{\gamma}^{(0)} f^{(ab)}_{\gamma\gamma}(Q^2)\,,
\end{eqnarray}
where
\begin{eqnarray*}
{\mathcal M}^{(0)}_\gamma \!\!&=&\!\! -\,
\frac{e^2}{Q^2} \big[\bar{u}(p^\prime)\gamma^\mu u(p) \big] 
\left[\chi^{\dagger}(p^\prime_p) v_\mu \chi(p_p)\right]\,,\quad \text{and}
\end{eqnarray*}

\vspace{-0.5cm}

{\small
\begin{eqnarray}
f^{(ab)}_{\gamma\gamma}(Q^2)\!\!&=&\!\! -\, 2e^2 \Big\{E I^{-}(p,0|1,0,1,1)+ E^\prime I^{-}(p,0|0,1,1,1) 
\nonumber\\
&&\!\!\! +\, E I^{+}(p^\prime,0|0,1,1,1)+ E^\prime I^{+}(p^\prime,0|1,0,1,1)\Big\}\,.
\nonumber\\
\label{eq:M0_LO}
\end{eqnarray}
}The relevant real parts of the 3-point loop-integrations $I^{\pm}$ contributing to the elastic cross-section were 
worked out in Ref.~\cite{Choudhary:2023rsz} up to ${\mathcal O}(1/M)$ accuracy. However, in this work, they are needed 
at the level of ${\mathcal O}(1/M^2)$ whose explicit expressions are provided in the appendix. It is notable that the 
functions $I^{-}(p,0|1,0,1,1)$ and $I^{+}(p^\prime,0|1,0,1,1)$ are IR-divergent, while $I^{-}(p,0|0,1,1,1)$ and 
$I^{+}(p^\prime,0|0,1,1,1)$ are finite. Moreover, it was shown in Ref.~\cite{Choudhary:2023rsz} that the {\it bona fide}
LO part [i.e., ${\mathcal O}(M^0)$] of such $I^{\pm}$ integrals are equal but of opposite signs, namely, 
\begin{eqnarray}
I^{-}(p,0|1,0,1,1)\!\!&\equiv&\!\!I^{(0)}(p,0|1,0,1,1)\,,
\nonumber\\
I^{+}(p^\prime,0|1,0,1,1)\!\!&=&\!\!-\,I^{(0)}(p,0|1,0,1,1)+ {\mathcal O}\left(M^{-1}\right)\,,
\nonumber\\
I^{-}(p,0|0,1,1,1)\!\!&=&\!\!I^{(0)}(p,0|0,1,1,1)+ {\mathcal O}\left(M^{-1}\right)\,,
\nonumber\\
 I^{+}(p^\prime,0|0,1,1,1)\!\!&=&\!\!-\,I^{(0)}(p,0|0,1,1,1)+ {\mathcal O}\left(M^{-1}\right)\,,\qquad\,
\end{eqnarray}
where the expressions for $I^{(0)}(p,0|1,0,1,1)$ and $I^{(0)}(p,0|0,1,1,1)$ are also provided in the appendix. 
Consequently, given that at the true LO in HB$\chi$PT, the incoming and outgoing lepton energies and velocities are the 
same, i.e., $E=E^\prime$ and $\beta=\beta^\prime$, respectively, (with $\beta=|{\bf p}|/E$ and 
$\beta^\prime=|{\bf p^\prime}|/E^\prime$), 
we find that at the true LO $f^{(0)}_{\gamma\gamma}$ completely vanishes in SPA, i.e.,
$f^{(0)}_{\gamma\gamma}(Q^2)\equiv \left[f^{(ab)}_{\gamma\gamma}(Q^2)\right]_{\rm LO}\!\!\!\!= 0$ as in 
Ref.~\cite{Talukdar:2019dko}. In other words, the ${\mathcal O}\left(M^{0}\right)$ IR-divergent terms arising from the 
genuine LO TPE diagrams (a) and (b) cancel exactly. However, the residual divergences of ${\mathcal O}(1/M)$ surviving 
must be considered commensurate with the subleading chiral order dynamical contributions from NLO or higher-order TPE 
amplitudes. The fractional TPE corrections to the ${\mathcal O}{(\alpha^2)}$ Born differential cross-section from the 
LO diagrams (a) and (b) are collected in the ${\mathcal O}{(\alpha)}$ term 
\begin{eqnarray}
\delta^{(ab)}_{\gamma \gamma}(Q^2) = 
\frac{2{\mathcal R}e\!\!\!\!\sum\limits_{spins}\!
\left[\mathcal{M}_{\gamma}^{(0)*}\mathcal{M}_{\gamma \gamma}^{(ab)}\right]}{\sum\limits_{spins} 
\left|\mathcal{M}_{\gamma}^{(0)}\right|^2}
\stackrel{\gamma_{\rm soft}}{\leadsto} 2{\mathcal R}e f^{(ab)}_{\gamma\gamma}(Q^2),
\nonumber\\
\label{eq:LOpart}
\end{eqnarray}
where the true LO contributions to the elastic cross-section exactly vanish, namely, $\delta^{(ab);(0)}_{\gamma \gamma}=0$. 
This is solely an artifact of the SPA approach which potentially spoils the HB$\chi$PT power counting. We know from the 
recent work of Choudhary {\it et al.}~\cite{Choudhary:2023rsz} that an exact analytical evaluation of the true LO diagrams 
in HB$\chi$PT leads to a non-zero ${\mathcal O}\left(M^{0}\right)$ result, namely,
\begin{eqnarray}
\delta^{\rm (0)}_{\gamma \gamma}(Q^2) =\pi \alpha \frac{\sqrt{-Q^2}}{2E}\left(\frac{1}{1+\frac{Q^2}{4E^2}}\right)\,.   
\label{eq:delta_LO}
\end{eqnarray} 
The result is analogous to the well-known McKinley-Feshbach contribution obtained using distorted Born-wave approximation to
the second-order in non-relativistic perturbation theory~\cite{McKinley:1948zz}. Thus, to restore the correct power counting 
we include the above {\it exact} result, Eq.~\eqref{eq:delta_LO}, as a part of the LO TPE contributions. Using the results of 
evaluating the 3-point functions $I^{\pm}$ to ${\mathcal O}(1/M^{2})$, where the IR-divergent loop-integrals are tackled by 
employing DR to isolate the IR singularities, we obtain the IR-finite fractional TPE contribution to the differential 
cross-section in the lab-frame:
\begin{widetext}
%%%%%%%%%%%%%%%%%%%%%%%%%%%%%%%%%%%%%%%%%%%%%%%%%%%%%%%%%%
\begin{eqnarray}
\overline{\delta^{(ab)}_{\gamma \gamma}}(Q^2) \!\!&=&\!\! \delta^{(ab)}_{\gamma \gamma}(Q^2) - \delta^{\rm (box)}_{\rm IR}(Q^2) 
\stackrel{\gamma^*_{\rm soft}}{\leadsto} \pi \alpha \frac{\sqrt{-Q^2}}{2E}\left(\frac{1}{1+\frac{Q^2}{4E^2}}\right)
- \frac{2\alpha Q^2}{\pi M E \beta^2} \Bigg[1 -\frac{\pi^2}{24}\left(\beta-\frac{1}{\beta}\right)
- \ln\sqrt{\frac{2\beta}{1-\beta}} 
\nonumber\\
&&\!\! +\, \frac{1}{2}\left(1-\frac{1}{\beta}\right) \ln\sqrt{\frac{1+\beta}{1-\beta}} 
+ \frac{1}{4}\left(\beta-\frac{1}{\beta}\right)\ln^2\sqrt{\frac{1+\beta}{1-\beta}} 
+ \frac{1}{4}\left(\beta-\frac{1}{\beta}\right){\rm Li}_2\bigg(\frac{1+\beta}{1-\beta}\bigg)
- \ln\sqrt{-\frac{Q^2}{2M E\beta }} 
\nonumber\\
&&\!\!\times\, \left\{1 +\left(\beta-\frac{1}{\beta}\right)\ln\sqrt{\frac{1+\beta}{1-\beta}}\,\right\}\Bigg] 
- \frac{\alpha Q^4}{16 \pi M^2 E^2 \beta^5} \Bigg[-\pi^2(1-\beta^2)+8\beta^3-8\beta(3+2\beta^2) \ln{\sqrt{\frac{2\beta}{1-\beta}}}
\nonumber\\
&&\!\! + 6(1-\beta^2)\,{\rm Li}_2\!\!\left(\frac{1+\beta}{1-\beta}\right)  
+6 \left(1-\beta^2\right)\ln^2\sqrt{\frac{1+\beta}{1-\beta}} +4(7+3\beta+2\beta^3)\ln{\sqrt{\frac{1+\beta}{1-\beta}}}
-4 \ln\sqrt{-\frac{Q^2}{2ME\beta}}
\nonumber\\
&&\!\! \times\,\Bigg\{2\beta (3+2\beta^2)+6(1-\beta^2)\ln{\sqrt{\frac{1+\beta}{1-\beta}}}\,\Bigg\}\,\Bigg]
+\mathcal{O}\left(\frac{1}{M^3}\right)\,. 
\label{eq:delta_ab}
\end{eqnarray}
The $\rm{Li}_2$ in the equation denotes the standard di-logarithm or Spence function [cf. Eq.~\eqref{eq:Li2}]. In SPA the 
residual IR-divergence from Lo diagrams (a) and (b) are of $\mathcal{O}(1/M)$ and given by  
\begin{eqnarray}
\delta^{\rm (box)}_{\rm IR}(Q^2) \!\!&=&\!\! 
\frac{\alpha}{\pi\beta}\left[\frac{1}{\epsilon}-\gamma_E+\ln\bigg(\frac{4\pi \mu^2}{m_l^2}\bigg) \right]
\left\{\ln \sqrt{\frac{1+\beta}{1-\beta}}-\frac{\beta}{\beta^\prime}
\ln \sqrt{\frac{1+\beta^\prime}{1-\beta^\prime}}\,  \right\}
\nonumber\\
&=&\!\! -\,\frac{\alpha Q^2}{2 \pi M E \beta^2}
\left[\frac{1}{\epsilon}-\gamma_E+\ln\bigg(\frac{4\pi \mu^2}{m_l^2}\bigg) \right] 
\left\{1+\bigg(\beta-\frac{1}{\beta}\bigg) \ln \sqrt{\frac{1+\beta}{1-\beta}}\, \right\} + 
\nonumber\\
&&\!\! +\,\frac{3\alpha Q^4}{8\pi M^2 E^2 \beta^4}
\left[\frac{1}{\epsilon}-\gamma_E+\ln\bigg(\frac{4\pi \mu^2}{m_l^2}\bigg) \right]
\left\{1-\frac{2}{3}\beta^2 +\bigg(\beta-\frac{1}{\beta}\bigg) \ln \sqrt{\frac{1+\beta}{1-\beta}}\,\right\}  
+ \mathcal{O}\left(\frac{1}{M^3}\right)\,,
\label{eq:delta_IR}
\end{eqnarray} 
%%%%%%%%%%%%%%%%%%%%%%%%%%%%%%%%%%%%%%%%%%%%%%%%%%%%%%%%%%
\end{widetext}
where $\epsilon=(4-D)/2<0$. Notably, the above-mentioned singular term $\delta^{\rm (box)}_{\rm IR}(Q^2)$ is identical to
the one obtained in the exact TPE evaluation of Choudhary {\it et al.}~\cite{Choudhary:2023rsz}. It will ultimately cancel 
the IR-divergent counterparts from charge-odd soft-photon ($\gamma^*_{\rm soft}$) bremsstrahlung process 
$\delta^{\rm (soft)}_{\rm IR}$, viz., those singular terms originating from the interference process between the LO 
lepton-bremsstrahlung and LO proton-bremsstrahlung diagrams contributing to the elastic cross-section, such that, 
$\delta^{\rm (soft)}_{\rm IR}+\delta^{\rm (box)}_{\rm IR}=0$~\cite{Bhoomika24}. Furthermore, it is notable that the
singularities from diagrams (a) and (b) are the only sources of IR-divergence in the TPE contributions. No 
IR-divergent contribution is generated from the subleading-order TPE diagrams. As already mentioned in the introduction, 
unlike Refs.~\cite{Talukdar:2019dko,Talukdar:2020aui}, we keep the {\it scheme-dependent} IR-divergent terms to be 
subtracted precisely as given in Eq.~\eqref{eq:delta_IR}, without re-adjustments with the finite TPE part unlike in 
Refs.~\cite{Talukdar:2019dko,Talukdar:2020aui}.

%%%%%%%%%%%%%%%%%%%%%%%%%%%%%%%%%%%%%%%%%%%%%%%%%%%FIGURE-3%%%%%%%%%%%%%%%%%%%%%%%%%%%%%%%%%%%%%%%%%%%%%%%%%%%%%
%%%%%%%%%%%%%%%%%%%%%%%%%%%%%%%%%%%%%%%%%%%%%%%%%%%%%%%%%%%%%%%%%%%%%%%%%%%%%%%%%%%%%%%%%%%%%%%%%%%%%%%%%%%%%%%%%

%%%%%%%%%%%%%%%%%%%%%%%%%%%%%%%%%%%%%%%%%%%%%%%%%%%%%%%%%%
\subsection{Next-to-leading order TPE contribution}
%%%%%%%%%%%%%%%%%%%%%%%%%%%%%%%%%%%%%%%%%%%%%%%%%%%%%%%%%%
We now turn to the evaluation of the seven NLO TPE diagrams (c)-(i), displayed in Fig.~\ref{fig:NLO_TPE}. Each amplitude 
either consists of one insertion of proton-photon interaction vertex arising from the NLO Lagrangian 
${\mathcal L}^{(1)}_{\pi N}$ or from the NLO proton propagator [i.e., the $\mathcal{O}(1/M)$ term in Eq.~\eqref{eq:p_prop}].
The TPE amplitudes are given by the following integral representations~\cite{Choudhary:2023rsz}:
\begin{widetext}
%%%%%%%%%%%%%%%%%%%%%%%%%%%%%%%%%%%%%%%%%%%%%%%%%%%%%%%%%%
\begin{eqnarray}
    {\mathcal{M}^{(c)}_{\rm box}} \!\!&=&\!\! \frac{e^4}{2 M} 
    \int \frac{{\rm d}^4  k}{(2\pi)^4 i}\frac{\left[\Bar{u}(p^\prime)\gamma^{\mu}(\slashed{p}-\slashed{k}+m_l)\gamma^{\nu}u(p)\right]\,
    \left[\chi^{\dagger}(p_p^\prime)\Big\{v_{\mu} (k+2p_p)_{\nu}- v_{\mu} v_{\nu} [v\cdot(k+2p_p)]\Big\}\chi(p_p)\right]}{(k^2+i0)\, 
    [(Q-k)^2+i0]\,(k^2-2k\cdot p+i0)\, (v\cdot k +v\cdot p_p +i0)} \,,\qquad\,
\\
%%%%%%%%%%%%%%%%%%%%%%%%%%%%%%%%%%%%%%%%%%%%%%%%%%%%%%%%%%
    {\mathcal{M}^{(d)}_{\rm xbox}} \!\!&=&\!\! \frac{e^4}{2M} 
    \int \frac{{\rm d}^4 k}{(2\pi)^4 i}\frac{\left[\Bar{u}(p^\prime)\gamma^{\mu}(\slashed{p}-\slashed{k}+m_l)\gamma^{\nu}u(p)\right]\,
    \left[\chi^{\dagger}(p_p^\prime)\Big\{v_{\nu}(p_p+p^\prime_p-k)_{\mu}-v_{\mu}v_{\nu} [v\cdot(p_p+p_p^\prime-k)]\Big\}\chi(p_p)\right]}{(k^2+i0)\,
    [(Q-k)^2+i0]\,(k^2-2k\cdot p+i0)\, (-v\cdot k +v\cdot p_p^\prime +i0)} \,,\qquad\,
\\
%%%%%%%%%%%%%%%%%%%%%%%%%%%%%%%%%%%%%%%%%%%%%%%%%%%%%%%%%%
    {\mathcal{M}^{(e)}_{\rm box}} \!\!&=&\!\! \frac{e^4}{2M} 
    \int \frac{{\rm d}^4 k}{(2\pi)^4 i}\frac{\left[\Bar{u}(p^\prime)\gamma^{\mu}(\slashed{p}-\slashed{k}+m_l)\gamma^{\nu}u(p)\right]\,
    \left[\chi^{\dagger}(p^\prime_p)\Big\{v_{\nu}(p_p+p^\prime_p+k)_{\mu}-v_{\mu}v_{\nu} [v\cdot(p_p+p^\prime_p+k)]\Big\}\chi(p_p)\right]}
    {(k^2+i0)\, [(Q-k)^2+i0]\,(k^2-2k\cdot p+i0)\, (v\cdot k +v\cdot p_p +i0)} \,,\qquad\,
\\
%%%%%%%%%%%%%%%%%%%%%%%%%%%%%%%%%%%%%%%%%%%%%%%%%%%%%%%%%%
    {\mathcal{M}^{(f)}_{\rm xbox}} \!\!&=&\!\! \frac{e^4}{2M} 
    \int \frac{{\rm d}^4 k}{(2\pi)^4 i}\frac{\left[\Bar{u}(p^\prime)\gamma^{\mu}(\slashed{p}-\slashed{k}+m_l)\gamma^{\nu}u(p)\right]\,
    \left[\chi^{\dagger}(p^\prime_p)\Big\{v_{\mu} (2p^\prime_p-k)_{\nu}-v_{\mu}v_{\nu} [v\cdot(2p^\prime_p-k)]\Big\}\chi(p_p)\right]}
    {(k^2+i0)\, [(Q-k)^2+i0]\,(k^2-2k\cdot p+i0)\, (-v\cdot k +v\cdot p^\prime_p +i0)} \,,\qquad\,
\\
%%%%%%%%%%%%%%%%%%%%%%%%%%%%%%%%%%%%%%%%%%%%%%%%%%%%%%%%%%
    {\mathcal{M}^{(g)}_{\rm box}} \!\!&=&\!\! \frac{e^4}{2M} 
    \int \frac{{\rm d}^4 k}{(2\pi)^4 i}\frac{\left[\Bar{u}(p^\prime)\gamma^{\mu}(\slashed{p}-\slashed{k}+m_l)\gamma^{\nu}u(p)\right]\,
    \left[\chi^{\dagger}(p^\prime_p)v_{\mu} v_{\nu} \chi(p_p)\right]}{(k^2+i0)\, [(Q-k)^2+i0]\, (k^2-2k\cdot p+i0)} 
    \left(1-\frac{(p_p+k)^2}{(v\cdot p_p +v\cdot k +i0)^2}\right)\,,\qquad\,
\\
%%%%%%%%%%%%%%%%%%%%%%%%%%%%%%%%%%%%%%%%%%%%%%%%%%%%%%%%%%
    {\mathcal{M}^{(h)}_{\rm xbox}} \!\!&=&\!\! \frac{e^4}{2M} 
    \int \frac{{\rm d}^4 k}{(2\pi)^4 i}\frac{\left[\Bar{u}(p^\prime)\gamma^{\mu}(\slashed{p}-\slashed{k}+m_l)\gamma^{\nu}u(p)\right]\,
    \left[\chi^{\dagger}(p^\prime_p)v_{\mu} v_{\nu} \chi(p_p)\right]}{(k^2+i0)\, [(Q-k)^2+i0]\,(k^2-2k\cdot p+i0)} 
    \left(1-\frac{(p^\prime_p-k)^2}{(v\cdot p^\prime_p -v\cdot k +i0)^2}\right) \,,\qquad\,
\\
%%%%%%%%%%%%%%%%%%%%%%%%%%%%%%%%%%%%%%%%%%%%%%%%%%%%%%%%%%
    {\mathcal{M}^{(i)}_{\rm seagull}} \!\!&=&\!\! \frac{e^4}{M} 
    \int \frac{{\rm d}^4 k}{(2\pi)^4 i}\frac{\left[\Bar{u}(p^\prime)\gamma^{\mu}(\slashed{p}-\slashed{k}+m_l)\gamma^{\nu}u(p)\right]\,
    \left[\chi^{\dagger}(p^\prime_p)(v_{\mu} v_{\nu}-g_{\mu \nu} )\chi(p_p)\right]}{(k^2+i0)\, [(Q-k)^2+i0]\,(k^2-2k\cdot p+i0)} \,.\qquad\,
\end{eqnarray}
%%%%%%%%%%%%%%%%%%%%%%%%%%%%%%%%%%%%%%%%%%%%%%%%%%%%%%%%%%
\end{widetext}
The following comments are in order: 
\begin{itemize}
\item As in Ref.~\cite{Talukdar:2019dko}, some of the $v_\mu v_\nu$ terms cancel between $\mathcal{M}^{(c)}_{\rm box}$, 
$\mathcal{M}^{(f)}_{\rm box}$, $\mathcal{M}^{(g)}_{\rm box}$, $\mathcal{M}^{(h)}_{\rm xbox}$ and 
$\mathcal{M}^{(i)}_{\rm seagull}$.
\item The expressions for $\mathcal{M}^{(h)}_{\rm xbox}$ is written differently from that of Ref.~\cite{Talukdar:2019dko}, 
as they do not include the ${\mathcal O}(1/M)$ kinematically suppressed terms arising from the expansion of the LO TPE 
diagram (b). In Ref.~\cite{Talukdar:2019dko} they were omitted in the LO contributions by relocating them alongside the 
NLO TPE diagram (h).
\item Unlike Ref.~\cite{Talukdar:2019dko}, we here preserve the kinetic energy terms such as $v\cdot p_p$ and $v\cdot p^\prime_p$ 
in the propagator denominators. Hence, $\mathcal{M}^{(g)}_{\rm box}$ and $\mathcal{M}^{(h)}_{\rm xbox}$ do not mutually yield a 
vanishing contribution.
\item Since $v\cdot p^\prime_p=v\cdot Q \sim \mathcal{O}(1/M)$, in the NLO analysis of Ref.~\cite{Talukdar:2019dko} such terms 
were neglected in the numerators of the NLO TPE diagrams since they yielded $\mathcal{O}(1/M^2)$ terms which are beyond the 
intended accuracy. However, in this TPE analysis, since we intend to include the proton's structure effects at NNLO which yield 
$\mathcal{O}(1/M^2)$ terms, it is important to retain all $\mathcal{O}(1/M^2)$ kinematically suppressed terms arising from the 
NLO, as well as the LO TPE analysis.
\end{itemize} 
After shifting the integration variable $k\rightarrow -k+Q$ in the NLO crossed-box diagrams and including the partial 
cancellation between $v_{\mu}v_{\nu}$ terms, with $v\cdot p_p=0$, yield the following {\it reduced} NLO amplitudes: 
\begin{widetext}
%%%%%%%%%%%%%%%%%%%%%%%%%%%%%%%%%%%%%%%%%%%%%%%%%%%%%%%%%%
\begin{eqnarray}
    {\widetilde{\mathcal{M}}^{(c)}_{\rm box}} \!\!&=&\!\! \frac{e^4}{2 M} \int \frac{{\rm d}^4 k}{(2\pi)^4 i}
    \frac{[\Bar{u}(p^\prime)\gamma^{\mu}(\slashed{p}-\slashed{k}+m_l)\gamma^{\nu}u(p)]
    \Big[\chi^{\dagger}(p^\prime_p)v_{\mu} k_{\nu}\chi(p_p)\Big]}{(k^2+i0)\, [(Q-k)^2+i0]\, (k^2-2k\cdot p+i0)\, (v\cdot k +i0)} \,,
\\
%%%%%%%%%%%%%%%%%%%%%%%%%%%%%%%%%%%%%%%%%%%%%%%%%%%%%%%%%%
    {\widetilde{\mathcal{M}}^{(d)}_{\rm xbox}} \!\!&=&\!\! \frac{e^4}{2M} \int \frac{{\rm d}^4 k}{(2\pi)^4 i}
    \frac{[\Bar{u}(p^\prime)\gamma^{\mu}(\slashed{p}+\slashed{k}-\slashed{Q}+m_l)\gamma^{\nu}u(p)]
    \Big[\chi^{\dagger}(p^\prime_p)v_{\nu} k_{\mu}\chi(p_p)\Big]}{(k^2+i0)\, [(Q-k)^2+i0]\, (k^2+2k\cdot p^\prime+i0)\, (v\cdot k+i0)} ,
\\
%%%%%%%%%%%%%%%%%%%%%%%%%%%%%%%%%%%%%%%%%%%%%%%%%%%%%%%%%%
    {\widetilde{\mathcal{M}}^{(e)}_{\rm box}} \!\!&=&\!\! \frac{e^4}{2M} \int \frac{{\rm d}^4 k}{(2\pi)^4 i}
    \frac{[\Bar{u}(p^\prime)\gamma^{\mu}(\slashed{p}-\slashed{k}+m_l)\gamma^{\nu}u(p)]
    \Big[\chi^{\dagger}(p^\prime_p)v_{\nu}(Q+k)_{\mu}\chi(p_p)\Big]}{(k^2+i0)\, [(Q-k)^2+i0]\, (k^2-2k\cdot p+i0)\, (v\cdot k +i0)} \,,
\\
%%%%%%%%%%%%%%%%%%%%%%%%%%%%%%%%%%%%%%%%%%%%%%%%%%%%%%%%%%
    {\widetilde{\mathcal{M}}^{(f)}_{\rm xbox}} \!\!&=&\!\! \frac{e^4}{2M} \int \frac{{\rm d}^4 k}{(2\pi)^4 i}
    \frac{[\Bar{u}(p^\prime)\gamma^{\mu}(\slashed{p}+\slashed{k}-\slashed{Q}+m_l)\gamma^{\nu}u(p)]
    \Big[\chi^{\dagger}(p^\prime_p)v_{\mu} (k+Q)_{\nu}\chi(p_p)\Big]}{(k^2+i0)\, [(Q-k)^2+i0]\, (k^2+2k\cdot p^\prime+i0)\, (v\cdot k +i0)} \,,
\\
%%%%%%%%%%%%%%%%%%%%%%%%%%%%%%%%%%%%%%%%%%%%%%%%%%%%%%%%%%
    {\widetilde{\mathcal{M}}^{(g)}_{\rm box}} \!\!&=&\!\! \frac{e^4}{2M} \int \frac{{\rm d}^4 k}{(2\pi)^4 i}
    \frac{[\Bar{u}(p^\prime)\gamma^{\mu}(\slashed{p}-\slashed{k}+m_l)\gamma^{\nu}u(p)]
    \Big[\chi^{\dagger}(p^\prime_p)v_{\mu} v_{\nu} \chi(p_p)\Big]}{(k^2+i0)\, [(Q-k)^2+i0]\, (k^2-2k\cdot p+i0)} 
    \left(-\frac{k^2}{(v\cdot k +i0)^2}\right) \,,
\\
%%%%%%%%%%%%%%%%%%%%%%%%%%%%%%%%%%%%%%%%%%%%%%%%%%%%%%%%%%
    {\widetilde{\mathcal{M}}^{(h)}_{\rm xbox}} \!\!&=&\!\! \frac{e^4}{2M} \int \frac{{\rm d}^4 k}{(2\pi)^4 i}
    \frac{[\Bar{u}(p^\prime)\gamma^{\mu}(\slashed{p}+\slashed{k}-\slashed{Q}+m_l)\gamma^{\nu}u(p)]
    \Big[\chi^{\dagger}(p^\prime_p)v_{\mu} v_{\nu} \chi(p_p)\Big] }{(k^2+i0)\, [(Q-k)^2+i0]\, (k^2+2k\cdot p^\prime+i0)}
    \left(-\frac{k^2}{(v\cdot k +i0)^2}\right) \,,
\\
%%%%%%%%%%%%%%%%%%%%%%%%%%%%%%%%%%%%%%%%%%%%%%%%%%%%%%%%%%
    {\widetilde{\mathcal{M}}^{(i)}_{\rm seagull}} \!\!&=&\!\! -\,\frac{e^4}{M} \int \frac{{\rm d}^4 k}{(2\pi)^4 i}
    \frac{[\Bar{u}(p^\prime)\gamma^{\mu}(\slashed{p}-\slashed{k}+m_l)\gamma_{\mu}u(p)]
    \Big[\chi^{\dagger}(p^\prime_p)\chi(p_p)\Big]}{(k^2+i0)\, [(Q-k)^2+i0]\, (k^2-2k\cdot p+i0) } \,.
\end{eqnarray}
%%%%%%%%%%%%%%%%%%%%%%%%%%%%%%%%%%%%%%%%%%%%%%%%%%%%%%%%%%
\end{widetext}
The residual part of the {\it seagull} amplitude ${\widetilde{\mathcal{M}}^{(i)}_{\rm seagull}}$ is already a 3-point integral, 
and thus, evaluated exactly. In fact, using SPA for the 3-point amplitudes like the seagull diagram leads to unphysical UV 
divergence, which should be absent for all one-loop TPE diagrams, without the inclusion of additional pion-loops. Hence, for 
the NLO diagrams, SPA is limited to the evaluation of the box and crossed amplitudes (c)-(h) to reduce the complicated 4-point
functions into 3-point functions. As seen in the ensuing analysis, all contributions are IR/UV finite at NLO. Thus, after 
invoking SPA we find that the amplitudes either factorize into the LO Born amplitude $\mathcal{M}_\gamma^{(0)}$, 
Eq.~\eqref{eq:M0_LO}, or into the amplitude 
\begin{eqnarray}
{\mathcal N}^{(1)}_\gamma=-\frac{e^2}{MQ^2} \big[\bar{u}(p^\prime)\gamma^\mu u(p) \big] 
\left[\chi^{\dagger}(p^\prime_p) Q_\mu \chi(p_p)\right]\,,
\label{eq:N1_NLO}
\end{eqnarray} 
which form a part of the NLO correction to the LO Born amplitude ${\mathcal M}^{(1)}_\gamma$ for elastic lepton-proton scattering 
in HB$\chi$PT expressed in the lab-frame, i.e., with $p^\prime_p=Q$:
\begin{widetext}
%%%%%%%%%%%%%%%%%%%%%%%%%%%%%%%%%%%%%%%%%%%%%%%%%%%%%%%%%% 
\begin{eqnarray}
{\mathcal M}^{(1)}_\gamma = \frac{1}{2}{\mathcal N}^{(1)}_\gamma + \frac{e^2}{2MQ^2} \big[\bar{u}(p^\prime)\gamma^\mu u(p) \big] 
\Big[\chi^{\dagger}(p^\prime_p)\Big\{v_\mu (v\cdot Q)-(2+\kappa_s+\kappa_v)[S_\mu,S\cdot Q]\Big\}\chi(p_p)\Big]\,.
\label{eq:M1_NLO}
\end{eqnarray} 
Thus, the NLO box and crossed-box TPE amplitudes reduce in SPA from the original 4-point to the 3-point functions: 
%%%%%%%%%%%%%%%%%%%%%%%%%%%%%%%%%%%%%%%%%%%%%%%%%%%%%%%%%%
\begin{eqnarray}
    {\widetilde{\mathcal{M}}^{(c)}_{\rm box}} 
    &\stackrel{\gamma_{\rm soft}}{\leadsto}& -\,e^2E^\prime {\mathcal N}^{(1)}_\gamma \int 
    \frac{{\rm d}^4 k}{(2\pi)^4 i}\frac{1}{[(Q-k)^2+i0](k^2-2k\cdot p+i0) (v\cdot k +i0)} \,,
    \label{ampc}
\\
%%%%%%%%%%%%%%%%%%%%%%%%%%%%%%%%%%%%%%%%%%%%%%%%%%%%%%%%%%
    {\widetilde{\mathcal{M}}^{(d)}_{\rm xbox}} 
    &\stackrel{\gamma_{\rm soft}}{\leadsto}& -\,e^2E {\mathcal N}^{(1)}_\gamma \int 
    \frac{{\rm d}^4 k}{(2\pi)^4 i}\frac{1}{[(Q-k)^2+i0]\, (k^2+2k\cdot p^\prime+i0)\, (v\cdot k +i0)} \,,
    \label{ampd}
\\
%%%%%%%%%%%%%%%%%%%%%%%%%%%%%%%%%%%%%%%%%%%%%%%%%%%%%%%%%%
    {\widetilde{\mathcal{M}}^{(e)}_{\rm box}} 
    &\stackrel{\gamma_{\rm soft}}{\leadsto}& -\,e^2E {\mathcal N}^{(1)}_\gamma \int 
    \frac{{\rm d}^4 k}{(2\pi)^4 i}\frac{1}{(k^2+i0)\, (k^2-2k\cdot p+i0)\, (v\cdot k +i0)}
\nonumber\\
    &&+\,\frac{e^2 Q^2}{M} \mathcal{M}_{\gamma}^{(0)}\int \frac{{\rm d}^4 k}{(2\pi)^4 i}
    \frac{1}{[(Q-k)^2+i0]\, (k^2-2k\cdot p+i0)\, (v\cdot k +i0)} \,,
    \label{ampe}
\\
%%%%%%%%%%%%%%%%%%%%%%%%%%%%%%%%%%%%%%%%%%%%%%%%%%%%%%%%%%
    {\widetilde{\mathcal{M}}^{(f)}_{\rm xbox}} 
    &\stackrel{\gamma_{\rm soft}}{\leadsto}& -\,e^2E^\prime {\mathcal N}^{(1)}_\gamma \int 
    \frac{{\rm d}^4 k}{(2\pi)^4 i}\frac{1}{(k^2+i0)\, (k^2+2k\cdot p^\prime+i0)\, (v\cdot k+i0)}
\nonumber\\
    &&-\,\frac{e^2Q^2}{M} \mathcal{M}_{\gamma}^{(0)}\int \frac{{\rm d}^4 k}{(2\pi)^4 i}
    \frac{1}{[(Q-k)^2+i0]\, (k^2+2k\cdot p^\prime+i0)\, (v\cdot k +i0)} \,,
    \label{ampf}
\end{eqnarray}
%%%%%%%%%%%%%%%%%%%%%%%%%%%%%%%%%%%%%%%%%%%%%%%%%%%%%%%%%%
\begin{eqnarray}
    {\widetilde{\mathcal{M}}^{(g)}_{\rm box}} 
    &\stackrel{\gamma_{\rm soft}}{\leadsto}& \frac{e^2E^\prime Q^2}{M} \mathcal{M}_{\gamma}^{(0)}\int 
    \frac{{\rm d}^4 k}{(2\pi)^4 i}\frac{1}{[(Q-k)^2+i0]\, (k^2-2k\cdot p+i0)\, (v\cdot k +i0)^2} \,,
\\
%%%%%%%%%%%%%%%%%%%%%%%%%%%%%%%%%%%%%%%%%%%%%%%%%%%%%%%%%%
    {\widetilde{\mathcal{M}}^{(h)}_{\rm xbox}} 
    &\stackrel{\gamma_{\rm soft}}{\leadsto}& \frac{e^2EQ^2}{M} \mathcal{M}_{\gamma}^{(0)}\int 
    \frac{{\rm d}^4 k}{(2\pi)^4 i}\frac{1}{[(Q-k)^2+i0]\, (k^2+2k\cdot p^\prime+i0)\, (v\cdot k +i0)^2} \,.
\end{eqnarray}
%%%%%%%%%%%%%%%%%%%%%%%%%%%%%%%%%%%%%%%%%%%%%%%%%%%%%%%%%%
\end{widetext} 
Since we go beyond the NLO evaluation of Ref.~\cite{Talukdar:2019dko} by including NNLO corrections, we refrain from 
substituting $E=E^\prime$ and $\beta=\beta^\prime$. However, we find it legitimate to use this kind of approximation to 
evaluate the genuine NNLO diagrams. Unlike Ref.~\cite{Talukdar:2019dko}, the sum of the TPE amplitudes (g) and (h) do 
not exactly cancel in our case and rather yield residual terms of ${\mathcal O}(1/M^2)$. In other words,
$\widetilde{\mathcal{M}}^{(g)}_{\rm box}+\widetilde{\mathcal{M}}^{(h)}_{\rm xbox} \sim {\mathcal O}(1/M^2)$, which is 
formally commensurate with the dynamical contributions from the NNLO TPE diagrams. Thus, the sum of the NLO amplitudes 
is summarized as 
\begin{eqnarray}
\mathcal{M}^{(cde\cdots i)}_{\gamma \gamma} \!\!&=&\!\! \widetilde{\mathcal{M}}_{\rm box}^{(c)} 
+ \widetilde{\mathcal{M}}_{\rm xbox}^{(d)} + \widetilde{\mathcal{M}}_{\rm box}^{(e)} + \widetilde{\mathcal{M}}_{\rm xbox}^{(f)} 
\nonumber \\
&&\!\! +\, \widetilde{\mathcal{M}}_{\rm box}^{(g)}  + \widetilde{\mathcal{M}}_{\rm xbox}^{(h)} 
+ {\widetilde{\mathcal{M}}^{(i)}_{\rm seagull}}
\nonumber\\
&\stackrel{\gamma_{\rm soft}}{\leadsto}& \mathcal{M}_{\gamma}^{(0)} f^{(cde\cdots h)}_{\gamma\gamma}(Q^2) 
+ {\widetilde{\mathcal{M}}^{(i)}_{\rm seagull}} \,, 
\end{eqnarray}
where
\begin{widetext}
%%%%%%%%%%%%%%%%%%%%%%%%%%%%%%%%%%%%%%%%%%%%%%%%%%%%%%%%%%
\begin{eqnarray}
\label{eq:ffNLO}
f^{(cde\cdots h)}_{\gamma \gamma}(Q^2) \!\!&=&\!\! -\,\frac{e^2Q^2}{M}\bigg\{- I^{-}(p,0|0,1,1,1) 
+  I^{+}(p^\prime,0|0,1,1,1) - E^\prime I^{-}(p,0|0,1,1,2) - E I^{+}(p^\prime,0|0,1,1,2) \bigg\}
\\
&&\!\! -\,\frac{e^2{\mathcal M}^{(0)*}_{\gamma}{\mathcal N}^{(1)}_{\gamma}}{\left|{\mathcal M}^{(0)}_{\gamma}\right|^2}
\bigg\{E^\prime I^{-}(p,0|0,1,1,1)+E I^{+}(p^\prime,0|0,1,1,1)+ EI^{-}(p,0|1,0,1,1) + E^\prime I^{+}(p,0|1,0,1,1)\bigg\}\,.
\nonumber
\end{eqnarray}
%%%%%%%%%%%%%%%%%%%%%%%%%%%%%%%%%%%%%%%%%%%%%%%%%%%%%%%%%%
\end{widetext}
Here we use the same definitions of the loop-integrals $I^{-}(p,0|1,0,1,1)$, $I^{-}(p,0|0,1,1,1)$, $I^{+}(p^\prime,0|0,1,1,1)$, 
$I^{+}(p^\prime,0|1,0,1,1)$, $I^{-}(p,0|0,1,1,2)$ and $I^{+}(p^\prime,0|0,1,1,2)$, as in Ref.~\cite{Choudhary:2023rsz}. The 
analytical expressions of these integrals evaluated to ${\mathcal O}(1/M^2)$ are displayed in the appendix. It is notable that 
even though the above amplitude contains IR-divergent integrals $I^{-}(p,0|1,0,1,1)$ and $I^{+}(p^\prime,0|1,0,1,1)$, in effect 
they do not contribute to the cross-section. The reason is that the pre-factor 
${\mathcal M}^{(0)*}_{\gamma}{\mathcal N}^{(1)}_{\gamma}$ causes the entire expression in the second-line of 
Eq.~\eqref{eq:ffNLO} to vanish identically when evaluating the cross-section, namely,
\begin{equation*}
\sum\limits_{spins}\left[{\mathcal M}^{(0)*}_{\gamma}{\mathcal N}^{(1)}_{\gamma}\right]=0\,.
\end{equation*}
The $\mathcal{O}(\alpha/M)$ fractional TPE corrections (relative to the LO) to the elastic differential cross-section from the 
NLO diagrams are then collected within the term
\begin{widetext}
%%%%%%%%%%%%%%%%%%%%%%%%%%%%%%%%%%%%%%%%%%%%%%%%%%%%%%%%%%
\begin{eqnarray}
\delta^{(cde\cdots i)}_{\gamma\gamma} (Q^2)=\frac{2{\mathcal R}e\sum\limits_{spins}
\left[\mathcal{M}_{\gamma}^{(0)*}{\mathcal M}_{\gamma \gamma}^{(cde\cdots i)}\right]}{\sum\limits_{spins} 
\left|\mathcal{M}_{\gamma}^{(0)}\right|^2} = \delta^{(cde\cdots h)}_{\gamma\gamma}(Q^2) 
+ \delta_{\gamma \gamma;{\rm NLO}}^{\rm (seagull)}(Q^2)\,,
\end{eqnarray}
where the contributions from the TPE box and crossed-box diagrams (c)-(h) are given as
%%%%%%%%%%%%%%%%%%%%%%%%%%%%%%%%%%%%%%%%%%%%%%%%%%%%%%%%%%
\begin{eqnarray}
\delta^{(cde\cdots h)}_{\gamma\gamma}(Q^2) &\stackrel{\gamma_{\rm soft}}{\leadsto}& 
2{\mathcal R}e f^{(cde\cdots h)}_{\gamma\gamma}(Q^2)
=\frac{\alpha Q^2}{\pi M E\beta^{2}}\Bigg[1 - \frac{\pi^2}{6} \beta 
- \left(\frac{1}{\beta}+\beta\right) \ln\sqrt{\frac{1+\beta}{1-\beta}} + 2 \beta \ln^2\!\!\sqrt{\frac{1+\beta}{1-\beta}}
- 4 \beta \ln\!\sqrt{-\frac{Q^2}{2M E\beta}}
\nonumber\\
&&\!\! \times\,\ln\sqrt{\frac{1+\beta}{1-\beta}}+\beta{\rm Li}_2\bigg(\frac{2 \beta}{1+\beta}\bigg) 
+ \beta{\rm Li}_2\bigg(\frac{1+\beta}{1-\beta}\bigg )\Bigg]
+ \frac{\alpha Q^4}{24\pi M^2 E^2 \beta^5}\Bigg[\pi^2\beta^2-54\beta-24\beta^3 \ln\sqrt{\frac{2\beta}{1-\beta}}
\nonumber\\
&&\!\! -12\beta^2\ln^2\sqrt{\frac{1+\beta}{1-\beta}} 
- 6\beta^2{\rm Li}_2\left(\frac{1+\beta}{1-\beta}\right) - 6\beta^2{\rm Li}_2\left(\frac{2\beta}{1+\beta}\right)
+6\left(3+3\beta^2+2\beta^3\right)\ln{\sqrt{\frac{1+\beta}{1-\beta}}}
\nonumber\\
&&\!\!
- 24\beta^2\!\ln\sqrt{-\frac{Q^2}{2M\beta E}}\left(\beta-\ln{\sqrt{\frac{1+\beta}{1-\beta}}}\,\right)\Bigg]
+{\mathcal O}\left(\frac{1}{M^3}\right)\,,
\label{eq:delta_c-h}
\end{eqnarray}
and that of the seagull diagram (i) is given as
\begin{eqnarray}
\label{eq:delta_seagull_NLO}
\delta_{\gamma \gamma;{\rm NLO}}^{\rm (seagull)}(Q^2)\!\!&=&\!\!\frac{4\pi \alpha }{M}\bigg[\frac{Q^2}{Q^2+4E E^\prime}\bigg]\,
{\mathcal R}e\bigg\{4(Q^2 v+2 E p +2 E p^\prime)\cdot I^{-}_1(p, 0|1, 1, 1, 0) +16 m_l^2 E I(Q|1,1,1,0)\bigg\}
\\
&=&\!\! -\,\frac{4\alpha E }{\pi M}\bigg[\frac{Q^2}{Q^2+4E^2}\bigg]\left(\frac{\nu_l^2-1}{\nu_l^2}\right)
\Bigg[\left(\frac{1+\nu_l^2}{2\nu_l}\right) \bigg\{\frac{\pi^2}{3}
+\ln^2{\sqrt{\frac{\nu_l+1}{\nu_l-1}}+{\rm Li}_2\left(\frac{\nu_l-1}{\nu_l+1}\right)}\bigg\}-\ln\sqrt{-\frac{Q^2}{m_l^2}}\,\,\Bigg]
\nonumber\\
&&\!\! +\,\frac{8\alpha E^2 }{\pi M^2}\bigg[\frac{Q^2}{Q^2+4E^2}\bigg]^2\left(\frac{\nu_l^2-1}{\nu_l^2}\right)
\Bigg[\left(\frac{1+\nu_l^2}{2\nu_l}\right)\bigg\{\frac{\pi^2}{3}
+\ln^2{\sqrt{\frac{\nu_l+1}{\nu_l-1}}+{\rm Li}_2\left(\frac{\nu_l-1}{\nu_l+1}\right)}\bigg\}-\ln\sqrt{-\frac{Q^2}{m_l^2}}\,\,\Bigg]
\nonumber\\
&&\!\! -\,\frac{\alpha Q^2 }{2\pi M^2 \nu_l^2}\bigg[\frac{Q^2}{Q^2+4E^2}\bigg]
\Bigg[\!\left(\frac{\nu_l^2-1}{\nu_l}\right) \! \bigg\{\frac{\pi^2}{3}
+\ln^2{\sqrt{\frac{\nu_l+1}{\nu_l-1}}+{\rm Li}_2\left(\frac{\nu_l-1}{\nu_l+1}\right)}\bigg\}+2\ln\sqrt{-\frac{Q^2}{m_l^2}}\,\Bigg]
\!+\!\mathcal{O}\left(\!\frac{1}{M^3}\!\right).
\nonumber
\end{eqnarray}
%%%%%%%%%%%%%%%%%%%%%%%%%%%%%%%%%%%%%%%%%%%%%%%%%%%%%%%%%%
\end{widetext} 
Here, $\nu_l=\sqrt{1-4m^2_l/Q^2}$ is a $Q^2$-dependent kinematical variable. The definition of the relativistic three-point 
scalar and tensor loop-integrals, $I(Q|1, 1, 1, 0)$ and $I^{-\mu}_1(p, 0|1, 1, 1, 0)$, respectively, are taken from 
Ref.~\cite{Choudhary:2023rsz}. For completeness, their analytical expressions are provided in the appendix. In particular, the
expression for the above seagull contribution is slightly more general than what was presented in Ref.~\cite{Talukdar:2019dko}, 
where the evaluation was carried out only to $\mathcal{O}(1/M)$. Retaining all $\mathcal{O}\left(1/M^2\right)$ kinematically 
suppressed terms arising beyond the NLO contributions are essential, as they must be considered formally at the same footing 
with the dynamical contributions from the genuine NNLO TPE diagrams which we consider next. 

%%%%%%%%%%%%%%%%%%%%%%%%%%%%%%%%%%%%%%%%%%%%%%%%%%%%%%%%%%
\subsection{Next-to-next-to-leading order TPE contribution}
\begin{figure*}[tbp]
\begin{center}
\includegraphics[scale=0.49]{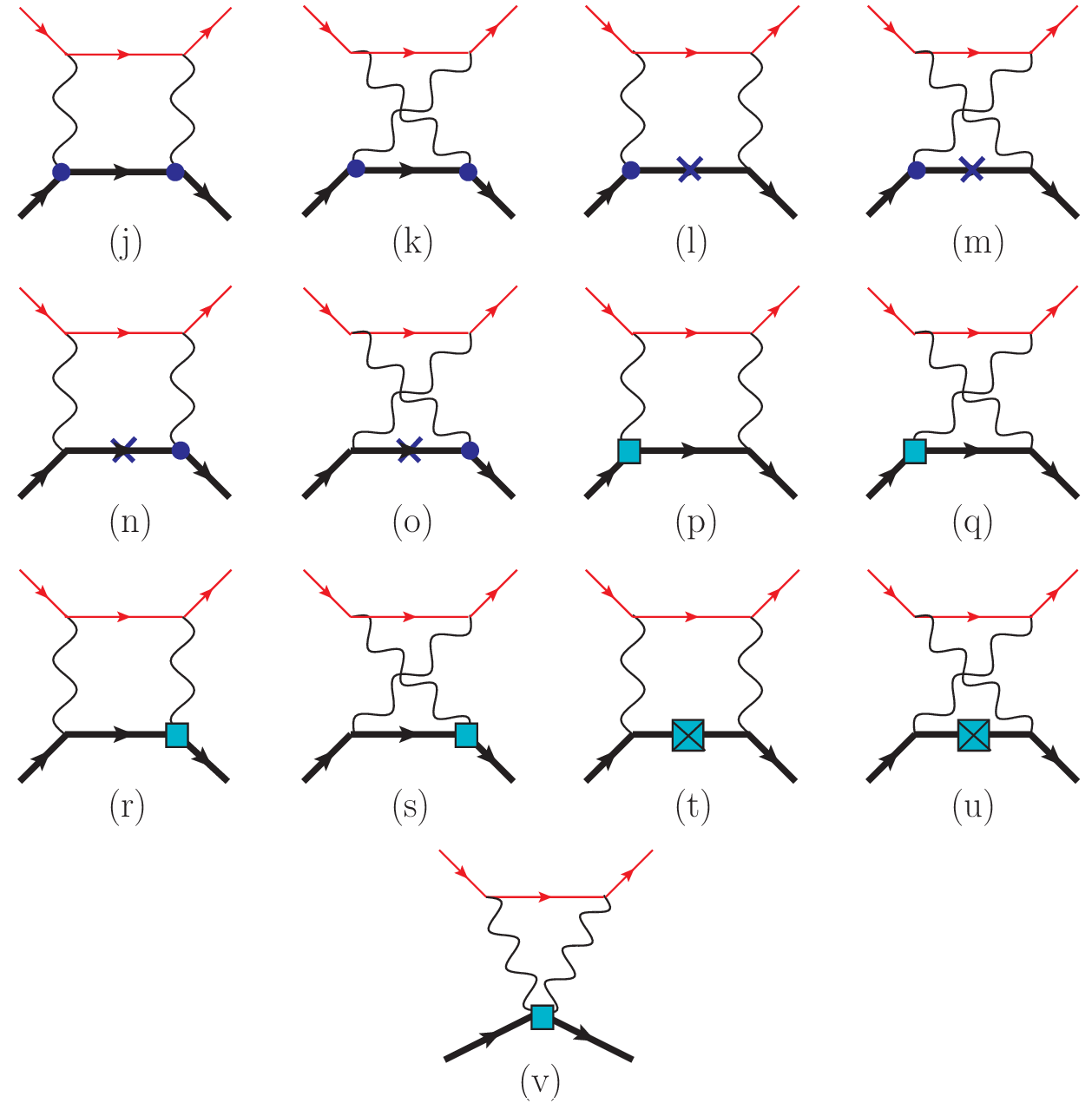}
	\caption{The subset of the NNLO TPE diagrams of $\mathcal{O}(\alpha^2/M^2)$ contributing to our 
             $\mathcal{O}(\alpha^3/M^2)$ elastic differential cross-section. The thick, thin, and wiggly 
             lines denote the proton, lepton, and photon propagators. The circular blobs represent 
             insertions of NLO proton-photon vertices and the crosses represent ${\mathcal O}(1/M)$ proton 
             propagators. The square blobs represent the insertions of renormalized NNLO proton-photon 
             effective vertices and those with crosses represent ${\mathcal O}(1/M^2)$ proton propagator 
             terms. } 
\label{fig:NNLO_TPE}
\end{center}
\end{figure*}

%%%%%%%%%%%%%%%%%%%%%%%%%%%%%%%%%%%%%%%%%%%%%%%%%%%%%%%%%%
Next, we evaluate the thirteen NNLO TPE one-loop diagrams (j)-(v), displayed in Fig~\ref{fig:NNLO_TPE}. 
The diagrams are constructed based on the following components: 
(i) two insertions of NLO proton-photon effective vertices, 
(ii) one NLO proton-photon vertex and an $\mathcal{O}(1/M)$ proton propagator, and (iii) one insertion of either an NNLO 
proton-photon vertex or $\mathcal{O}(1/M^2)$ proton propagator. We must emphasize that the effective proton-photon vertices
include all possible UV-divergent pion-loops with counterterms relevant to this order that also contribute to the iso-scalar 
and iso-vector nucleon form factors (see e.g., Refs.~\cite{Bernard:1995dp,Bernard:1998gv}). Thus, diagrams (p)-(s) are 
constituted essentially from two-loop TPE graphs which {\it factorize} into a photon- and a pion-loop contributing to the 
renormalized proton vertices. Besides such ``form factor" type TPE graphs, there are other two-loop TPE graphs including 
complicated {\it non-factorizable} photon and pion loops contributing to the NNLO corrections (cf. Fig.~\ref{fig:pi-loops}). 
In this work, however, we shall restrict our analysis only to the ``effective" one-loop accuracy, namely, including only the
form factor type contributions from the proton's structure to TPE. To this end, the effective renormalized NNLO proton-photon 
vertex in HB$\chi$PT attributed to pion-loops and LECs like the proton's anomalous magnetic moment 
$\kappa_p=(\kappa_s+\kappa_v)/2=1.795$, is given by~\cite{Talukdar:2020aui}
\begin{widetext}
%%%%%%%%%%%%%%%%%%%%%%%%%%%%%%%%%%%%%%%%%%%%%%%%%%%%%%%%%%
\begin{eqnarray}
\mathcal{V}_{\mu}^{(2);{\rm ren}} \!\!&=&\!\! (F_1^p -1)v_{\mu} +\frac{1}{M}\bigg\{ (F_1^p-1)\left(Q_{\mu}+\frac{Q^2}{2M} v_{\mu}\right)
+2(F_1^p+F_2^p-1-\kappa_p)[S_{\mu},S\cdot Q]\bigg\} 
\nonumber\\
&&\!\! -\,\frac{Q^2}{8M^2}(F_1^p-2F_2^p-1)v_{\mu} + {\mathcal O}\left(\frac{1}{M^3}\right)\,,
\end{eqnarray}
%%%%%%%%%%%%%%%%%%%%%%%%%%%%%%%%%%%%%%%%%%%%%%%%%%%%%%%%%%
\end{widetext} 
where the proton's Dirac and Pauli form factors $F_{1,2}^p$, to $\mathcal{O}(1/M^2)$, can be Taylor expanded at low-energies in 
terms of the corresponding Dirac and Pauli mean square radii, $\langle r^2_{1,2}\rangle\sim \mathcal{O}(1/M^{2})$, and the 
anomalous magnetic moment, $\kappa_p\sim \mathcal{O}(M^{0})$ (see Ref.~\cite{Bernard:1998gv} for details):
\begin{eqnarray}
F_1^p(Q^2) \!\!&=&\!\! 1+ \frac{Q^2}{6}\langle r_1^2\rangle +\mathcal{O}(M^{-3})\,, \quad \text{and}
\nonumber\\
F_2^p(Q^2) \!\!&=&\!\! \kappa_p+ \frac{Q^2}{6}\langle r^2_{2}\rangle +\mathcal{O}(M^{-3})\,.
\end{eqnarray} 
It turns out that for the NNLO TPE and other radiative corrections up to $\mathcal{O}(1/M^2)$ suppressed terms, only the Dirac
radius contributes at this order, which in turn can be related to the proton's {\it root mean square} (rms) electric or charge 
radius $r_p\equiv \sqrt{\langle r^2_{E}\rangle}$, namely,
\begin{eqnarray}
\langle r^2_1 \rangle=\langle r^2_E \rangle-\frac{3\kappa_p}{2M^2}  +\mathcal{O}(M^{-3})\,.
\end{eqnarray}
We now spell out the NNLO amplitudes given by the following integral representations:
\begin{widetext} 
%%%%%%%%%%%%%%%%%%%%%%%%%%%%%%%%%%%%%%%%%%%%%%%%%%%%%%%%%%
\begin{eqnarray}
{\mathcal{M}^{(j)}_{\rm box}} \!\!&=&\!\!  \frac{e^4}{4M^2} \int \frac{{\rm d}^4 k}{(2\pi)^4 i}
\frac{\left[\Bar{u}(p^\prime)\gamma^{\mu}(\slashed{p}-\slashed{k}+m_l)\gamma^{\nu}u(p)\right]
}{(k^2+i0)\, [(Q-k)^2+i0]\,(k^2-2k\cdot p+i0)\,(v\cdot p_p+v \cdot k+i0)} 
\nonumber \\
&&\hspace{1cm}\times \left[\chi^{\dagger}(p^\prime_p)\Big\{(p_p+p_p^\prime +k)_{\mu}-v_{\mu} [v \cdot (p_p+p_p^\prime+k)]\Big\}\,
\Big\{(2p_p+k)_{\nu}-v_{\nu} [v\cdot (2p_p+k)]\Big\}\chi(p_p)\right]\,,
\\
%%%%%%%%%%%%%%%%%%%%%%%%%%%%%%%%%%%%%%%%%%%%%%%%%%%%%%%%%%
{\mathcal{M}^{(k)}_{\rm xbox}} \!\!&=&\!\!  \frac{e^4}{4M^2} \int \frac{{\rm d}^4 k}{(2\pi)^4 i}
\frac{\left[\Bar{u}(p^\prime)\gamma^{\mu}(\slashed{p}-\slashed{k}+m_l)\gamma^{\nu}u(p)\right]
}{(k^2+i0)\, [(Q-k)^2+i0]\,(k^2-2k\cdot p+i0)\,(v\cdot p_p^\prime-v \cdot k+i0)}
\nonumber \\
&&\hspace{1cm}\times \left[\chi^{\dagger}(p^\prime_p)\Big\{(2p_p^\prime - k)_{\nu}-v_{\nu} v \cdot (2 p_p^\prime -k)\Big\}\,
\Big\{(p_p^\prime +p_p-k)_{\mu}-v_{\mu} [v\cdot (p_p^\prime+ p_p-k)]\Big\}\chi(p_p)\right]\,,
\\
%%%%%%%%%%%%%%%%%%%%%%%%%%%%%%%%%%%%%%%%%%%%%%%%%%%%%%%%%%
{\mathcal{M}^{(l)}_{\rm box}} \!\!&=&\!\!  \frac{e^4}{4M^2} \int \frac{{\rm d}^4 k}{(2\pi)^4 i}
\frac{\left[\Bar{u}(p^\prime)\gamma^{\mu}(\slashed{p}-\slashed{k}+m_l)\gamma^{\nu}u(p)\right]
\left[\chi^{\dagger}(p^\prime_p)\Big\{v_{\mu}(2p_p+k)_{\nu}-v_{\mu}v_{\nu} [v\cdot (2p_p+k)]\Big\}\chi(p_p)\right]}
{(k^2+i0)\, [(Q-k)^2+i0]\,(k^2-2k\cdot p+i0)} 
\nonumber \\
&&\hspace{1cm}\times \left(1-\frac{(p_p+k)^2}{(v\cdot p_p+v\cdot k +i0)^2}\right) \,,
\\
%%%%%%%%%%%%%%%%%%%%%%%%%%%%%%%%%%%%%%%%%%%%%%%%%%%%%%%%%%
{\mathcal{M}^{(m)}_{\rm xbox}} \!\!&=&\!\!  \frac{e^4}{4M^2} \int \frac{{\rm d}^4 k}{(2\pi)^4 i}
\frac{\left[\Bar{u}(p^\prime)\gamma^{\mu}(\slashed{p}-\slashed{k}+m_l)\gamma^{\nu}u(p)\right]
\left[\chi^{\dagger}(p^\prime_p)\Big\{v_{\nu}(p_p+p_p^\prime-k)_{\mu}-v_{\mu}v_{\nu} [v\cdot (p_p+p_p^\prime -k)]\Big\}\chi(p_p)\right]}
{(k^2+i0)\, [(Q-k)^2+i0] \,(k^2-2k\cdot p+i0)} 
\nonumber \\
&&\hspace{1cm}\times \left(1-\frac{(p_p^\prime-k)^2}{(v\cdot p_p^\prime-v\cdot k +i0)^2}\right)\,,
\end{eqnarray}
%%%%%%%%%%%%%%%%%%%%%%%%%%%%%%%%%%%%%%%%%%%%%%%%%%%%%%%%%%
\begin{eqnarray}
{\mathcal{M}^{(n)}_{\rm box}} \!\!&=&\!\!  \frac{e^4}{4M^2} \int \frac{{\rm d}^4 k}{(2\pi)^4 i}
\frac{\left[\Bar{u}(p^\prime)\gamma^{\mu}(\slashed{p}-\slashed{k}+m_l)\gamma^{\nu}u(p)\right]
\left[\chi^{\dagger}(p^\prime_p)\Big\{(p_p+p_p^\prime+k)_{\mu} v_{\nu}-v_{\mu}v_{\nu} [v\cdot (p_p+p_p^\prime+k)]\Big\}\chi(p_p)\right]}
{(k^2+i0)\, [(Q-k)^2+i0]\,(k^2-2k\cdot p+i0)} 
\nonumber\\
&&\hspace{1cm}\times \left(1-\frac{(p_p+k)^2}{(v\cdot p_p+v\cdot k +i0)^2}\right) \,,
\\
%%%%%%%%%%%%%%%%%%%%%%%%%%%%%%%%%%%%%%%%%%%%%%%%%%%%%%%%%
{\mathcal{M}^{(o)}_{\rm xbox}} \!\!&=&\!\!  \frac{e^4}{4M^2} \int \frac{{\rm d}^4 k}{(2\pi)^4 i}
\frac{\left[\Bar{u}(p^\prime)\gamma^{\mu}(\slashed{p}-\slashed{k}+m_l)\gamma^{\nu}u(p)\right]
\left[\chi^{\dagger}(p^\prime_p)\Big\{v_{\mu}(2p_p^\prime-k)_{\nu}-v_{\mu}v_{\nu} [v\cdot (2 p_p^\prime -k)]\Big\}\chi(p_p)\right]}
{(k^2+i0)\, [(Q-k)^2+i0]\,(k^2-2k\cdot p+i0)} 
\nonumber \\
&&\hspace{1cm} \times \left(1-\frac{(p_p^\prime-k)^2}{(v\cdot p_p^\prime-v\cdot k +i0)^2}\right)\,,
\\
%%%%%%%%%%%%%%%%%%%%%%%%%%%%%%%%%%%%%%%%%%%%%%%%%%%%%%%%%%
{\mathcal{M}^{(p)}_{\rm box}} \!\!&=&\!\! e^4 Q^2\left(\frac{\langle r_1^2\rangle }{6}\!+\!\frac{\kappa_p}{4 M^2}\right) \!\! 
\int \! \frac{{\rm d}^4 k}{(2\pi)^4 i}\frac{\left[\Bar{u}(p^\prime)\gamma^{\mu}(\slashed{p}-\slashed{k}+m_l)\gamma^{\nu}u(p)\right]
\Big[\chi^{\dagger}(p^\prime_p)v_{\mu}v_{\nu}\chi(p_p)\Big]}
{(k^2+i0) [(Q-k)^2+i0] (k^2-2k\cdot p+i0) (v\cdot k+ v\cdot p_p +i0)} \!+\! {\mathcal O}\left(M^{-3}\right),\quad\, 
\\
%%%%%%%%%%%%%%%%%%%%%%%%%%%%%%%%%%%%%%%%%%%%%%%%%%%%%%%%%%
{\mathcal{M}^{(q)}_{\rm xbox}} \!\!&=&\!\! e^4 Q^2\left(\frac{\langle r_1^2\rangle }{6}\!+\!\frac{\kappa_p}{4 M^2}\right) \!\!
\int \! \frac{{\rm d}^4 k}{(2\pi)^4 i}\frac{\left[\Bar{u}(p^\prime)\gamma^{\mu}(\slashed{p}-\slashed{k}+m_l)\gamma^{\nu}u(p)\right]\!
\Big[\chi^{\dagger}(p^\prime_p)v_{\mu}v_{\nu}\chi(p_p)\Big]}
{(k^2+i0) [(Q-k)^2+i0] (k^2-2k\cdot p+i0) (-v\cdot k\!+\!v\cdot p^\prime_p \!+\!i0)} \!+\! {\mathcal O}\left(M^{-3}\right),\quad\, 
\\
%%%%%%%%%%%%%%%%%%%%%%%%%%%%%%%%%%%%%%%%%%%%%%%%%%%%%%%%%%
{\mathcal{M}^{(r)}_{\rm box}} \!\!&=&\!\!  e^4 Q^2\left(\frac{\langle r_1^2\rangle }{6}\!+\!\frac{\kappa_p}{4 M^2}\right) \!\! 
\int\! \frac{{\rm d}^4 k}{(2\pi)^4 i}\frac{\left[\Bar{u}(p^\prime)\gamma^{\mu}(\slashed{p}-\slashed{k}+m_l)\gamma^{\nu}u(p)\right]\!
\Big[\chi^{\dagger}(p^\prime_p)v_{\mu}v_{\nu}\chi(p_p)\Big]}
{(k^2+i0) [(Q-k)^2+i0] (k^2-2k\cdot p+i0) (v\cdot k+ v\cdot p_p +i0)} \!+\! {\mathcal O}\left(M^{-3}\right),\quad\, 
\\
%%%%%%%%%%%%%%%%%%%%%%%%%%%%%%%%%%%%%%%%%%%%%%%%%%%%%%%%%%
{\mathcal{M}^{(s)}_{\rm xbox}} \!\!&=&\!\!  e^4 Q^2\left(\frac{\langle r_1^2\rangle }{6}\!+\!\frac{\kappa_p}{4 M^2}\right) \!\!
\int \! \frac{{\rm d}^4 k}{(2\pi)^4 i}\frac{\left[\Bar{u}(p^\prime)\gamma^{\mu}(\slashed{p}-\slashed{k}+m_l)\gamma^{\nu}u(p)\right]\!
\Big[\chi^{\dagger}(p^\prime_p)v_{\mu}v_{\nu}\chi(p_p)\Big]}
{(k^2+i0) [(Q-k)^2+i0](k^2-2k\cdot p+i0)(-v\cdot k\!+\!v\cdot p^\prime_p\!+\!i0)} \!+\! {\mathcal O}\left(M^{-3}\right),\quad\,
\\
%%%%%%%%%%%%%%%%%%%%%%%%%%%%%%%%%%%%%%%%%%%%%%%%%%%%%%%%%%
{\mathcal{M}^{(t)}_{\rm box}} \!\!&=&\!\!  \frac{e^4}{4M^2}\!\! \int \! \frac{{\rm d}^4 k}{(2\pi)^4 i}
\frac{\left[\Bar{u}(p^\prime)\gamma^{\mu}(\slashed{p}-\slashed{k}+m_l)\gamma^{\nu}u(p)\right]\!
\Big[\chi^{\dagger}(p^\prime_p)v_{\mu}v_{\nu}\chi(p_p)\Big]}
{(k^2+i0) [(Q-k)^2+i0](k^2-2k\cdot p+i0)} \!\! \left\{\!\frac{(v\cdot p_p\!+\!v\cdot k)^3\!-\!(p_p+k)^2 (v \cdot p_p\!+\!v\cdot k)}
{(v \cdot p_p+v \cdot k+i0)^2}\!\right\}\!,
\nonumber\\
\\
%%%%%%%%%%%%%%%%%%%%%%%%%%%%%%%%%%%%%%%%%%%%%%%%%%%%%%%%%%
{\mathcal{M}^{(u)}_{\rm xbox}} \!\!&=&\!\!  \frac{e^4}{4M^2}\!\! \int \! \frac{{\rm d}^4 k}{(2\pi)^4 i}
\frac{\left[\Bar{u}(p^\prime)\gamma^{\mu}(\slashed{p}-\slashed{k}+m_l)\gamma^{\nu}u(p)\right]\!
\Big[\chi^{\dagger}(p^\prime_p)v_{\mu}v_{\nu}\chi(p_p)\Big]}
{(k^2+i0)\, [(Q-k)^2+i0]\,(k^2-2k\cdot p+i0)} \!\! \left\{\!\frac{(v\cdot p_p^\prime\!-\!v\cdot k)^3
\!-\!(p_p^\prime\!-\!k)^2 (v\cdot p_p^\prime\!-\!v\cdot k)}{(v\cdot p_p^\prime -v\cdot k+i0)^2}\!\right\},\quad\,
\nonumber\\
\\
%%%%%%%%%%%%%%%%%%%%%%%%%%%%%%%%%%%%%%%%%%%%%%%%%%%%%%%%%%%
{\mathcal{M}^{(v)}_{\rm seagull}} \!\!&=&\!\! \frac{e^4}{4 M^2}\!\! \int \! \frac{{\rm d}^4 k}{(2\pi)^4 i}
\frac{\left[\Bar{u}(p^\prime)\gamma^{\mu}(\slashed{p}-\slashed{k}+m_l)\gamma^{\nu}u(p)\right]\!
\left[\chi^{\dagger}(p^\prime_p)\Big\{Q_{\mu} v_{\nu}+ v_{\mu} Q_{\nu}+(v\cdot Q)\left[g_{\mu \nu}
-3(v \cdot k)v_{\mu}v_{\nu}\right]\Big\}\chi(p_p)\right]}
{(k^2+i0)\, [(Q-k)^2+i0]\,(k^2-2k\cdot p+i0)}\,.
\nonumber\\
\end{eqnarray}
%%%%%%%%%%%%%%%%%%%%%%%%%%%%%%%%%%%%%%%%%%%%%%%%%%%%%%%%%%
\end{widetext}
%%%%%%%%%%%%%%%%%%%%%%%%%%%%%%%%%%%%%%%%%%%%%%%%%%%%%FIGURE-4%%%%%%%%%%%%%%%%%%%%%%%%%%%%%%%%%%%%%%%%%%%%%%%%%%%%
%%%%%%%%%%%%%%%%%%%%%%%%%%%%%%%%%%%%%%%%%%%%%%%%%%%%%%%%%%%%%%%%%%%%%%%%%%%%%%%%%%%%%%%%%%%%%%%%%%%%%%%%%%%%%%%%%
In particular, for the diagrams (p)-(s) with the NNLO proton-photon vertex insertion, only the $\mathcal{O}(1/M^2)$ part of 
the amplitudes are taken into consideration, omitting possible $\mathcal{O}(1/M^3)$ terms. Again, as in the two previous 
orders, some of the terms proportional to $v_\mu v_\nu$ cancel between the amplitudes $\mathcal{M}^{(j)}_{\rm box}$, 
$\mathcal{M}^{(k)}_{\rm xbox}$, $\mathcal{M}^{(l)}_{\rm box}$, $\mathcal{M}^{(m)}_{\rm xbox}$, $\mathcal{M}^{(n)}_{\rm box}$, 
$\mathcal{M}^{(o)}_{\rm xbox}$, $\mathcal{M}^{(t)}_{\rm box}$ and $\mathcal{M}^{(u)}_{\rm xbox}$, where we use 
$v\cdot p_p=0$ (lab frame) and $v\cdot p^\prime_p = - \frac{Q^2}{2 M}$ which contribute at ${\cal O} (1/M^3)$, and hence, 
dropped to NNLO. In particular, the amplitudes for $\mathcal{M}^{(t)}_{\rm box}$  and $\mathcal{M}^{(u)}_{\rm xbox}$ cancel 
completely when adding the $v_\mu v_\nu$ terms. The non-vanishing, reduced amplitudes (using $p^\prime_p = Q$ in lab-frame) 
are given as 
\begin{widetext}
%%%%%%%%%%%%%%%%%%%%%%%%%%%%%%%%%%%%%%%%%%%%%%%%%%%%%%%%%%
\begin{eqnarray}
 {\widetilde{\mathcal{M}}^{(j)}_{\rm box}}
\!\!&=&\!\!  \frac{e^4}{4M^2} \int \frac{{\rm d}^4 k}{(2\pi)^4 i}
\frac{\left[\Bar{u}(p^\prime)\gamma^{\mu}(\slashed{p}-\slashed{k}+m_l)\gamma^{\nu}u(p)\right]}
{(k^2+i0)\, [(Q-k)^2+i0]\,(k^2-2k\cdot p+i0)\,(v \cdot k+i0)} 
\nonumber \\
&&\hspace{1cm} \times \left[\chi^{\dagger}(p^\prime_p)\Big\{(Q+k)_{\mu} k_{\nu}-v_{\mu}k_{\nu} [v \cdot (Q+k)] 
- (Q+k)_{\mu} v_{\nu} (v\cdot k)\Big\}\chi(p_p)\right]\,,
\\
%%%%%%%%%%%%%%%%%%%%%%%%%%%%%%%%%%%%%%%%%%%%%%%%%%%%%%%%%%
{\widetilde{\mathcal{M}}^{(k)}_{\rm xbox}} \!\!&=&\!\!  \frac{e^4}{4M^2} \int \frac{{\rm d}^4 k}{(2\pi)^4 i}
\frac{\left[\Bar{u}(p^\prime)\gamma^{\mu}(\slashed{p}+\slashed{k}-\slashed{Q}+m_l)\gamma^{\nu}u(p)\right]}
{(k^2+i0)\, [(Q-k)^2+i0]\,(k^2+2k\cdot p^\prime+i0)\,(v \cdot k+i0)}
\nonumber \\
&&\hspace{1cm} \times \left[\chi^{\dagger}(p^\prime_p)\Big\{k_{\mu}(Q+k)_{\nu}- k_{\mu} v_{\nu} [v\cdot (Q+k)] 
- v_{\mu}(Q+k)_{\nu} (v\cdot k)\Big\}\chi(p_p)\right]\,,
\end{eqnarray}
%%%%%%%%%%%%%%%%%%%%%%%%%%%%%%%%%%%%%%%%%%%%%%%%%%%%%%%%%%
\begin{eqnarray}
{\widetilde{\mathcal{M}}^{(l)}_{\rm box}} \!\!&=&\!\!  \frac{e^4}{4M^2} \int \frac{{\rm d}^4 k}{(2\pi)^4 i}
\frac{\left[\Bar{u}(p^\prime)\gamma^{\mu}(\slashed{p}-\slashed{k}+m_l)\gamma^{\nu}u(p)\right]
\Big[\chi^{\dagger}(p^\prime_p)v_{\mu}k_{\nu}\chi(p_p)\Big]}
{(k^2+i0)\, [(Q-k)^2+i0]\,(k^2-2k\cdot p+i0)} \left(1-\frac{k^2}{(v\cdot k +i0)^2}\right) \,,
\\
%%%%%%%%%%%%%%%%%%%%%%%%%%%%%%%%%%%%%%%%%%%%%%%%%%%%%%%%%%
{\widetilde{\mathcal{M}}^{(m)}_{\rm xbox}} \!\!&=&\!\!  \frac{e^4}{4M^2} \int \frac{{\rm d}^4 k}{(2\pi)^4 i}
\frac{\left[\Bar{u}(p^\prime)\gamma^{\mu}(\slashed{p}+\slashed{k}-\slashed{Q}+m_l)\gamma^{\nu}u(p)\right]
\Big[\chi^{\dagger}(p^\prime_p)k_{\mu}v_{\nu}\chi(p_p)\Big]}
{(k^2+i0)\, [(Q-k)^2+i0]\,(k^2+2k\cdot p^\prime+i0)} \left(1-\frac{k^2}{(v\cdot k +i0)^2}\right) \,,
\\
%%%%%%%%%%%%%%%%%%%%%%%%%%%%%%%%%%%%%%%%%%%%%%%%%%%%%%%%%%
{\widetilde{\mathcal{M}}^{(n)}_{\rm box}} \!\!&=&\!\!  \frac{e^4}{4M^2} \int \frac{{\rm d}^4 k}{(2\pi)^4 i}
\frac{\left[\Bar{u}(p^\prime)\gamma^{\mu}(\slashed{p}-\slashed{k}+m_l)\gamma^{\nu}u(p)\right]
\Big[\chi^{\dagger}(p^\prime_p)(Q+k)_{\mu}v_{\nu}\chi(p_p)\Big]}
{(k^2+i0)\, [(Q-k)^2+i0]\,(k^2-2k\cdot p+i0)} \left(1-\frac{k^2}{(v\cdot k +i0)^2}\right) 
\nonumber\\
&&+\,\frac{e^4}{4M^2} \int \frac{{\rm d}^4 k}{(2\pi)^4 i}
\frac{\left[\Bar{u}(p^\prime)\gamma^{\mu}(\slashed{p}-\slashed{k}+m_l)\gamma^{\nu}u(p)\right]
\Big[\chi^{\dagger}(p^\prime_p)v_{\mu}v_{\nu}\chi(p_p)\Big]}
{(k^2+i0)\, [(Q-k)^2+i0]\,(k^2-2k\cdot p+i0)}\cdot \frac{k^2}{(v\cdot k +i0)}\,,
\\
%%%%%%%%%%%%%%%%%%%%%%%%%%%%%%%%%%%%%%%%%%%%%%%%%%%%%%%%%%
{\widetilde{\mathcal{M}}^{(o)}_{\rm xbox}} \!\!&=&\!\!  \frac{e^4}{4M^2} \int \frac{{\rm d}^4 k}{(2\pi)^4 i}
\frac{\left[\Bar{u}(p^\prime)\gamma^{\mu}(\slashed{p}+\slashed{k}-\slashed{Q}+m_l)\gamma^{\nu}u(p)\right]
\Big[\chi^{\dagger}(p^\prime_p)v_{\mu}(Q+k)_{\nu}\chi(p_p)\Big]}
{(k^2+i0)\, [(Q-k)^2+i0]\,(k^2+2k\cdot p^\prime+i0)} \left(1-\frac{k^2}{(v\cdot k +i0)^2}\right) 
\nonumber\\
&&+\,\frac{e^4}{4M^2} \int \frac{{\rm d}^4 k}{(2\pi)^4 i}
\frac{\left[\Bar{u}(p^\prime)\gamma^{\mu}(\slashed{p}+\slashed{k}-\slashed{Q}+m_l)\gamma^{\nu}u(p)\right]
\Big[\chi^{\dagger}(p^\prime_p)v_{\mu}v_{\nu}\chi(p_p)\Big]}
{(k^2+i0)\, [(Q-k)^2+i0]\,(k^2+2k\cdot p^\prime+i0)}\cdot \frac{k^2}{(v\cdot k +i0)}\,.
\end{eqnarray}
%%%%%%%%%%%%%%%%%%%%%%%%%%%%%%%%%%%%%%%%%%%%%%%%%%%%%%%%%%
\end{widetext}
Since terms proportional to $v_\mu v_\nu$ do not cancel for the amplitudes (p)-(s), as well as in the seagull amplitude 
(v), their expressions are left unmodified. Next, we implement SPA for all the NNLO box and crossed-box diagrams which 
are explicit functions of the four-point loop-integrals. In contrast, the seagull diagram (v), being a three-point 
loop-function will be evaluated exactly without implementing SPA. We find that the NNLO box and crossed-box amplitudes
factorize either into the LO Born amplitude $\mathcal{M}^{(0)}_\gamma$, Eq.~\eqref{eq:M0_LO}, or the NLO amplitude 
$\mathcal{N}^{(1)}_\gamma$, Eq.~\eqref{eq:N1_NLO}, to yield the following reduced amplitudes with each being the sum of 
2- and 3-point functions: 
\begin{widetext} 
%%%%%%%%%%%%%%%%%%%%%%%%%%%%%%%%%%%%%%%%%%%%%%%%%%%%%%%%%%
\begin{eqnarray}
{\widetilde{\mathcal{M}}^{(j)}_{\rm box}} 
&\stackrel{\gamma_{\rm soft}}{\leadsto}& \frac{e^2 E}{2M} {\mathcal N}^{(1)}_\gamma \!\!\int 
\frac{{\rm d}^4 k}{(2\pi)^4 i}\frac{1}{(k^2+i0)\,(k^2-2k\cdot p+i0)} 
+\frac{e^2 E^\prime}{2M} {\mathcal N}^{(1)}_\gamma \!\!\int 
\frac{{\rm d}^4 k}{(2\pi)^4 i}\frac{1}{[(Q-k)^2+i0]\,(k^2-2k\cdot p+i0)}
\nonumber\\
&&+\, \frac{e^2 Q^2}{2 M}  {\mathcal N}^{(1)}_\gamma \int 
\frac{{\rm d}^4 k}{(2\pi)^4 i}\frac{1}{[(Q-k)^2+i0]\,(k^2-2k\cdot p+i0)\,(v\cdot k +i0)}
\nonumber\\
&&-\,\frac{e^2 Q^2}{2M^2} {\mathcal M}^{(0)}_\gamma \int 
\frac{{\rm d}^4 k}{(2\pi)^4 i}\frac{1}{[(Q-k)^2+i0]\,(k^2-2k\cdot p+i0)} 
+\mathcal{O}\left(\frac{1}{M^3}\right)\,,
\\
%%%%%%%%%%%%%%%%%%%%%%%%%%%%%%%%%%%%%%%%%%%%%%%%%%%%%%%%%%
{\widetilde{\mathcal{M}}^{(k)}_{\rm xbox}} 
&\stackrel{\gamma_{\rm soft}}{\leadsto}& \frac{e^2 E^\prime}{2M} {\mathcal N}^{(1)}_\gamma \!\!\int 
\frac{{\rm d}^4 k}{(2\pi)^4 i}\frac{1}{(k^2+i0)\,(k^2+2k\cdot p^\prime +i0)} 
+\frac{e^2 E}{2M} {\mathcal N}^{(1)}_\gamma \!\!\int 
\frac{{\rm d}^4 k}{(2\pi)^4 i}\frac{1}{[(Q-k)^2+i0]\,(k^2+2k\cdot p^\prime+i0)}
\nonumber\\
&&-\, \frac{e^2 Q^2}{2 M}  {\mathcal N}^{(1)}_\gamma \int 
\frac{{\rm d}^4 k}{(2\pi)^4 i}\frac{1}{[(Q-k)^2+i0]\,(k^2+2k\cdot p^\prime  +i0)\,(v\cdot k +i0)}
\nonumber\\
&&+\,\frac{e^2 Q^2}{2M^2} {\mathcal M}^{(0)}_\gamma \int 
\frac{{\rm d}^4 k}{(2\pi)^4 i}\frac{1}{[(Q-k)^2+i0]\,(k^2+2k\cdot p^\prime+i0)} 
+\mathcal{O}\left(\frac{1}{M^3}\right)\,,
\\
%%%%%%%%%%%%%%%%%%%%%%%%%%%%%%%%%%%%%%%%%%%%%%%%%%%%%%%%%%
{\widetilde{\mathcal{M}}^{(l)}_{\rm box}} 
&\stackrel{\gamma_{\rm soft}}{\leadsto}& -\, \frac{e^2 E^\prime}{2M}  {\mathcal N}^{(1)}_\gamma \int 
\frac{{\rm d}^4 k}{(2\pi)^4 i}\frac{1}{[(Q-k)^2+i0]\,(k^2-2k\cdot p+i0)} \,,
\nonumber \\
&&+\,  \frac{e^2 Q^2 E^\prime}{2M}  {\mathcal N}^{(1)}_\gamma \int 
\frac{{\rm d}^4 k}{(2\pi)^4 i}\frac{1}{[(Q-k)^2+i0]\,(k^2-2k\cdot p+i0)\,(v \cdot k+i0)^2} \,,
\\
%%%%%%%%%%%%%%%%%%%%%%%%%%%%%%%%%%%%%%%%%%%%%%%%%%%%%%%%%%
{\widetilde{\mathcal{M}}^{(m)}_{\rm xbox}} 
&\stackrel{\gamma_{\rm soft}}{\leadsto}& -\, \frac{e^2 E}{2M} {\mathcal N}^{(1)}_\gamma \int 
\frac{{\rm d}^4 k}{(2\pi)^4 i}\frac{1}{[(Q-k)^2+i0]\,(k^2+2k\cdot p^\prime+i0)} \,,
\nonumber \\
&&+\,  \frac{e^2 Q^2 E}{2M} {\mathcal N}^{(1)}_\gamma \int 
\frac{{\rm d}^4 k}{(2\pi)^4 i}\frac{1}{[(Q-k)^2+i0]\,(k^2+2k\cdot p^\prime+i0)\,(v \cdot k+i0)^2} \,,
\\
%%%%%%%%%%%%%%%%%%%%%%%%%%%%%%%%%%%%%%%%%%%%%%%%%%%%%%%%%%
{\widetilde{\mathcal{M}}^{(n)}_{\rm box}} 
&\stackrel{\gamma_{\rm soft}}{\leadsto}& -\frac{e^2 E}{2M} {\mathcal N}^{(1)}_\gamma \!\!\int 
\frac{{\rm d}^4 k}{(2\pi)^4 i}\frac{1}{(k^2+i0)\,(k^2-2k\cdot p+i0)} 
+\frac{e^2 Q^2}{2M^2} {\mathcal M}^{(0)}_\gamma \!\!\int 
\frac{{\rm d}^4 k}{(2\pi)^4 i}\frac{1}{[(Q-k)^2+i0]\,(k^2-2k\cdot p+i0)}
\nonumber\\
&&-\, \frac{e^2 Q^4}{2 M^2}  {\mathcal M}^{(0)}_\gamma \int 
\frac{{\rm d}^4 k}{(2\pi)^4 i}\frac{1}{[(Q-k)^2+i0]\,(k^2-2k\cdot p  +i0)\,(v\cdot k +i0)^2}
\nonumber\\
&&-\,\frac{e^2 Q^2 E^\prime}{2M^2} {\mathcal M}^{(0)}_\gamma \int 
\frac{{\rm d}^4 k}{(2\pi)^4 i}\frac{1}{[(Q-k)^2+i0]\,(k^2-2k\cdot p+i0)\,(v\cdot k +i0)} \,,
\\
%%%%%%%%%%%%%%%%%%%%%%%%%%%%%%%%%%%%%%%%%%%%%%%%%%%%%%%%%%
{\widetilde{\mathcal{M}}^{(o)}_{\rm xbox}} 
&\stackrel{\gamma_{\rm soft}}{\leadsto}& -\frac{e^2 E^\prime}{2M} {\mathcal N}^{(1)}_\gamma \!\!\int 
\frac{{\rm d}^4 k}{(2\pi)^4 i}\frac{1}{(k^2+i0)\,(k^2+2k\cdot p^\prime+i0)} 
-\frac{e^2 Q^2}{2M^2} {\mathcal M}^{(0)}_\gamma \!\!\int 
\frac{{\rm d}^4 k}{(2\pi)^4 i}\frac{1}{[(Q-k)^2+i0]\,(k^2+2k\cdot p^\prime+i0)}
\nonumber\\
&&+\, \frac{e^2 Q^4}{2 M^2}  {\mathcal M}^{(0)}_\gamma \int 
\frac{{\rm d}^4 k}{(2\pi)^4 i}\frac{1}{[(Q-k)^2+i0]\,(k^2+2k\cdot p^\prime  +i0)\,(v\cdot k +i0)^2}
\nonumber\\
&&-\,\frac{e^2 Q^2 E}{2M^2} {\mathcal M}^{(0)}_\gamma \int 
\frac{{\rm d}^4 k}{(2\pi)^4 i}\frac{1}{[(Q-k)^2+i0]\,(k^2+2k\cdot p^\prime +i0)\,(v\cdot k +i0)} \,,
\\
%%%%%%%%%%%%%%%%%%%%%%%%%%%%%%%%%%%%%%%%%%%%%%%%%%%%%%%%%%
{\mathcal M}^{(p)}_{\rm box}
&\stackrel{\gamma_{\rm soft}}{\leadsto}& -\,2e^2Q^2E 
\left(\frac{\langle r_1^2\rangle}{6} +\frac{\kappa_p}{4M^2}\right) \mathcal{M}_{\gamma}^{(0)}\int 
\frac{{\rm d}^4 k}{(2\pi)^4 i}\frac{1}{(k^2+i0)\,(k^2-2k\cdot p +i0)\, (v\cdot k +i0)}
\nonumber\\
&&-\,2e^2Q^2E^\prime  
\left(\frac{\langle r_1^2\rangle}{6} +\frac{\kappa_p}{4M^2}\right) \mathcal{M}_{\gamma}^{(0)}\int 
\frac{{\rm d}^4 k}{(2\pi)^4 i}\frac{1}{[(Q-k)^2+i0]\,(k^2-2k\cdot p +i0)\, (v\cdot k +i0)} 
+\mathcal{O}\left(\frac{1}{M^3}\right),\,\,
\\
%%%%%%%%%%%%%%%%%%%%%%%%%%%%%%%%%%%%%%%%%%%%%%%%%%%%%%%%%%
{\mathcal M}^{(q)}_{\rm xbox}
&\stackrel{\gamma_{\rm soft}}{\leadsto}& -\,2e^2Q^2E  
\left(\frac{\langle r_1^2\rangle}{6} +\frac{\kappa_p}{4M^2}\right) \mathcal{M}_{\gamma}^{(0)} \int 
\frac{{\rm d}^4 k}{(2\pi)^4 i}\frac{1}{[(Q-k)^2+i0]\,(k^2+2k\cdot p^\prime +i0)\, (v\cdot k +i0)}
\nonumber \\
&&-\,2e^2Q^2E^\prime  
\left(\frac{\langle r_1^2\rangle}{6} +\frac{\kappa_p}{4M^2}\right) \mathcal{M}_{\gamma}^{(0)} \int 
\frac{{\rm d}^4 k}{(2\pi)^4 i}\frac{1}{(k^2+i0)\, (k^2+2k\cdot p^\prime +i0)\, (v\cdot k +i0)}
+\mathcal{O}\left(\frac{1}{M^3}\right)\,,
\\
%%%%%%%%%%%%%%%%%%%%%%%%%%%%%%%%%%%%%%%%%%%%%%%%%%%%%%%%%%
{\mathcal M}^{(r)}_{\rm box} 
&\stackrel{\gamma_{\rm soft}}{\leadsto}& -\,2e^2Q^2E  
\left(\frac{\langle r_1^2\rangle}{6} +\frac{\kappa_p}{4M^2}\right) \mathcal{M}_{\gamma}^{(0)}\int 
\frac{{\rm d}^4 k}{(2\pi)^4 i}\frac{1}{(k^2+i0)\,(k^2-2k\cdot p +i0)\, (v\cdot k +i0)}
\nonumber\\
&&-\,2e^2Q^2E^\prime  
\left(\frac{\langle r_1^2\rangle}{6} +\frac{\kappa_p}{4M^2}\right) \mathcal{M}_{\gamma}^{(0)}\int 
\frac{{\rm d}^4 k}{(2\pi)^4 i}\frac{1}{[(Q-k)^2+i0]\,(k^2-2k\cdot p +i0)\, (v\cdot k +i0)} 
+\mathcal{O}\left(\frac{1}{M^3}\right),\,\,
\end{eqnarray}
%%%%%%%%%%%%%%%%%%%%%%%%%%%%%%%%%%%%%%%%%%%%%%%%%%%%%%%%%%
\begin{eqnarray}
{\mathcal M}^{(s)}_{\rm xbox}
&\stackrel{\gamma_{\rm soft}}{\leadsto}&-\,2e^2Q^2E  
\left(\frac{\langle r_1^2\rangle}{6} +\frac{\kappa_p}{4M^2}\right) \mathcal{M}_{\gamma}^{(0)} \int 
\frac{{\rm d}^4 k}{(2\pi)^4 i}\frac{1}{ [(Q-k)^2+i0]\,(k^2+2k\cdot p^\prime +i0)\, (v\cdot k +i0)}
\nonumber \\
&&-\,2e^2Q^2E^\prime 
\left(\frac{\langle r_1^2\rangle}{6} +\frac{\kappa_p}{4M^2}\right) \mathcal{M}_{\gamma}^{(0)} \int 
\frac{{\rm d}^4 k}{(2\pi)^4 i}\frac{1}{(k^2+i0)\, (k^2+2k\cdot p^\prime +i0)\, (v\cdot k +i0)} 
+\mathcal{O}\left(\frac{1}{M^3}\right)\,.
\end{eqnarray}
%%%%%%%%%%%%%%%%%%%%%%%%%%%%%%%%%%%%%%%%%%%%%%%%%%%%%%%%%%
%%%%%%%%%%%%%%%%%%%%%%%%%%%%%%%%%%%%%%%%%%%%%%%%%%FIGURE-5%%%%%%%%%%%%%%%%%%%%%%%%%%%%%%%%%%%%%%%%%%%%%%%%%%%%%%
\begin{figure*}[tbp]
\begin{center}
\includegraphics[scale=0.48]{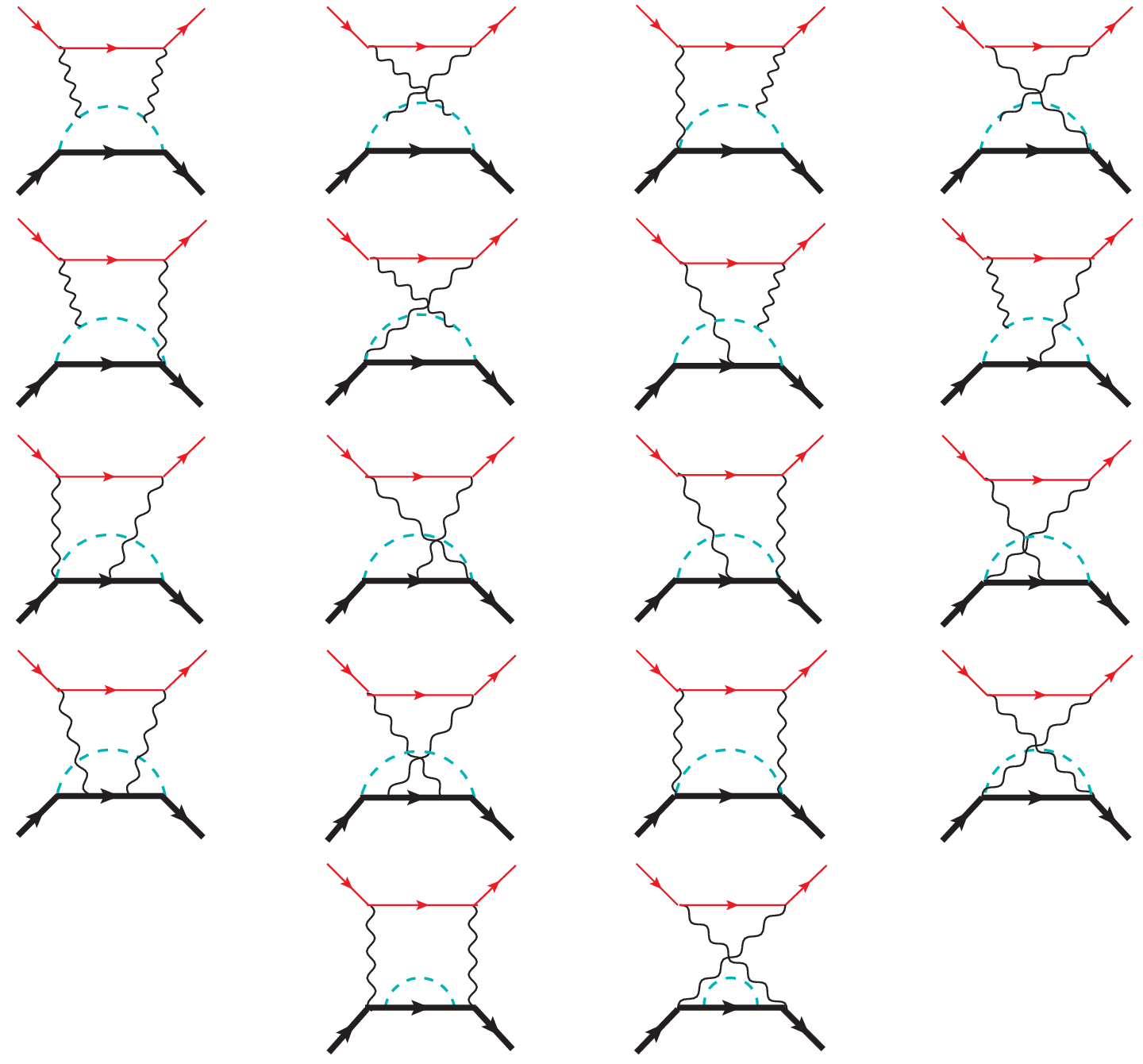}
\caption{The subset of two-loop NNLO TPE diagrams of $\mathcal{O}(\alpha^2/M^2)$ with non-factorizable 
         pion-loops, which in principle contribute to the $\mathcal{O}(\alpha^3/M^2)$ elastic 
         differential cross-section. The thick and thin solid lines denote the proton and lepton 
         propagators, whereas the dashed and wiggly lines denote the pion and photon propagators. To 
         one-loop approximation in the perturbative expansion, the contribution from these diagrams in 
         combination with the appropriate counter-term diagrams are excluded in our analysis since they 
         are expected to yield rather tiny contributions. } 
\label{fig:pi-loops}
\end{center}
\end{figure*}
%%%%%%%%%%%%%%%%%%%%%%%%%%%%%%%%%%%%%%%%%%%%%%%%%%%%%%%%%%%%%%%%%%%%%%%%%%%%%%%%%%%%%%%%%%%%%%%%%%%%%%%%%%%%%%%%%
Hence, the sum of the NNLO amplitudes can be summarized as
\begin{eqnarray}
\mathcal{M}_{\gamma \gamma}^{(jkl\cdots v)} \!\!&=&\!\! \widetilde{\mathcal{M}}_{\rm box}^{(j)}+\widetilde{\mathcal{M}}_{\rm xbox}^{(k)}
+\widetilde{\mathcal{M}}_{\rm box}^{(l)}+\widetilde{\mathcal{M}}_{\rm xbox}^{(m)}+\widetilde{\mathcal{M}}_{\rm box}^{(n)}
+\widetilde{\mathcal{M}}_{\rm xbox}^{(o)}+{\mathcal M}_{\rm box}^{(p)}+{\mathcal M}_{\rm xbox}^{(q)}+{\mathcal M}_{\rm box}^{(r)}
+{\mathcal M}_{\rm xbox}^{(s)}+{\mathcal{M}^{(\rm v)}_{\rm seagull}}
\nonumber\\
&\stackrel{\gamma_{\rm soft}}{\leadsto}&  \mathcal{M}_{\gamma}^{(0)} f^{(jkl\cdots u)}_{\gamma\gamma}(Q^2) 
+ {\mathcal{M}^{(v)}_{\rm seagull}}\,, \qquad \text{where}
\end{eqnarray}

\vspace{-0.2cm}

\begin{eqnarray}
f^{(jkl\cdots u)}_{\gamma \gamma}(Q^2) \!\!&=&\!\! -\,\frac{e^2Q^2}{2M^2}\bigg\{E^\prime \bcancel{I^{-}(p,0|0,1,1,1)} 
+  E \bcancel{I^{+}(p^\prime,0|0,1,1,1)} +Q^2 I^{-}(p,0|0,1,1,2) - Q^2 I^{+}(p^\prime,0|0,1,1,2) \bigg\}
\nonumber\\
&&\!\! +\,\frac{e^2 Q^2{\mathcal M}^{(0)*}_{\gamma}{\mathcal N}^{(1)}_{\gamma}}{2 M \left |{\mathcal M}^{(0)}_{\gamma}\right|^2}
\bigg\{I^{-}(p,0|0,1,1,1) - I^{+}(p^\prime,0|0,1,1,1) + E^\prime I^{-}(p,0|0,1,1,2) + E I^{+}(p,0|0,1,1,2)\bigg\}
\nonumber\\
&&\!\! -\,4 e^2 Q^2 \left(\frac{\langle r_1^2\rangle}{6} +\frac{\kappa_p}{4M^2}\right)\bigg\{E \cancel{I^{-}(p,0|1,0,1,1)} 
+  E^\prime \cancel{I^{+}(p^\prime,0|1,0,1,1)} + E^\prime \bcancel{I^{-}(p,0|0,1,1,1)} 
\nonumber\\
&& \hspace{3.9cm} + E \bcancel{I^{+}(p^\prime,0|0,1,1,1) \bigg\}} + {\mathcal O}\left(\frac{1}{M^3}\right)\,.
\label{eq:fNNLO}
\end{eqnarray}
%%%%%%%%%%%%%%%%%%%%%%%%%%%%%%%%%%%%%%%%%%%%%%%%%%%%%%%%%%
\end{widetext}
The loop-integrals in the above expression already appeared in the NLO amplitude $f^{(cde\cdots h)}_{\gamma\gamma}$, 
Eq.~\eqref{eq:ffNLO} whose analytical expressions of ${\mathcal O}(1/M^2)$ are provided in the appendix. Just as in the 
NLO case, the IR-divergent integrals $I^{-}(p,0|1,0,1,1)$ and $I^{+}(p^\prime,0|1,0,1,1)$ do not contribute to the 
cross-section since the pre-factor ${\mathcal M}^{(0)*}_{\gamma}{\mathcal N}^{(1)}_{\gamma}$ causes the entire 
second-line expression in Eq.~\eqref{eq:fNNLO} to vanish identically when calculating the elastic cross-section.  
Moreover, in this work, since we intend to preserve the accuracy of all analytical expressions to $\mathcal{O}(1/M^2)$, 
it is legitimate at this point to substitute $E=E^\prime$ and $\beta=\beta^\prime$ in all terms of $\mathcal{O}(1/M^2)$. 
Consequently, with the integrals $I^{-}(p,0|0,1,1,1)$ and $I^{-}(p,0|1,0,1,1)$ effectively canceling with 
$I^{+}(p^\prime,0|0,1,1,1)$ and $I^{+}(p^\prime,0|1,0,1,1)$, respectively, the proton's structure effects in TPE drop 
out up to $\mathcal{O}(1/M^2)$. The only terms in the amplitude that will effectively contribute to the cross-section 
are the 3-point integral functions $I^{-}(p,0|0,1,1,2)$ and $I^{+}(p^\prime,0|0,1,1,2)$ appearing in the first line of 
$f^{(jkl\cdots u)}_{\gamma\gamma}$, Eq.~\eqref{eq:fNNLO}. Notably, as seen from the $\mathcal{O}(1/M^2)$ expressions 
of these 3-point functions provided in the appendix, they are each of $\mathcal{O}(M)$, and therefore, generate 
kinematically enhanced contributions of $\mathcal{O}(1/M)$ to the amplitude $f^{(jkl\cdots u)}_{\gamma\gamma}$ which 
do not cancel. 

Finally, the $\mathcal{O}(\alpha/M^2)$ fractional contribution (relative to the LO) to the elastic differential 
cross-section from the NNLO diagrams are collected within the term
\begin{eqnarray}
\delta^{(jkl\cdots v)}_{\gamma\gamma} (Q^2) \!\!&=&\!\! \frac{2{\mathcal R}e\sum\limits_{spins}
\left[\mathcal{M}_{\gamma}^{(0)*}{\mathcal M}_{\gamma \gamma}^{(jkl\cdots v)}\right]}{\sum\limits_{spins} 
\left|\mathcal{M}_{\gamma}^{(0)}\right|^2} 
\nonumber\\
&=&\!\! \delta^{(jkl\cdots u)}_{\gamma\gamma}(Q^2) 
+ \delta_{\gamma \gamma;{\rm NNLO}}^{\rm (seagull)}(Q^2)\,,
\end{eqnarray}
where the contributions from the TPE box and crossed-box diagrams (j)-(u), and the NNLO seagull diagram (v), are 
respectively, given as
\begin{widetext}
%%%%%%%%%%%%%%%%%%%%%%%%%%%%%%%%%%%%%%%%%%%%%%%%%%%%%%%%%%
\begin{eqnarray}
\delta^{(jkl\cdots u)}_{\gamma\gamma}(Q^2) 
\!\!&\stackrel{\gamma_{\rm soft}}{\leadsto}&\!\! \frac{2 \alpha Q^2}{\pi M E\beta} 
\ln{\sqrt{\frac{1+\beta}{1-\beta}}} + \frac{\alpha Q^4}{2 \pi M^2 E^2 \beta^3}\left(\beta 
- \ln{\sqrt{\frac{1+\beta}{1-\beta}}}\,\,\right) + {\mathcal O}\left(\frac{1}{M^3}\right)\,, \quad \text{and}
\label{eq:delta_j-u}
\\
\delta_{\gamma \gamma;{\rm NNLO}}^{\rm (seagull)}(Q^2)\!\!&=&\!\!\frac{\pi \alpha Q^2 }{4 M^2}
{\mathcal R}e\bigg\{\cancel{2 Q\cdot I^{-}_1(p, 0|1, 1, 1, 0)} - \cancel{Q^2 I(Q|1,1,1,0)}\bigg\} 
+ {\mathcal O}\left(\frac{1}{M^3}\right)
\sim {\mathcal O}\left(\frac{1}{M^3}\right)\,,
\label{eq:delta_seagull_NNLO}
\end{eqnarray} 
%%%%%%%%%%%%%%%%%%%%%%%%%%%%%%%%%%%%%%%%%%%%%%%%%%%%%%%%%%
\end{widetext} 
where, as said, the seagull terms have been evaluated exactly. Besides, they yield kinematically suppressed terms of
${\mathcal O}\left(1/M^3\right)$, which are dropped from our analytical results with the intended accuracy of 
${\mathcal O}\left(1/M^2\right)$. Thus, on the one hand, we see that the contributions arising from the NNLO seagull 
diagram (v), as well as the proton's structure-dependent form factor diagrams (p)-(s), are all of 
${\mathcal O}\left(1/M^3\right)$ [cf. Eq.~\eqref{eq:delta_seagull_NNLO} and the last line of Eq.~\eqref{eq:fNNLO}], 
namely, beyond our working accuracy, and hence, excluded from our results. On the other hand, it is interesting to 
find a kinematically enhanced ${\mathcal O}\left(1/M\right)$ term in our NNLO expression of 
$\delta^{(jkl\cdots u)}_{\gamma\gamma}$ which is formally commensurate with NLO. This is attributed to the 
contribution of the 3-point function $I^{(0)}(p,0|0,1,1,2)\sim {\mathcal O}(M)$ constituting the leading order part 
of the loop-integrals $I^{-}(p,0|0,1,1,2)$ and $I^{+}(p^\prime,0|0,1,1,2)$ whose expressions are provided in the 
appendix. Therefore, the only proton's structure-dependent contributions at this order can potentially arise from the
non-factorizable two-loop (i.e., a photon loop and a pion-loop) diagrams (cf. Fig.~\ref{fig:pi-loops}), along with 
appropriate counter-terms, which we have excluded in our analysis owing to the rather intricate nature of their 
analytical evaluations necessary to express their contributions to the cross-sections in algebraic closed form. In a
naive perturbative counting scheme, such two-loop contributions are expected to be subdominant compared to their 
one-loop counterparts. Nonetheless, this calls for a more rigorous systematic evaluation of the TPE including all NNLO 
contributions, without recourse to any type of approximations, including SPA. Such an endeavor is a subject matter for
a future publication. 
%%%%%%%%%%%%%%%%%%%%%%%%%%%%%%%%%%%%%%%%%%%%%%%%%%%%%%%%%% 

%%%%%%%%%%%%%%%%%%%%%%%%%%%%%%%%%%%%%%%%%%%%%%%%%%%%%%%%%% 
\subsection{A Summary of the TPE Expression} 
%%%%%%%%%%%%%%%%%%%%%%%%%%%%%%%%%%%%%%%%%%%%%%%%%%%%%%%%%% 
In the lab-frame the total fractional TPE contribution (relative to the LO) including up to the 
${\mathcal O}(\alpha/M^2)$ or NNLO corrections to the unpolarized $\ell$-p elastic differential cross-section, after 
isolating the residual ${\mathcal O}(1/M)$ IR-divergence $\delta^{\rm (box)}_{\rm IR}$ from diagrams (a) and (b), 
Eq.~\eqref{eq:delta_IR}, is given by
\begin{eqnarray}
\overline{\delta^{\rm (TPE)}_{\gamma \gamma}}(Q^2) \!\!&=&\!\! \frac{2{\mathcal R}e\sum\limits_{spins}
\left[\mathcal{M}_{\gamma}^{(0)*}\mathcal{M}_{\gamma \gamma}^{\rm (TPE)}\right]}{\sum\limits_{spins} 
\left|\mathcal{M}_{\gamma}^{(0)}\right|^2}-\delta^{\rm (box)}_{\rm IR}(Q^2)\,.
\nonumber\\
\end{eqnarray}
With the sum of all the 22 TPE amplitudes (a)-(v), given by 
$\mathcal{M}_{\gamma \gamma}^{\rm (TPE)} = {\mathcal{M}^{(a)}_{\rm box}} + \cdots + {\mathcal{M}^{(v)}_{\rm seagull}}$,    
the total IR-finite part of the TPE contributions can be expressed as
\begin{eqnarray}
\overline{\delta^{\rm (TPE)}_{\gamma \gamma}}(Q^2) \!\!&=&\!\!  \delta^{\rm (LO)}_{\gamma \gamma}(Q^2) 
+ \delta^{\rm (NLO)}_{\gamma \gamma}(Q^2) 
\nonumber\\
&&\!\! +\, \delta^{\rm (NNLO)}_{\gamma \gamma}(Q^2) + {\mathcal O}\left(M^{-3}\right)\,,
\end{eqnarray}
where $\delta^{\rm (LO)}_{\gamma \gamma}\sim {\mathcal O}(M^0)$, $\delta^{\rm (NLO)}_{\gamma \gamma}\sim {\mathcal O}(1/M)$
and $\delta^{\rm (NNLO)}_{\gamma \gamma}\sim {\mathcal O}(1/M^2)$. 
As mentioned earlier, our TPE evaluation employing SPA 
runs into a technical problem, namely, that the strict LO [i.e., ${\mathcal O}(M^0)$] contribution vanishes completely, 
thereby potentially spoiling the HB$\chi$PT power counting with a non-vanishing NLO contribution. To rectify this 
artifact, we have, therefore, included the exact LO TPE result, Eq.~\eqref{eq:delta_LO}, taken from the recent 
work~\cite{Choudhary:2023rsz} on the exact analytical TPE evaluation: 
\begin{eqnarray}
\delta^{\rm (LO)}_{\gamma \gamma}(Q^2) = \delta^{\rm (0)}_{\gamma \gamma}(Q^2)= 
\pi \alpha \frac{\sqrt{-Q^2}}{2E}\left(\frac{1}{1+\frac{Q^2}{4E^2}}\right)\,.
\label{eq:LO}
\end{eqnarray}
Next, by collecting all the NLO [i.e., ${\mathcal O}(1/M)$] contributions from Eqs.~\eqref{eq:delta_ab}, 
\eqref{eq:delta_c-h}, \eqref{eq:delta_seagull_NLO} and \eqref{eq:delta_j-u}, yields our total NLO fractional TPE contributions
to the elastic cross-section, relative to the LO Born term [cf. Eq.~\eqref{eq:diff_LO}]: 
\begin{widetext}
%%%%%%%%%%%%%%%%%%%%%%%%%%%%%%%%%%%%%%%%%%%%%%%%%%%%%%%%%%
\begin{eqnarray}
\delta^{\rm (NLO)}_{\gamma \gamma}(Q^2) &\stackrel{\gamma_{\rm soft}}{\leadsto}& -\frac{\alpha Q^2}{\pi ME \beta^2}
\Bigg[1+\frac{\pi^2}{12}\left(\beta+\frac{1}{\beta}\right)+(1-\beta)\ln\sqrt{\frac{1+\beta}{1-\beta}} 
- 2\ln\sqrt{\frac{2\beta}{1-\beta}} - 2\ln\sqrt{-\frac{Q^2}{2M E\beta}}
\nonumber\\
&&\hspace{1.75cm} +\,2\left(\beta+\frac{1}{\beta}\right)\ln\sqrt{-\frac{Q^2}{2M E\beta}}\ln\sqrt{\frac{1+\beta}{1-\beta}}
-\frac{1}{2}\left(3\beta+\frac{1}{\beta}\right)\ln^2\sqrt{\frac{1+\beta}{1-\beta}}-\beta {\rm Li}_2\left(\frac{2\beta}{1+\beta}\right)
\nonumber\\
&&\hspace{1.75cm} -\, \frac{1}{2}\left(\beta+\frac{1}{\beta}\right){\rm Li}_2\left(\frac{1+\beta}{1-\beta}\right)\Bigg]
- \frac{4\alpha E}{\pi M}\left[\frac{Q^2}{Q^2+4E^2}\right]\left(\frac{\nu_l^2-1}{\nu_l^2}\right)
\Bigg[\left(\frac{1+\nu_l^2}{2\nu_l}\right)\bigg\{\frac{\pi^2}{3}
\nonumber\\
&&\hspace{1.75cm} +\,\ln^2\sqrt{\frac{\nu_l+1}{\nu_l-1}}
+{\rm Li}_2\left(\frac{\nu_l-1}{\nu_l+1}\right)\bigg\}-\ln\sqrt{-\frac{Q^2}{m_l^2}}\,\,\Bigg]\,.
\label{eq:NLO_sum}
\end{eqnarray}
Likewise, collecting all the NNLO [i.e., ${\mathcal O}(1/M^2)$] contributions from Eqs.~\eqref{eq:delta_ab}, 
\eqref{eq:delta_c-h}, \eqref{eq:delta_seagull_NLO}, \eqref{eq:delta_j-u} and \eqref{eq:delta_seagull_NNLO}, yields our 
NNLO fractional TPE contributions, relative to the LO Born term [cf. Eq.~\eqref{eq:diff_LO}]:
\begin{eqnarray}
\delta^{\rm (NNLO)}_{\gamma \gamma}(Q^2)&\stackrel{\gamma_{\rm soft}}{\leadsto}&\frac{\alpha Q^4}{8 \pi M^2 E^2 \beta^5}
\Bigg[\frac{\pi^2}{6} (3-\beta^2)-18\beta+12\beta \ln{\sqrt{\frac{2\beta}{1-\beta}}}
-\left(3+\beta^2\right)\ln^2{\sqrt{\frac{1+\beta}{1-\beta}}}-(3-\beta^2)
\nonumber\\
&&\hspace{1.9cm} \times\,{\rm Li}_2\left(\frac{1+\beta}{1-\beta}\right)-2 \beta^2 {\rm Li}_2\left(\frac{2\beta}{1+\beta}\right)
+2 \ln{\sqrt{-\frac{Q^2}{2ME\beta}}}\left(6\beta+2(3-\beta^2)\ln{\sqrt{\frac{1+\beta}{1-\beta}}}\,\right)
\nonumber\\
&&\hspace{1.9cm} \,-2\left(4+3\beta-\beta^2 \right)\ln{\sqrt{\frac{1+\beta}{1-\beta}}}\,\,\Bigg]
+ \frac{8\alpha E^2}{\pi M^2}\bigg[\frac{Q^2}{Q^2+4E^2}\bigg]^2\left(\frac{\nu_l^2-1}{\nu_l^2}\right)
\Bigg[\left(\frac{1+\nu_l^2}{2\nu_l}\right)\bigg\{\frac{\pi^2}{3}
\nonumber\\
&&\hspace{1.9cm} +\,\ln^2{\sqrt{\frac{\nu_l+1}{\nu_l-1}}} + {\rm Li}_2\left(\frac{\nu_l-1}{\nu_l+1}\right)\bigg\}
- \ln\sqrt{-\frac{Q^2}{m_l^2}}\,\,\Bigg]-\frac{\alpha Q^2 }{2\pi M^2 \nu_l^2}\bigg[\frac{Q^2}{Q^2+4E^2}\bigg]
\Bigg[\left(\frac{\nu_l^2-1}{\nu_l}\right) 
\nonumber\\
&&\hspace{1.9cm} \times\,\bigg\{\frac{\pi^2}{3}+\ln^2{\sqrt{\frac{\nu_l+1}{\nu_l-1}}+{\rm Li}_2\left(\frac{\nu_l-1}{\nu_l+1}\right)}\bigg\} 
+ 2\ln\sqrt{-\frac{Q^2}{m_l^2}}\,\,\Bigg] + o\left(\frac{1}{M^2}\right)\,.
\label{eq:NNLO_sum}
\end{eqnarray}
%%%%%%%%%%%%%%%%%%%%%%%%%%%%%%%%%%%%%%%%%%%%%%%%%%%%%%%%%%
\end{widetext}
The $o\left(\frac{1}{M^2}\right)$ symbol\,\footnote{The symbol ``$o(\cdots)$" must be distinguished from the 
analogous symbol ``${\mathcal O}(\cdots)$", both denoting terms omitted in a given perturbative expression of a
certain order displayed within the parentheses. While the former denotes omitted terms of the same order as those 
appearing explicitly in the expression, the latter represents all possible higher-order omitted terms in the 
expression not relevant to the intended accuracy.} used above denotes all possible NNLO terms that may arise from 
the excluded non-factorizable two-loop contributions, as displayed in Fig.~\ref{fig:pi-loops}. The resultant 
magnitudes of the individual LO, NLO, and NNLO contributions, as well as their sum are shown in 
Fig.~\ref{fig:delta_TPE}. Furthermore, in Fig.~\ref{fig:delta_TPE_compare}, we present a comparison our SPA-based
TPE prediction with those of the earlier SPA works of Talukdar 
{\it et al.}~\cite{Talukdar:2019dko,Talukdar:2020aui}, as well as the {\it exact} TPE prediction of Choudhary 
{\it et al.}~\cite{Choudhary:2023rsz}.    

%%%%%%%%%%%%%%%%%%%%%%%%%%%%%%%%%%%%%%%%%%%%%%%%%%%%%%%%%%%%%%%%%%%%%%%%%%%%%%%%%%%%%%%%%%%%%%%%%%%%%%%
\section{Charge Asymmetry up to Next-to-leading Order} 
\label{sec:asym}
%%%%%%%%%%%%%%%%%%%%%%%%%%%%%%%%%%%%%%%%%%%%%%%%%%%%%%%%%%%%%%%%%%%%%%%%%%%%%%%%%%%%%%%%%%%%%%%%%%%%%%%
To evaluate the charge asymmetry to NLO accuracy, we include the HB$\chi$PT results for the various higher-order 
contributions (radiative and hadronic) to the unpolarized elastic lepton-proton scattering cross-section, along the 
SPA-based evaluated TPE results of Ref.~\cite{Talukdar:2020aui}. The differential cross-sections for lepton and 
anti-lepton scattering with the proton is expressed as
\begin{eqnarray}
\frac{{\rm d}\sigma^{(\mp)}_{el}(Q^2)}{{\rm d}\Omega^\prime_l} \!\!&=&\!\!
\left[\frac{{\rm d}\sigma_{el}(Q^2)}{{\rm d}\Omega^\prime_l}\right]_0
\nonumber\\
&&\times\,\left[1+\delta_{\rm even}(Q^2)\pm\delta_{\rm odd}(Q^2)\right]\,,
\end{eqnarray}
where the LO Born contribution is given by~\cite{Talukdar:2020aui}
\begin{eqnarray}
\left[\frac{{\rm d}\sigma_{el}(Q^2)}{{\rm d}\Omega^\prime_l}\right]_0 \!\!&=&\!\! 
\frac{\alpha^2 E^\prime\beta^\prime}{E\beta Q^2}\left(1-\frac{Q^2}{4M^2}\right)
\left[1+\frac{4EE^\prime}{Q^2}\right]\,.\quad\,
\label{eq:diff_LO}
\end{eqnarray}
Here, we recognize the charge-odd (charge-even) corrections $\pm\delta_{\rm odd}$ ($\delta_{\rm even}$), relative 
to the LO Born contribution, as those which change (do not change) signs when interchanging the polarities of the 
lepton charge. We first consider the parts of the TPE corrections up-to-and-including NLO expressed as
\begin{eqnarray}
\widetilde{\delta}^{\rm (TPE)}_{\gamma \gamma}(Q^2) \equiv \delta^{\rm (box)}_{\rm IR}(Q^2) 
+ \delta^{\rm (LO)}_{\gamma \gamma}(Q^2) 
+\delta^{\rm (NLO)}_{\gamma \gamma}(Q^2)\,.\quad\,
\label{eq:delta_tilde_TPE}
\end{eqnarray} 
Next by adding the low-energy NLO charge-odd soft-photon bremsstrahlung corrections 
$\delta^{\rm (odd;\,1)}_{\gamma \gamma^*}\sim {\mathcal O}(1/M)$ which were evaluated in Ref.~\cite{Talukdar:2020aui}, 
we obtain our total charge-odd radiative corrections $\delta_{\rm odd}$ up to NLO accuracy, expressed in the form: 
\begin{eqnarray}
\delta_{\rm odd}(Q^2)  = \widetilde{\delta}^{\rm (TPE)}_{\gamma \gamma}(Q^2) 
+ \delta^{\rm (odd;\,1)}_{\gamma \gamma^*}(Q^2) + o\left(\frac{1}{M^3}\right)\,, 
\label{eq:odd} 
\end{eqnarray}
where $o\left(\frac{1}{M^3}\right)$ denote neglected terms due to dynamical contributions from $N{}^3$LO charge-odd 
radiative corrections. The analytical expression for $\delta_{\rm odd}$ is given as
\begin{widetext}
%%%%%%%%%%%%%%%%%%%%%%%%%%%%%%%%%%%%%%%%%%%%%%%%%%%%%%%%%%
\begin{eqnarray}
\label{eq:delta_brem_odd}
\delta^{\rm (odd;\,1)}_{\gamma \gamma^*} (Q^2) \!\!&=&\!\! \delta^{\rm (soft)}_{\rm IR} (Q^2)
+ \frac{2\alpha}{\pi\beta} \ln\sqrt{\frac{-Q^2}{m^2_l}} \left\{\ln{\sqrt{\frac{1+\beta}{1-\beta}}}
-\frac{\beta}{\beta^\prime} \ln{\sqrt{\frac{1+\beta^\prime}{1-\beta^\prime}}}\,\right\}
\nonumber\\
&&-\,\frac{\alpha Q^2}{2\pi M E\beta}\left[\ln{\left(\frac{4\Delta^2_{\gamma^*}E^2}{-Q^2{E^\prime}^2}\right)}
\ln{\sqrt{\frac{1+\beta}{1-\beta}}}-\frac{1}{2}{\rm Li}_2\left(\frac{2\beta}{1+\beta}\right)
+\frac{1}{2} {\rm Li}_2\left(\frac{2\beta}{\beta-1}\right)\right]
\nonumber\\
&&-\,\frac{\alpha Q^2}{2\pi M\beta^\prime E^\prime}\left[\ln{\left(\frac{4\Delta^2_{\gamma^*}E^2}{-Q^2{E^\prime}^2}\right)}
\ln{\sqrt{\frac{1+\beta^\prime}{1-\beta^\prime}}}-\frac{1}{2}{\rm Li}_2\left(\frac{2\beta^\prime}{1+\beta^\prime}\right)
+\frac{1}{2} {\rm Li}_2\left(\frac{2\beta^\prime}{\beta^\prime-1}\right)\right]
\nonumber\\
&&+\,\frac{8\alpha \Delta_{\gamma^*} {E^\prime}^2 }{\pi M E^ 2 (Q^2+4 E E^\prime)}
\Bigg[-\Delta_{\gamma^*}(E+ E^\prime)+\Delta_{\gamma^*}\frac{E^2}{E^\prime \beta^\prime}
\left\{\left(1-\frac{\beta}{\beta^\prime}\right)\ln{\sqrt{\frac{1+\beta^\prime}{1-\beta^\prime}}}-\beta\right\}
\nonumber\\
&&\hspace{2.55cm}+\,\Delta_{\gamma^*}\frac{{E^\prime}^2}{E\beta} \left\{\left(1-\frac{\beta^\prime}{\beta}\right)
\ln{\sqrt{\frac{1+\beta}{1-\beta}}}-\beta^\prime \right\}+\frac{1}{E\beta}\Bigg\{2m^2_l\left(\frac{EE^\prime-E^{\prime 2}}{E}\right)
\nonumber\\
&&\hspace{2.55cm}+\,\Delta_{\gamma^*}\left(m^2_l-\frac{1}{2}Q^2\right)\Bigg\}\left\{\ln{\sqrt{\frac{1+\beta}{1-\beta}}}
-\frac{E\beta}{E^\prime \beta^\prime}\ln{\sqrt{\frac{1+\beta^\prime}{1-\beta^\prime}}}\,\right\}\Bigg]\,,
\end{eqnarray} 
%%%%%%%%%%%%%%%%%%%%%%%%%%%%%%%%%%%%%%%%%%%%%%%%%%%%%%%%%%
\end{widetext} 
where the term $\delta^{\rm (soft)}_{\rm IR}$ constitutes the IR-divergent part of the charge-odd soft-photon 
bremsstrahlung contributions~\cite{Talukdar:2020aui}. This IR-divergence is exactly equal but of opposite sign to the
TPE IR-divergent counterpart, i.e.,  $\delta^{\rm (soft)}_{\rm IR}=-\delta^{\rm (box)}_{\rm IR}$ 
[cf. Eq.~\eqref{eq:delta_IR}]. In other words, this implies that the two singular terms get eliminated in their sum in
$\delta_{\rm odd} (Q^2)$, Eq.~\eqref{eq:odd}. We should emphasize that the above result for the bremsstrahlung 
contributions borrowed from Ref.~\cite{Talukdar:2020aui} pertains to the use of the Coulomb gauge, whereas the TPE 
calculations invoking SPA in this work were carried out in the Lorentz gauge. Currently, work is in progress for a 
future publication~\cite{Bhoomika24}, where we shall consider the bremsstrahlung calculations (both charge-even and 
-odd) adopting the Lorentz gauge at NLO accuracy. In this process, possible sources of systematic errors due to the 
mismatch between the different gauge-dependent terms should get eliminated. It is nonetheless important to stress that
$\delta^{\rm (soft)}_{\rm IR}$ has precisely the same structure irrespective of the choice of the gauge. The NLO
expression for the total charge-odd radiative correction $\delta_{\rm odd}$ in Eq.~\eqref{eq:odd} includes the 
IR-finite combination of our SPA-based TPE corrections $\tilde{\delta}^{\rm (TPE)}_{\gamma \gamma}$, 
Eq.~\eqref{eq:delta_tilde_TPE}, as well as the associated charge-odd soft-photon bremsstrahlung counterparts 
$\delta^{\rm (odd;\,1)}_{\gamma \gamma^*}$, Eq.~\eqref{eq:delta_brem_odd}, considering all dynamical corrections terms
up-to-and-including NLO. 

We now consider the total LO charge-even radiative corrections (both real and virtual), 
$\delta^{\rm (even;\,0)}_{2\gamma}\sim {\mathcal O}(M^0)$, as well as the leading hadronic corrections to the OPE, 
$\delta^{(2)}_{\chi}\sim {\mathcal O}(1/M^2)$. Together, these constitute our total charge-even contributions 
$\delta_{\rm even}$ to NLO accuracy. Thus, we can express the charge-even contributions to the cross-section in the 
form\footnote{Notably, in HB$\chi$PT there are no ${\mathcal O}(1/M)$ charge-even corrections (either hadronic or 
radiative) to the elastic OPE lepton-proton cross-section arising up-to-and-including NLO 
accuracy~\cite{Talukdar:2020aui}. The subleading corrections begin contributing at ${\mathcal O}(1/M^2)$. }:
\begin{eqnarray}
\delta_{\rm even}(Q^2)= \delta^{\rm (even;\,0)}_{2\gamma}(Q^2) + \delta^{(2)}_{\chi}(Q^2) + o\left(\frac{1}{M^3}\right)\,,
\end{eqnarray}
where $o\left(\frac{1}{M^3}\right)$ denotes neglected terms due to N${}^3$LO charge-even radiative corrections. Below, 
we spell out the analytical results of Ref.~\cite{Talukdar:2020aui} which we borrow in our work to obtain our results
for the total charge-even fractional corrections. The following two results are considered: 

First, the expression for NLO charge-even radiative corrections $\delta^{\rm (even;\,0)}_{2\gamma}$ is given by the 
following expression\footnote{In Eqs.~\eqref{eq:delta_red_even} and \eqref{eq:var_kinetics}, a few typographical
errors which were found to occur in the corresponding expression, Eq.~(67) of Ref.~\cite{Talukdar:2020aui}, have been
rectified here.}:
\begin{widetext}
%%%%%%%%%%%%%%%%%%%%%%%%%%%%%%%%%%%%%%%%%%%%%%%%%%%%%%%%%%
\begin{eqnarray}
\delta^{\rm (even;\,0)}_{2\gamma} (Q^2) \!\!&=&\!\! \frac{\alpha}{\pi}\Bigg[\frac{\nu_l^2+1}{4\nu_l}
\ln{\left(\frac{\nu_l+1}{\nu_l-1}\right)}\ln{\left(\frac{\nu_l^2-1}{4\nu_l^2}\right)}+\frac{2\nu_l^2+1}{2\nu_l}
\ln{\left(\frac{\nu_l+1}{\nu_l-1}\right)}-\frac{\nu_l^2+1}{2\nu_l}\bigg\{{\rm Li}_2\left(\frac{\nu_l-1}{2\nu_l}\right)
-{\rm Li}_2\left(\frac{\nu_l+1}{2\nu_l}\right)\bigg\}
\nonumber\\
&&\hspace{0.4cm}-\,2+\sum_{f=e,\mu,\tau} \Bigg\{\frac{2}{3}\left(\nu_f^2-\frac{8}{3}\right)+\nu_f\left(\frac{3-\nu_f^2}{3}\right)
\ln{\left(\frac{\nu_f+1}{\nu_f-1}\right)}\Bigg\}-\frac{2}{3}\left(\nu_{\pi}^2+\frac{1}{3}\right)+\frac{\nu_{\pi}^2}{3}
\ln{\left(\frac{\nu_{\pi}+1}{\nu_{\pi}-1}\right)}
\nonumber\\
&&\hspace{0.4cm}+\,\frac{1}{\nu_l}\left(\frac{2m_l^2}{Q^2+4E E^\prime}\right)\left(1-\frac{Q^2}{4M^2}\right)\ln{\left(\frac{\nu_l+1}{\nu_l-1}\right)}
+\ln{\left(\frac{4\Delta^2_{\gamma^*}E^2}{m_l^2E^{\prime 2}}\right)}\bigg[\frac{\nu_l^2+1}{2\nu_l}\ln{\left(\frac{\nu_l+1}{\nu_l-1}\right)}-1\bigg]
\nonumber\\
&&\hspace{0.4cm}+\,\frac{1}{\beta}\ln{\sqrt{\frac{1+\beta}{1-\beta}}}+\frac{1}{\beta^\prime}\ln{\sqrt{\frac{1+\beta^\prime}{1-\beta^\prime}}}
-\frac{\nu_l^2+1}{2\nu_l}\Bigg\{\ln^2{\sqrt{\frac{1+\beta}{1-\beta}}}-\ln^2{\sqrt{\frac{1+\beta^\prime}{1-\beta^\prime}}}
-{\rm Li}_2\left(1+\frac{E-\lambda_\nu E^\prime}{E^\prime(1-\beta^\prime)\xi_\nu}\right)
\nonumber\\
&&\hspace{0.4cm}-\,{\rm Li}_2\left(1+\frac{E-\lambda_\nu E^\prime}{E^\prime(1+\beta^\prime)\xi_\nu}\right)+{\rm Li}_2
\left(1+\frac{E-\lambda_\nu E^\prime}{E (1-\beta)\lambda_\nu\xi_\nu}\right)
+{\rm Li}_2\left(1+\frac{E- \lambda_\nu E^\prime }{E (1+\beta)\lambda_\nu\xi_\nu}\right)\Bigg\}\Bigg]\,,
\label{eq:delta_red_even}
\end{eqnarray}
where, the kinematic factors $\nu_f, \nu_\pi, \xi_\nu$, and $\lambda_\nu$, are respectively, defined as 
\begin{eqnarray}
\nu_f=\sqrt{1-\frac{4m^2_f}{Q^2}}\,, \quad  \nu_\pi=\sqrt{1-\frac{4m^2_\pi}{Q^2}}\,, \quad \xi_\nu 
= \frac{2 \nu_l}{(\nu_l + 1)(\nu_l -1)}\,, \quad \text{and} \quad \lambda_\nu = \frac{\nu_l-1}{\nu_l+1}\,. 
\label{eq:var_kinetics}
\end{eqnarray}
%%%%%%%%%%%%%%%%%%%%%%%%%%%%%%%%%%%%%%%%%%%%%%%%%%%%%%%%%%    
\end{widetext} 
In the above expressions $\Delta_{\gamma^*}$ is an arbitrarily introduced free parameter~\cite{Mo:1968cg}. It 
represents a small-scale parameter stemming from the bremsstrahlung contributions that distinguishes a detectable 
hard-photon ($\gamma^*_{\rm hard}$) with a photon energy, $E_{\gamma^*}\gtrsim \Delta_{\gamma^*}$, from a 
soft-photon ($\gamma^*_{\rm soft}$) with energy, $E_{\gamma^*}\lesssim\Delta_{\gamma^*}$. The parameter 
$\Delta_{\gamma^*}$ has a magnitude smaller than any (energy/momentum) scale in the elastic scattering process. 
The inelastic soft-photon emissions are coherent with our elastic scattering process of interest. Therefore, such
a parameter serves as a detector {\it veto} in a practical experimental set-up, eliminating ``contamination" due 
to inelastic hard-photon emission background events ($\ell+{\rm p}\to \ell+{\rm p}+\gamma^*_{\rm hard}$). Since the
soft-photons go undetected, it is necessary to integrate the part of the {\it elastic radiative-tail} 
distribution~\cite{Mo:1968cg} up to photon energies of the order of $\Delta_{\gamma^*}$. Hence, its appearance in 
the aforementioned expressions for the radiative corrections to the elastic cross-section. 

Second, the result for the charge-even hadronic corrections in terms of the proton's electric radius 
$r_p=\sqrt{\langle r^2_E \rangle}$~\cite{Talukdar:2020aui} is given by
\begin{widetext}

\vspace{-0.2cm}

%%%%%%%%%%%%%%%%%%%%%%%%%%%%%%%%%%%%%%%%%%%%%%%%%%%%%%%%%%   
\begin{eqnarray}
\delta^{(2)}_{\chi}(Q^2) = \frac{Q^2}{3}\left[\langle r^2_E \rangle-\frac{3\kappa_p}{2M^2}\right] 
+ \frac{Q^2}{4M^2}\Bigg[1+2\kappa_p+(1+\kappa_p)^2\left(\frac{Q^2+4m_l^2- 4E^2}{Q^2+4E^2}\right) \Bigg] 
+ \mathcal{O}\left(\frac{1}{M^3}\right)\,.
\label{eq:hadronic}
\end{eqnarray}
%%%%%%%%%%%%%%%%%%%%%%%%%%%%%%%%%%%%%%%%%%%%%%%%%%%%%%%%%%    
\end{widetext} 
Both the analytical expressions for the charge-even corrections to the LO cross-section were worked out in HB$\chi$PT
by Talukdar {\it et al.}~\cite{Talukdar:2020aui}. 

Finally, we remark that the measurement of the charge asymmetry observable provides the most viable method to 
accurately pin down the TPE contributions ({\it via} the extraction of the charge-odd contributions) from elastic 
lepton-proton and anti-lepton-proton scattering cross-sections. The following ratio proportional to the difference 
of the differential cross-sections for lepton and anti-lepton scattering with the proton, defines the asymmetry 
${\cal A}^{(1)}$ observable up to NLO accuracy considered in this work:
\begin{eqnarray}
{\mathcal A}^{(1)}_{\ell^\pm}(Q^2) \!\!&=&\!\! \left[\frac{\frac{{\rm d}\sigma^{(-)}_{el}(Q^2)}{{\rm d}\Omega^\prime_l}
-\frac{{\rm d}\sigma^{(+)}_{el}(Q^2)}{{\rm d}\Omega^\prime_l}}{\frac{{\rm d}\sigma^{(-)}_{el}(Q^2)}{{\rm d}\Omega^\prime_l}
+\frac{{\rm d}\sigma^{(+)}_{el}(Q^2)}{{\rm d}\Omega^\prime_l}}\right]
\nonumber\\
&=&\!\!\frac{\delta_{\rm odd}(Q^2)}{1+ \delta_{\rm even}(Q^2)} + o\left(\frac{1}{M^2}\right)\,.
\label{eq:A_asy}
\end{eqnarray} 
As stated, the term $\delta_{\rm even}$ includes the renormalized NNLO hadronic corrections $\delta^{(2)}_{\chi}$, and
the subset of the UV/IR-finite LO radiative corrections $\delta^{\rm (even;\,0)}_{2\gamma}$ to the OPE process, namely, 
the virtual corrections (self-energy, vertex, and vacuum polarization) and their associated soft-photon 
bremsstrahlung counterparts. As noted in Ref.~\cite{Talukdar:2020aui}), the NLO charge-even radiative corrections 
vanish. Also note that the IR-divergent terms of $\delta^{\rm (even;\,0)}_{2\gamma}$, Eq.~\eqref{eq:delta_red_even}, 
cancel (not shown explicitly) in the evaluation of the charge-even radiative corrections between the charge-even 
virtual photon-loop corrections and the corresponding soft-photon bremsstrahlung corrections~\cite{Talukdar:2020aui}.

%%%%%%%%%%%%%%%%%%%%%%%%%%%%%%%%%%%%%%%%%%%%%%%%%%%%%%%%%%%%%%%%%%%%%%%%%%%%%%%%%%%%%%%%%%%%%%%%%%%%%%%
\section{Numerical Results and Discussion}
\label{sec:results}
%%%%%%%%%%%%%%%%%%%%%%%%%%%%%%%%%%%%%%%%%%%%%%%%%%%%%%%%%%%%%%%%%%%%%%%%%%%%%%%%%%%%%%%%%%%%%%%%%%%%%%%
We present our numerical results pertaining to the analytical TPE evaluations invoking SPA, as well as their 
contribution to the charge asymmetry between the lepton-proton and anti-lepton-proton scattering, obtained in the
previous sections. Notably, the numerical results are presented in view of the low-energy kinematical region of 
the planned MUSE experiment. The main purpose of our results is to obtain an improved version of the SPA-based 
prediction of the TPE, which was derived in the previous works of Talukdar 
{\it et al.}~\cite{Talukdar:2019dko,Talukdar:2020aui}. In addition, we analyze the discrepancy introduced by 
invoking SPA in comparison with the {\it exact} analytical TPE estimation in HB$\chi$PT in the recent work of 
Choudhary {\it et al.}~\cite{Choudhary:2023rsz}.    

\begin{figure*}[tbp]
\begin{center}
    \includegraphics[width=0.48\linewidth]{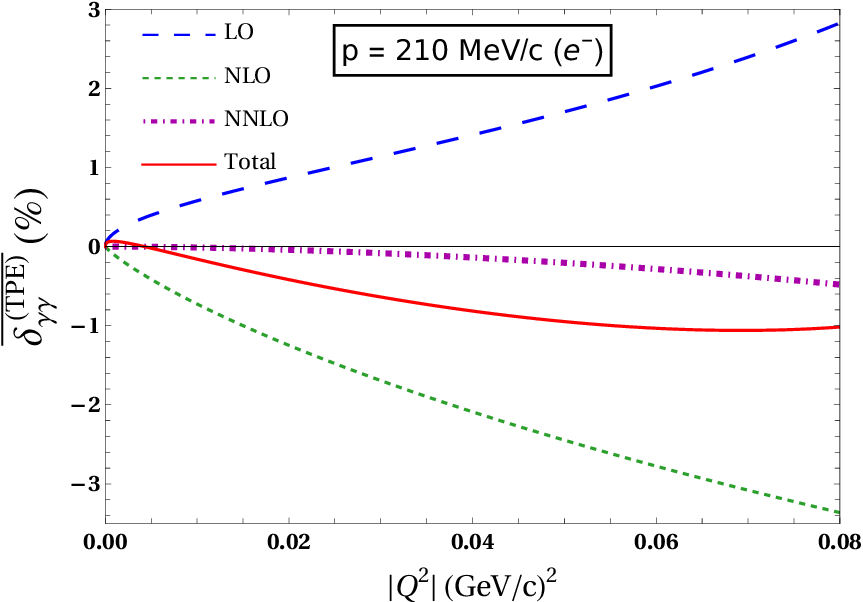}~\quad~\includegraphics[width=0.48\linewidth]{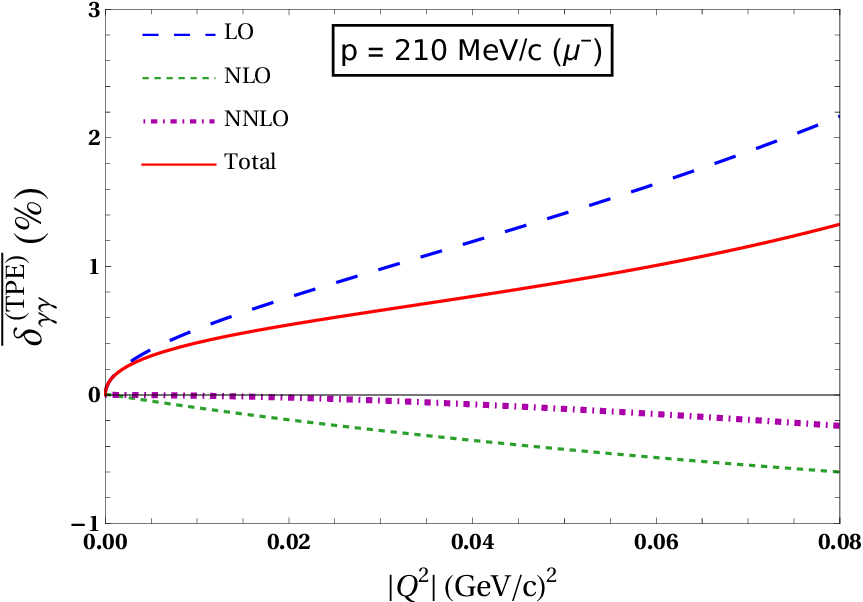}

\vspace{0.6cm}
    
    \includegraphics[width=0.48\linewidth]{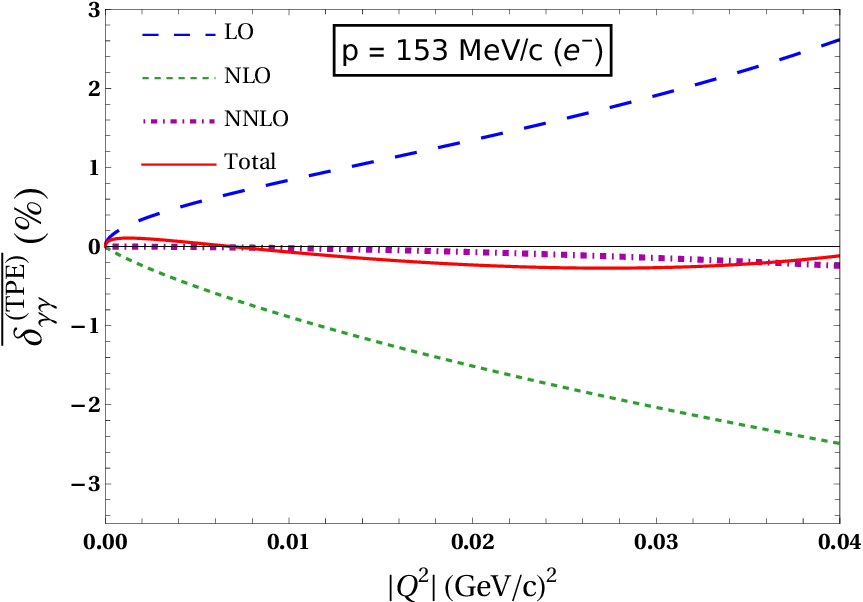}~\quad~\includegraphics[width=0.48\linewidth]{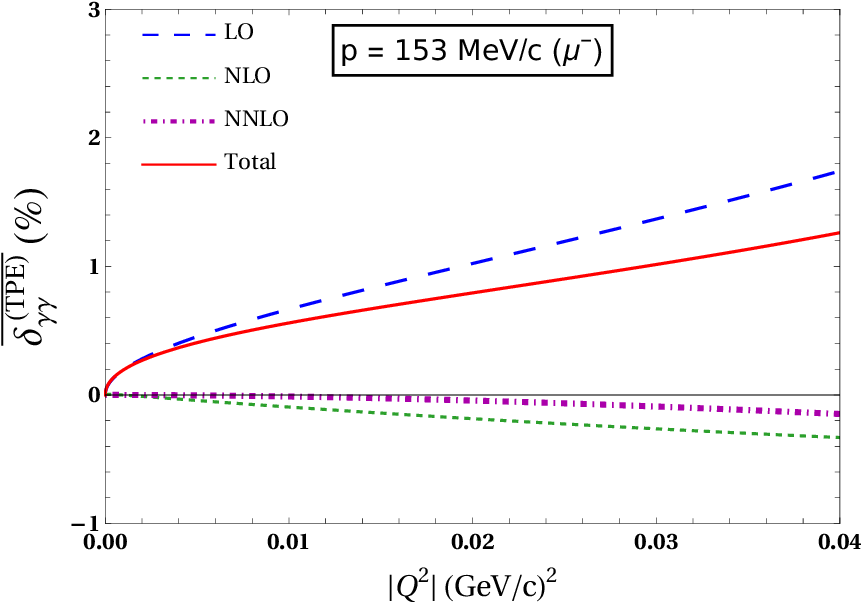}

\vspace{0.6cm}
    
\includegraphics[width=0.48\linewidth]{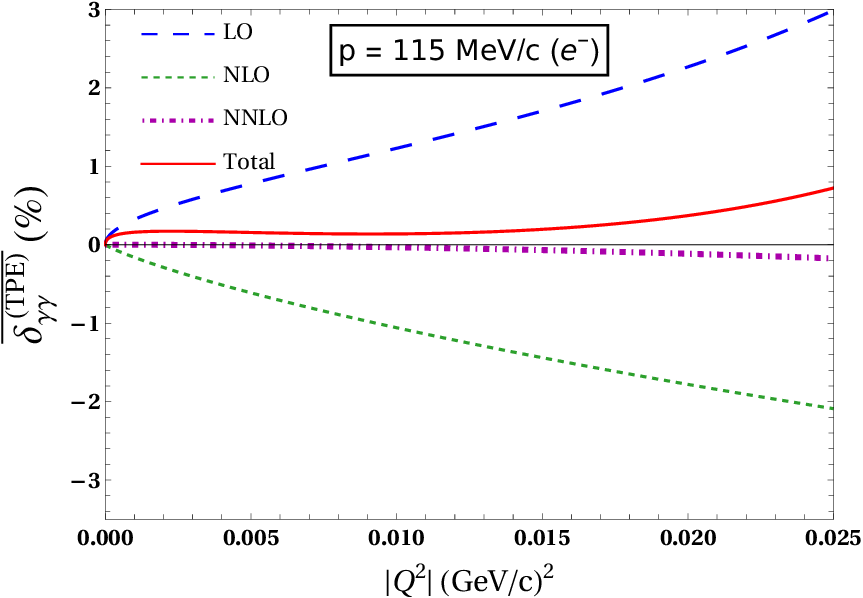}~\quad~\includegraphics[width=0.48\linewidth]{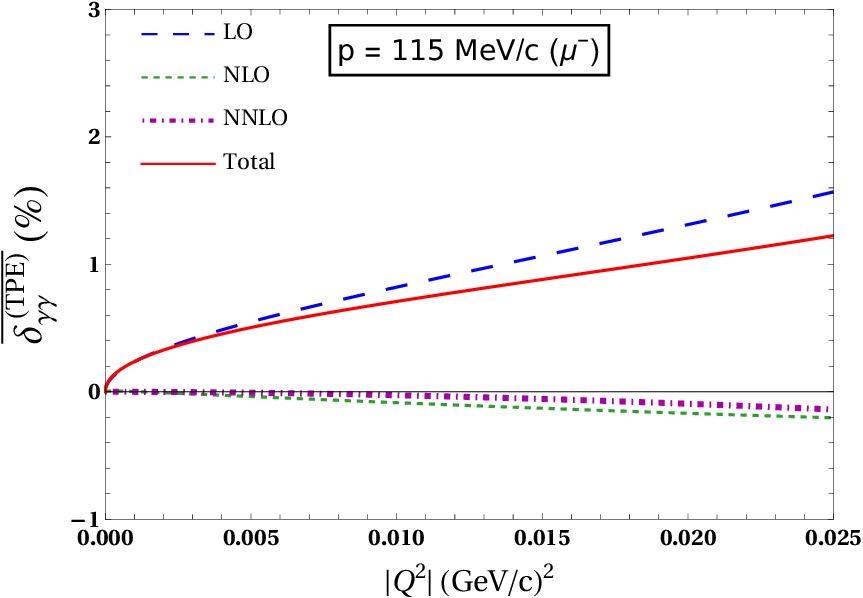}
\caption{Comparison of the IR-finite fractional TPE contributions (invoking SPA) at various orders (LO, NLO and
         NNLO) in HB$\chi$PT to the e-p (left panel) and $\mu$-p (right panel) elastic scattering cross-sections 
         as a function of the squared four-momentum transfer $|Q^2|$ given at the three MUSE proposed incoming 
         lepton beam three-momenta: 210~MeV/c, 153~MeV/c, and 115~MeV/c. The LO and NLO corrections are 
         represented by Eqs.~\eqref{eq:LO} and \eqref{eq:NLO_sum}, respectively. The NNLO corrections, 
         Eqs.~\eqref{eq:NNLO_sum}, include only the dominant proton's form factor contribution, excluding 
         contributions from the two-loop topologies with non-factorizable pion-loops (cf. Fig.~\ref{fig:pi-loops}).
         All the fractional contributions are considered relative to the LO Born differential cross-section [cf. 
         Eq.~\eqref{eq:diff_LO}].}
\label{fig:delta_TPE}
\end{center}
\end{figure*}

First, we display in Fig.~\ref{fig:delta_TPE} the dependence of the three different orders up-to-and-including the
NNLO corrections for the finite fractional TPE contributions in SPA to the e-p and $\mu$-p elastic scattering 
cross-sections as a function of the squared four-momentum transfer $|Q^2|$ corresponding to the MUSE kinematics (cf.
Table I of Ref.~\cite{Talukdar:2019dko}). In our formalism, LO, NLO, and NNLO refer to the collection of all terms 
contributing to the elastic cross-section which scale as $\mathcal{O}(M^0)$, $\mathcal{O}(1/M)$ and 
$\mathcal{O}(1/M^2)$, respectively. These three orders encompass both dynamical and kinematical contributions 
originating from TPE diagrams, as described in Sec.~\ref{sec:theory}. Although using SPA the true LO contribution 
vanishes, we nevertheless enforce the correct HB$\chi$PT power counting ``by hand" by including the true LO part 
following from the exact analytical TPE results of Ref.~\cite{Choudhary:2023rsz}, Eq.~\eqref{eq:delta_LO}, akin to 
the well-known McKinley-Feshbach contribution~\cite{McKinley:1948zz}. Hence, in Fig.~\ref{fig:delta_TPE} we display
this LO TPE result along with our subleading contributions obtained in SPA. We re-emphasize here that in the LO TPE
diagrams (a) and (b), there are additional kinematically suppressed corrections that scale as $\mathcal{O}(1/M)$. 
Such terms formally contribute to the NLO cross-section, Eq.~\eqref{eq:delta_ab}, along with the dynamical 
contributions stemming from the NLO TPE box- and crossed-box diagrams (c)-(h), Eq.~(\ref{eq:delta_c-h}), and the 
seagull contribution, Eq.~\eqref{eq:delta_seagull_NLO}. In addition, we also include within the NLO TPE fractional
corrections, the (unusual) kinematically enhanced contributions scaling as $\mathcal{O}(1/M)$ which stem from 
genuine NNLO TPE diagrams (j)-(u), Eq.~(\ref{eq:delta_j-u}). All such NLO TPE contributions are consolidated in 
Eq.~\eqref{eq:NLO_sum} and plotted in Fig.~\ref{fig:delta_TPE}. Likewise, kinematically suppressed terms that 
scale as $\mathcal{O}(1/M^2)$ have multiple origins: (i) the LO TPE diagrams, Eq.~(\ref{eq:delta_ab}), (ii) the 
NLO TPE diagrams, Eqs.~\eqref{eq:delta_c-h} and ~\eqref{eq:delta_seagull_NLO}, and (iii) the contributions from the
genuine NNLO TPE diagrams, Eqs.~\eqref{eq:delta_j-u} and \eqref{eq:delta_seagull_NNLO}. These are consolidated as a
part of our NNLO TPE corrections, Eq.~\eqref{eq:NNLO_sum}, and plotted in the Fig.~\ref{fig:delta_TPE}. It is to be
noted that the NNLO corrections, in principle, should also include dynamical corrections arising from complicated 
two-loop NNLO TPE diagrams with non-factorizable pion-loops, as shown in Fig.~\ref{fig:pi-loops}. However, these 
contributions are neglected in this work on the rationale that they are tiny compared to the rest of the NNLO TPE 
diagrams considered in this work, Fig.~\ref{fig:NNLO_TPE}. Figure~\ref{fig:delta_TPE} also displays our SPA-based 
TPE predictions for elastic lepton-proton scattering, as obtained by summing the NLO and NNLO corrections to the exact
LO Feshbach-like result, Eq.~\eqref{eq:delta_LO} (solid red curves labeled as ``Total"). It goes without mentioning,
that the TPE contributions vanish as $|Q^2|\rightarrow0$, just as we expect of all virtual radiative corrections to 
the elastic cross-section. In addition, for the electron scattering, we also observe large cancellations between the
positive (true) LO and the negative NLO contributions, significantly reducing the magnitude of the total fractional 
TPE corrections invoking SPA. 

%%%%%%%%%%%%%%%%%%%%%%%%%%%%%%%%%%%%%%%%%%%%%%%%%%FIGURE-6%%%%%%%%%%%%%%%%%%%%%%%%%%%%%%%%%%%%%%%%%%%%%%%%%%%%%%%
%%%%%%%%%%%%%%%%%%%%%%%%%%%%%%%%%%%%%%%%%%%%%%%%RESULT-PLOT-1%%%%%%%%%%%%%%%%%%%%%%%%%%%%%%%%%%%%%%%%%%%%%%%%%%%%
\begin{figure*}[tbp]
\begin{center}
    \includegraphics[width=0.48\linewidth]{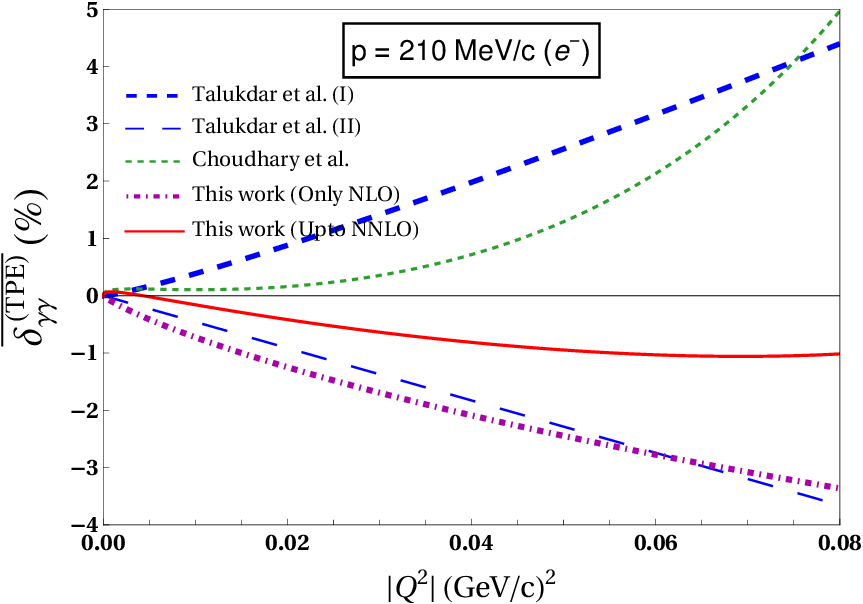}~\quad~\includegraphics[width=0.48\linewidth]{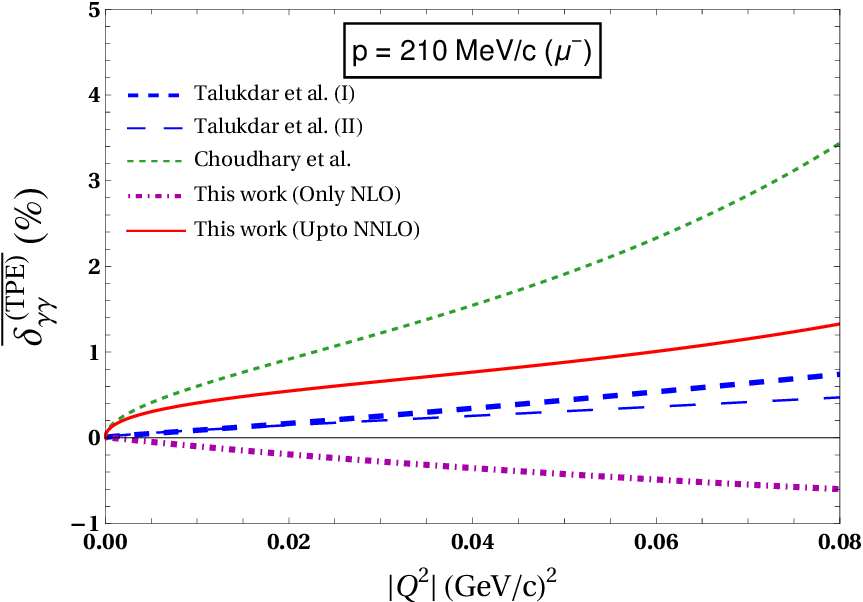}

\vspace{0.6cm}
    
    \includegraphics[width=0.48\linewidth]{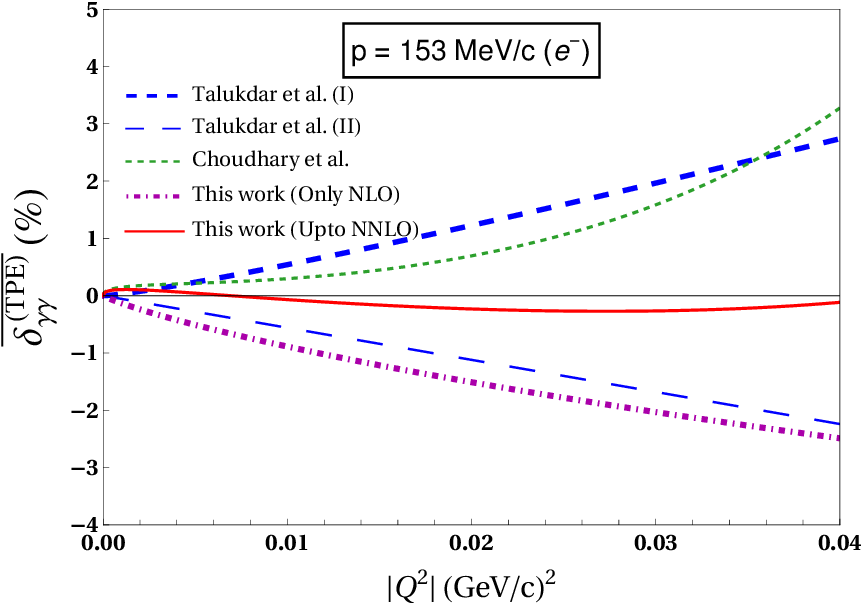}~\quad~\includegraphics[width=0.48\linewidth]{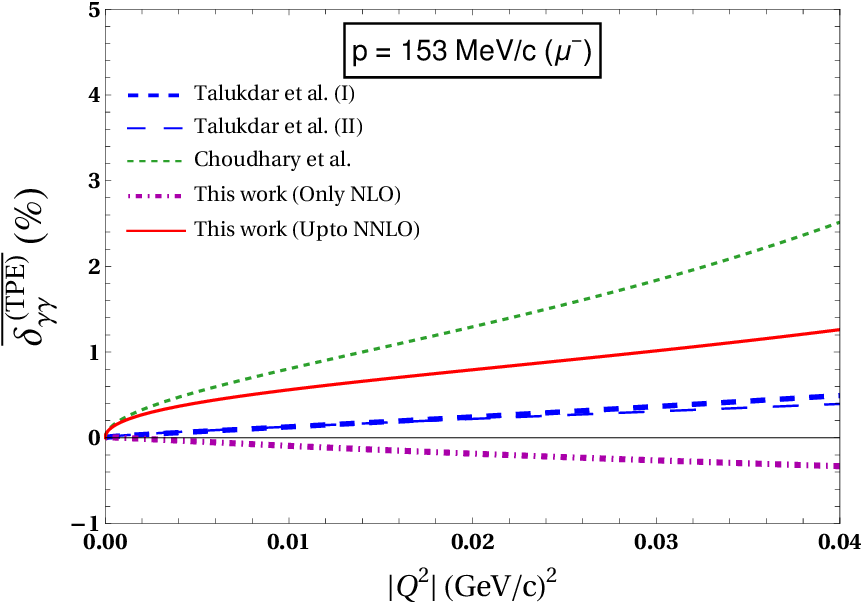} 

\vspace{0.6cm}

    \includegraphics[width=0.48\linewidth]{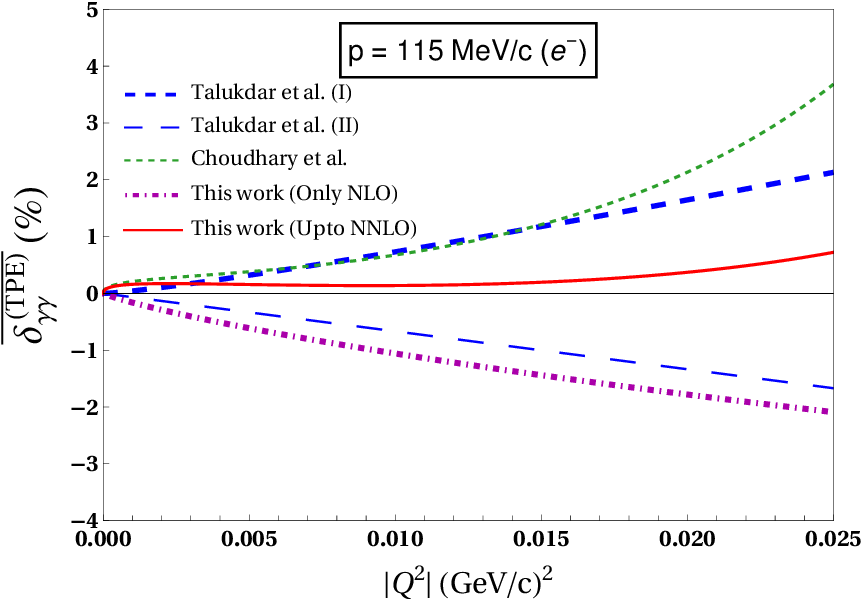}~\quad~\includegraphics[width=0.48\linewidth]{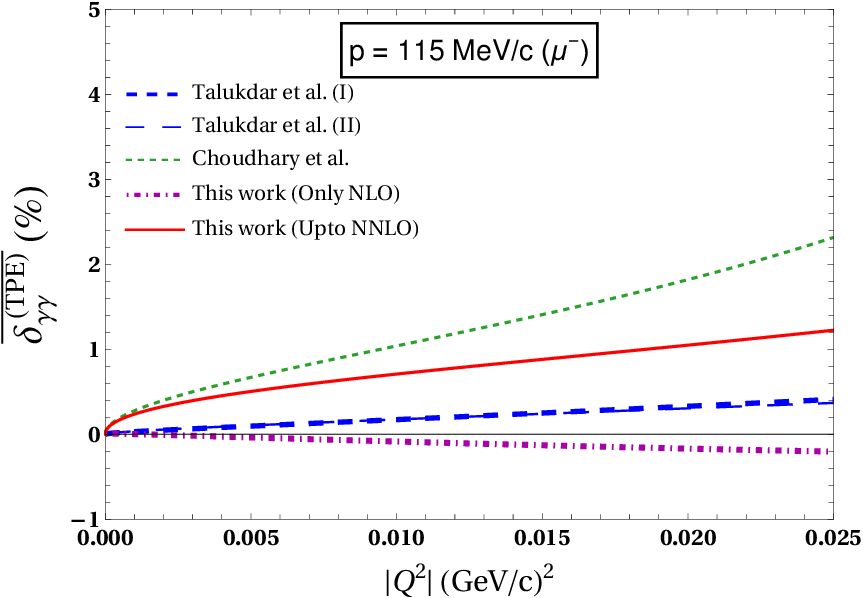}
    \caption{Comparison of our complete SPA results up to NNLO for the IR-finite TPE contributions with the 
             analogous SPA-based NLO HB$\chi$PT calculation of Talukdar {\it et al.}~\cite{Talukdar:2019dko}
             where the LO contributions were found to vanish completely, i.e., without Feshbach-like term. Two 
             types of plots for Talukdar {\it et al.}~\cite{Talukdar:2019dko} are displayed: plot (I) represents
             the original results of Ref.~\cite{Talukdar:2019dko}, while plot (II) represents modifications we 
             made in the original expression to match the IR-subtraction scheme adopted in this work. In 
             addition, the comparison is also shown with the exact analytical TPE evaluation in HB$\chi$PT in 
             the work of Choudhary et al.~\cite{Choudhary:2023rsz}. The e-p ($\mu$-p) elastic scattering results
             are displayed on the left (right) panel as a function of the squared four-momentum transfer
             $|Q^2|$ at the three MUSE proposed incoming lepton beam three-momenta, 210~MeV/c, 153~MeV/c, and 
             115~MeV/c. The fractional contributions are considered relative to the LO Born differential 
             cross-section [cf. Eq.~\eqref{eq:diff_LO}].}
\label{fig:delta_TPE_compare} 
\end{center}
\end{figure*}

%%%%%%%%%%%%%%%%%%%%%%%%%%%%%%%%%%%%%%%%%%%%%%%%%%%%%%%%%%%%%%%%%%%%%%%%%%%%%%%%%%%%%%%%%%%%%%%%%%%%%%%%%%%%%%%%%
For the NLO TPE contributions in Eq.~\eqref{eq:NLO_sum}, one finds terms proportional to the so-called 
{\it chiral-logarithm}, $\ln{\sqrt{-\frac{Q^2}{2M \beta E}}}$, which are responsible for the significant negative 
enhancement of the electron results. This contrasts the previous SPA results of Talukdar 
{\it et al.}~\cite{Talukdar:2019dko,Talukdar:2020aui}, where such enhancements were missing. Moreover, as already 
discussed in the introduction, the IR-divergence subtraction scheme used in those SPA-based works was $Q^2$-dependent,
which contrasts with our improved $Q^2$-independent scheme [see Eq.~\eqref{eq:delta_IR}] (also see 
Ref.~\cite{Choudhary:2023rsz}). Notably, modifying the TPE result of Ref.~\cite{Talukdar:2019dko} by incorporating the
same subtraction scheme as used here, yields results similar to our NLO corrections for the electron scattering case,
[see Fig.~\ref{fig:delta_TPE_compare}]. We also find that our NLO SPA results for both the electron and muon are 
negative, decreasing with decreasing incoming lepton-beam three-momentum. This behavior contrasts the exact NLO TPE 
results of Choudhary {\it et al}.~\cite{Choudhary:2023rsz}, where the contributions were obtained strictly positive for
the muon, while for electrons they were predominantly negative for most of the low-$Q^2$ MUSE domain. 

Regarding the NNLO corrections in Fig.~\ref{fig:delta_TPE}, besides the kinematically suppressed $\mathcal{O}(1/M^2)$
terms stemming from the LO and the NLO TPE diagrams (a)-(i), we also include the dynamical contributions from the 
genuine NNLO TPE amplitudes (j)-(v), some of which contain contributions from the proton's structure. However, we 
find that all such structure-dependent terms, along with the NNLO seagull contribution, are kinematically suppressed
to $\mathcal{O}(1/M^3)$, and hence, dropped explicitly in the NNLO fractional TPE contributions displayed in 
Eqs.~\eqref{eq:delta_j-u} and \eqref{eq:delta_seagull_NNLO}. As seen from Fig.~\ref{fig:delta_TPE}, the NNLO 
contributions are relatively small compared to the more important LO and NLO contributions (less than $\sim 15\%$ of
the NLO corrections in the MUSE regime). The negative nature of the NNLO contributions is reminiscent of the typical 
corrections induced by incorporating phenomenological form factors, which are known to suppress the magnitude of the 
overall TPE and other radiative corrections. Their magnitude increases with increasing incoming lepton momentum both 
for the electron and muon. Also apparent in the figure is that our NNLO results compared to the NLO are much less 
sensitive to the lepton mass $m_l$ variation. 
%%%%%%%%%%%%%%%%%%%%%%%%%%%%%%%%%%%%%%%%%%%%%%%%%%FIGURE-7%%%%%%%%%%%%%%%%%%%%%%%%%%%%%%%%%%%%%%%%%%%%%%%%%%%%%%%
%%%%%%%%%%%%%%%%%%%%%%%%%%%%%%%%%%%%%%%%%%%%%%%%RESULT-PLOT-2%%%%%%%%%%%%%%%%%%%%%%%%%%%%%%%%%%%%%%%%%%%%%%%%%%%%
%%%%%%%%%%%%%%%%%%%%%%%%%%%%%%%%%%%%%%%%%%%%%%%%%%%%%%%%%%%%%%%%%%%%%%%%%%%%%%%%%%%%%%%%%%%%%%%%%%%%%%%%%%%%%%%%%

Next in Fig.~\ref{fig:delta_TPE_compare}, we display a comparison between our SPA-based TPE results with the analogous 
HB$\chi$PT results evaluated earlier by Talukdar {\it et al.}~\cite{Talukdar:2019dko}, and with the exact TPE results 
of Choudhary {\it et al.}~\cite{Choudhary:2023rsz}. The blue thick dashed curves, labeled ``Talukdar et al. (I)", 
represent the results of Talukdar {\it et al.} with a $Q^2$-dependent IR cancellation scheme which differs from the 
$Q^2$-independent scheme used in this work. The blue thin long-dashed curves, labeled ``Talukdar et al. (II)", 
represent the results of Talukdar {\it et al.} which we have modified by incorporating the same IR-subtraction scheme
used by Choudhary {\it et al.}, and also adopted in this work.\footnote{The finite part of the TPE corrections in 
Talukdar {\it et al.} had to be re-adjusted with a term $\ln\left(-Q^2/m_l^2\right)$ for the sake of comparison with 
other TPE works in the literature with $Q^2$-independent IR-divergent terms. Nevertheless, we emphasize that the total 
radiative corrections (including the bremsstrahlung) remain unaffected by such a subtraction scheme ambiguity.} 
%
% I must insist on keeping this footnote because we are emphasizing this crucial change in our summarizing discussion. Moreover, we have never said earlier that the ambiguity of the total radiative correction remains unaffected by the subtraction scheme.
%
Also displayed in Fig.~\ref{fig:delta_TPE_compare} are our NLO TPE corrections to the elastic cross-section which
include all $\mathcal{O}(1/M)$ terms represented by the violet dot-dashed curves labeled ``This work (Only NLO)". 
This, in particular, facilitates a direct comparison with our modified ``Talukdar et al. (II)" results of 
Ref.~\cite{Talukdar:2019dko}, which rely on the same $Q^2$-independent IR-subtraction scheme while excluding the 
Feshbach-like term, Eq.~\eqref{eq:delta_LO}. Finally, our SPA results for the total fractional TPE contributions 
including the partial NNLO corrections (i.e., LO+NLO+NNLO) are displayed in the figure, as represented by the red 
solid-line curves, labeled ``This work (Upto NNLO)". We find that our SPA results differ significantly from the
corresponding exact TPE results of Choudhary {\it et al.}.~\cite{Choudhary:2023rsz}. Notably, in the case of NLO 
results for electron scattering, both our SPA and the exact TPE results of Choudhary {\it et al.} yield negative 
contributions at the low-$|Q^2|$ MUSE kinematics, with the latter NLO results being much more non-linear. In 
contrast, for muon scattering, our SPA versus the exact TPE evaluations yield negative versus positive NLO 
contributions, with the magnitudes of the SPA results smaller than the exact ones. The following comments highlight
the crucial differences between our SPA and the exact TPE results of Choudhary 
{\it et al.}~\cite{Choudhary:2023rsz}, which justify the complementary features of each TPE analysis:  
\begin{itemize}
\item Choudhary {\it et al.} features HB$\chi$PT calculations up-to-and-including NLO where it is found that the NLO 
corrections are numerically almost as large as the LO for electron scattering at low-$|Q^2|$ values. This naturally 
raises concerns regarding the magnitude of the NNLO corrections for the validity of the EFT convergence, which was 
not addressed in that work. In this regard, by extending to the dominant class of NNLO corrections which are found 
substantially smaller than the NLO, our SPA results validate the perturbative convergence of the HB$\chi$PT analysis 
employed in both works, albeit the numerical differences in the respective findings. 
\item  Within our version of a one-loop approximation for the genuine NNLO diagrams contributing to the TPE, we 
include the dominant contribution from pion-loops renormalizing the proton-photon vertices, which can, in principle, 
bring about proton's structure-dependent (form factors or charge radius) modifications to the TPE loops [although
in our case such ${\mathcal O}(1/M^2)$ terms cancel out exactly, with residual kinematically suppressed 
${\mathcal O}(1/M^3)$ terms neglected, cf. Eq.~\eqref{eq:fNNLO}]. This indicates that the negative NNLO TPE 
contributions could suppress the overall magnitude of the TPE corrections obtained in the work of Choudhary 
{\it et al.} 
\item 
A detailed investigation reveals that our SPA-based analysis misses the contributions from loop-integrations, such as
$I(Q|1,1,1,0)$, $I(Q|1,1,0,1)$,\footnote{The loop-function $I(Q|1,1,0,1)$ appears in the contributions from the LO 
TPE diagrams (a) and (b) evaluated in the exact TPE analysis of Choudhary {\it et al.} and leads to the Feshbach-like 
term, Eq.~\eqref{eq:delta_LO}.} $I^{-}(p,0|1,1,1,1)$ and $I^{+}(p^\prime,0|1,1,1,1)$, which are present in the exact 
TPE analysis of Choudhary {\it et al}~\cite{Choudhary:2023rsz}.   
\end{itemize} 

%%%%%%%%%%%%%%%%%%%%%%%%%%%%%%%%%%%%%%%%%%%%%%%%%%FIGURE-8%%%%%%%%%%%%%%%%%%%%%%%%%%%%%%%%%%%%%%%%%%%%%%%%%%%%%%%
%%%%%%%%%%%%%%%%%%%%%%%%%%%%%%%%%%%%%%%%%%%%%%%%RESULT-PLOT-3%%%%%%%%%%%%%%%%%%%%%%%%%%%%%%%%%%%%%%%%%%%%%%%%%%%%
\begin{figure*}[tbp]
\begin{center}
    \includegraphics[width=0.48\linewidth]{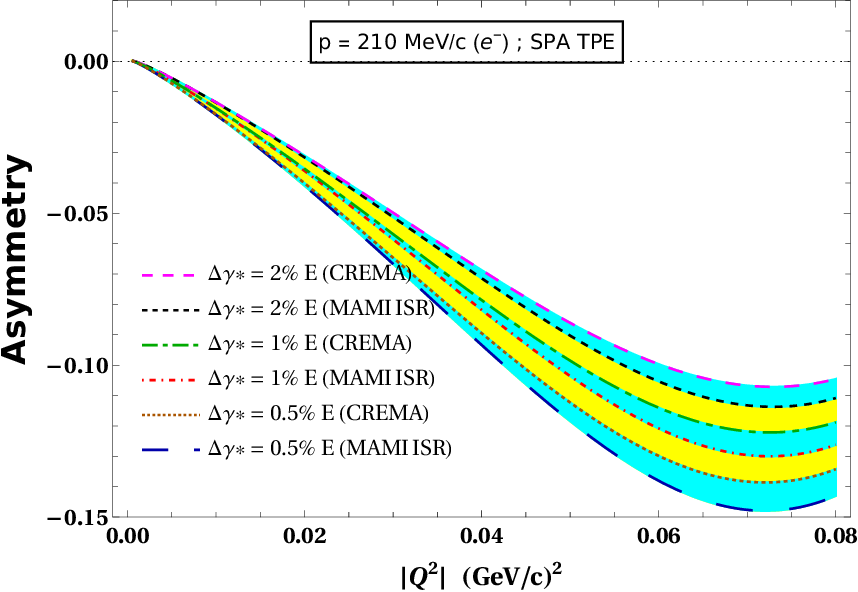}~\quad~\includegraphics[width=0.48\linewidth]{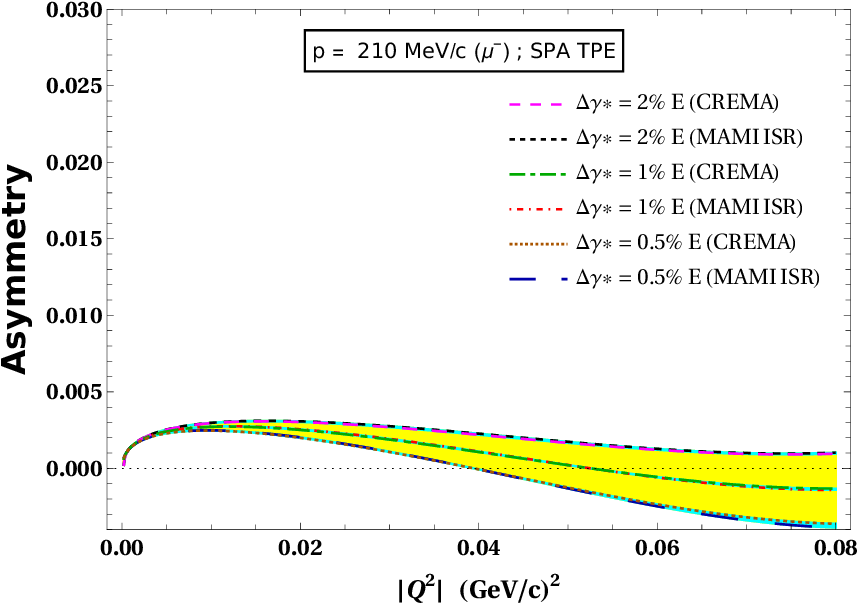}

\vspace{0.6cm}
    
    \includegraphics[width=0.48\linewidth]{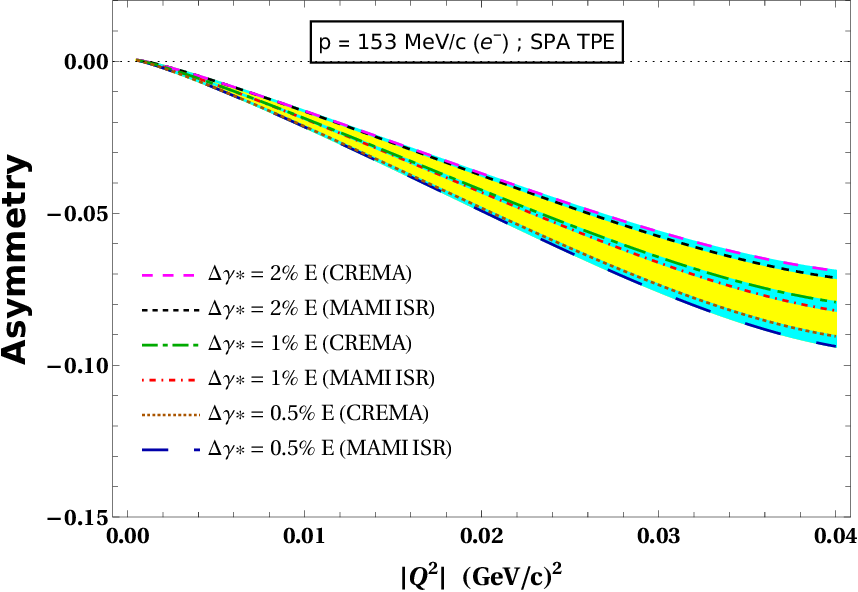}~\quad~\includegraphics[width=0.48\linewidth]{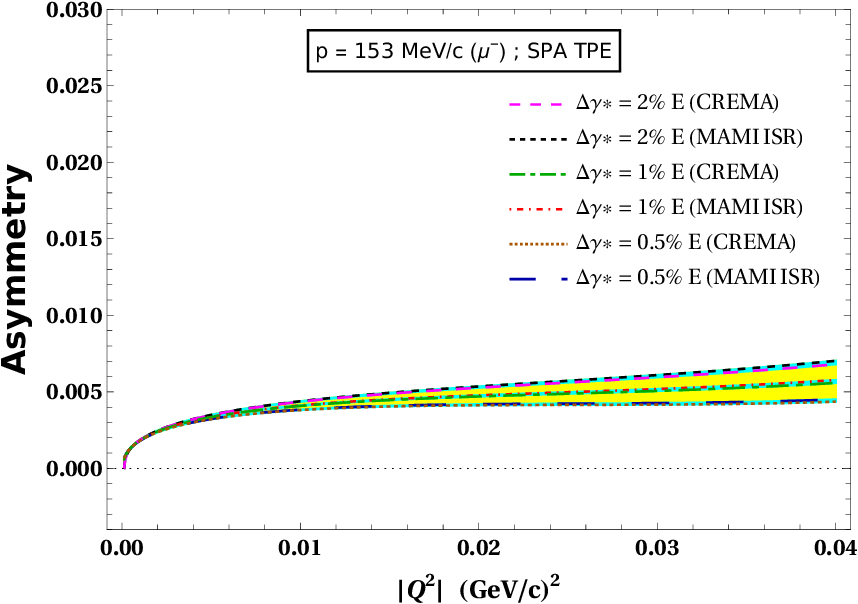}
     
\vspace{0.6cm}
    
\includegraphics[width=0.48\linewidth]{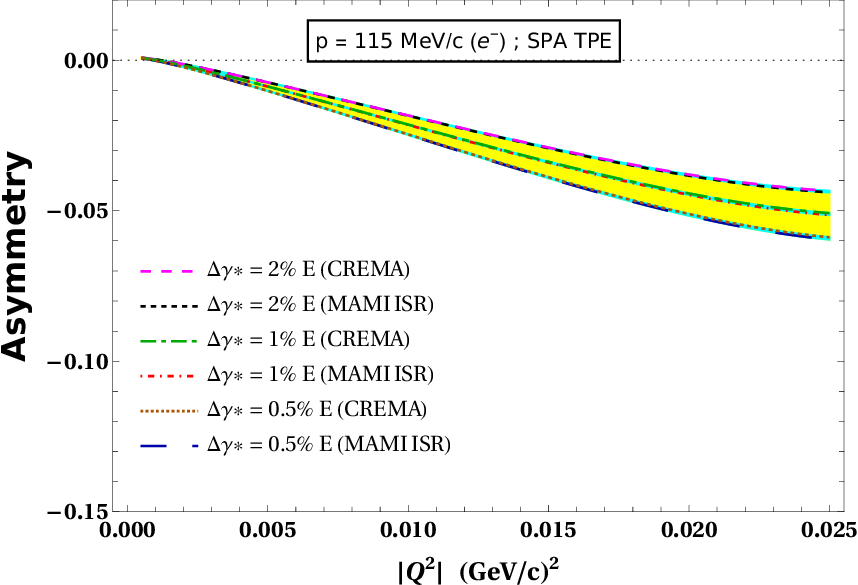}~\quad~\includegraphics[width=0.48\linewidth]{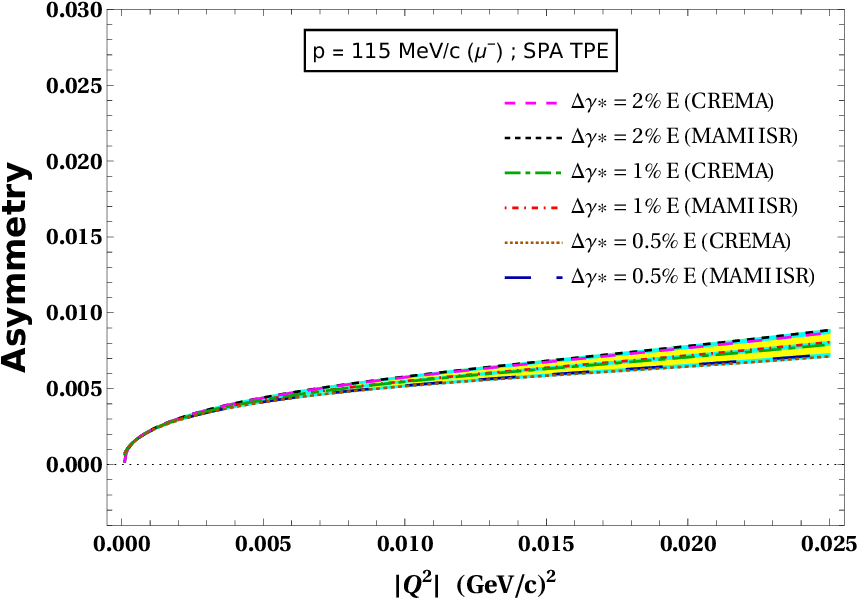}
\caption{Our results for the charge asymmetry (${\mathcal A}^{(1)}_{\ell^\pm}$) between lepton-proton 
        ($\ell^-$-p) and anti-lepton-proton ($\ell^+$-p) scattering cross-sections up-to-and-including NLO
        accuracy in HB$\chi$PT, where our SPA-based TPE contributions (LO and NLO) are taken as the input. 
        The e-p ($\mu$-p) elastic scattering results are displayed on the left (right) panel as a function of 
        the squared four-momentum transfer $|Q^2|$ at the three MUSE proposed incoming lepton beam 
        three-momenta: 210~MeV/c, 153~MeV/c, and 115~MeV/c. The yellow bands depict the uncertainty in our
        predictions owing to the variation of the detector threshold, namely, $0.5\%<\Delta_{\gamma^*}<2\%$
        of the incoming lepton energy $E$ stemming from the soft-photon bremsstrahlung contributions. 
        Likewise, the cyan bands depict the uncertainty due to the variation of the proton's rms charge 
        radius within the range of the extracted values from the recent precision e-p scattering 
        measurements by the A1 Collaboration~\cite{Mihovilovic:2019jiz}, and that from the erstwhile 
        high-precision muonic hydrogen atomic-spectroscopy measurements by the CREMA 
        Collaboration~\cite{{Antognini:2013txn,Pohl:2013yb}}.}
\label{fig:asymmetry1}
\end{center}
\end{figure*}
%%%%%%%%%%%%%%%%%%%%%%%%%%%%%%%%%%%%%%%%%%%%%%%%%%%%%%%%%%%%%%%%%%%%%%%%%%%%%%%%%%%%%%%%%%%%%%%%%%%%%%%%%%%%%%%%% 

%%%%%%%%%%%%%%%%%%%%%%%%%%%%%%%%%%%%%%%%%%%%%%%%%%FIGURE-9%%%%%%%%%%%%%%%%%%%%%%%%%%%%%%%%%%%%%%%%%%%%%%%%%%%%%%%
%%%%%%%%%%%%%%%%%%%%%%%%%%%%%%%%%%%%%%%%%%%%%%%%RESULT-PLOT-4%%%%%%%%%%%%%%%%%%%%%%%%%%%%%%%%%%%%%%%%%%%%%%%%%%%%
\begin{figure*}[tbp]
\begin{center}
    \includegraphics[width=0.48\linewidth]{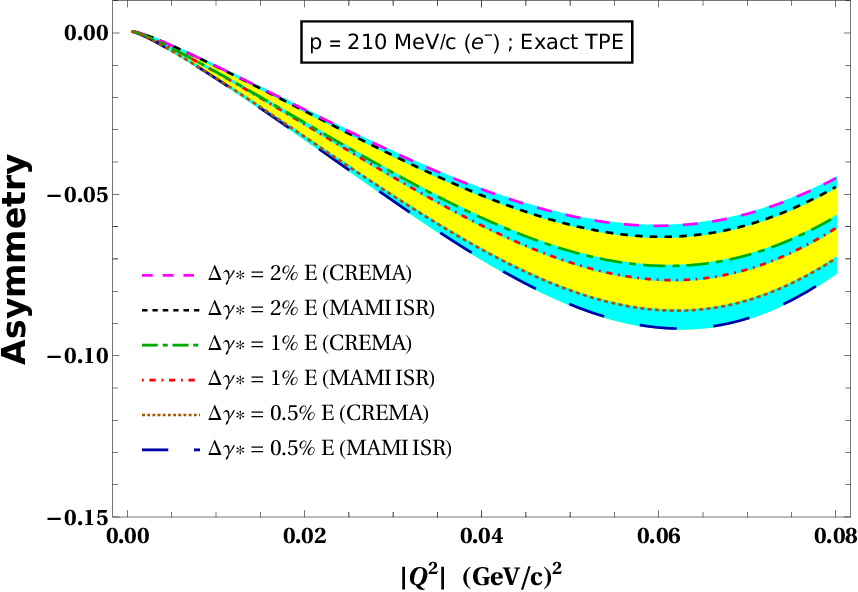}~\quad~\includegraphics[width=0.48\linewidth]{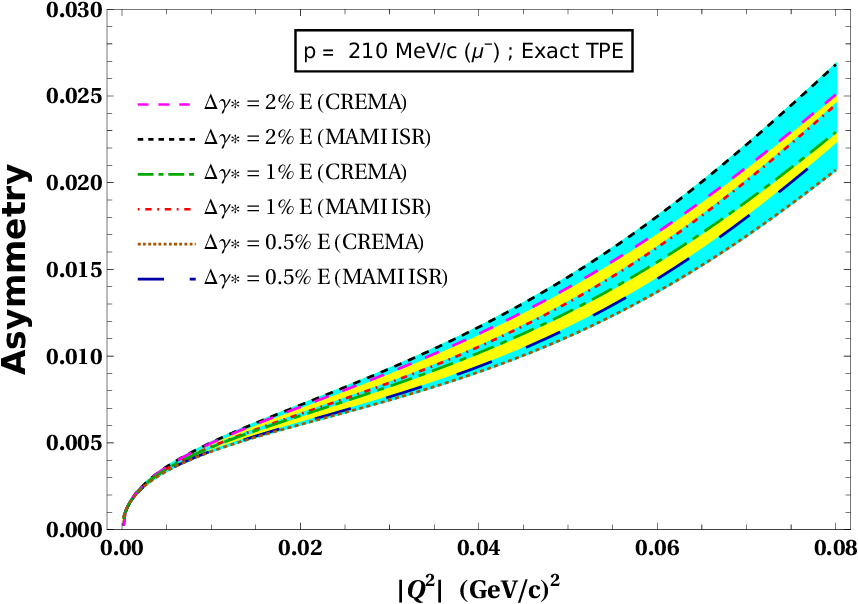}

\vspace{0.6cm}
    
    \includegraphics[width=0.48\linewidth]{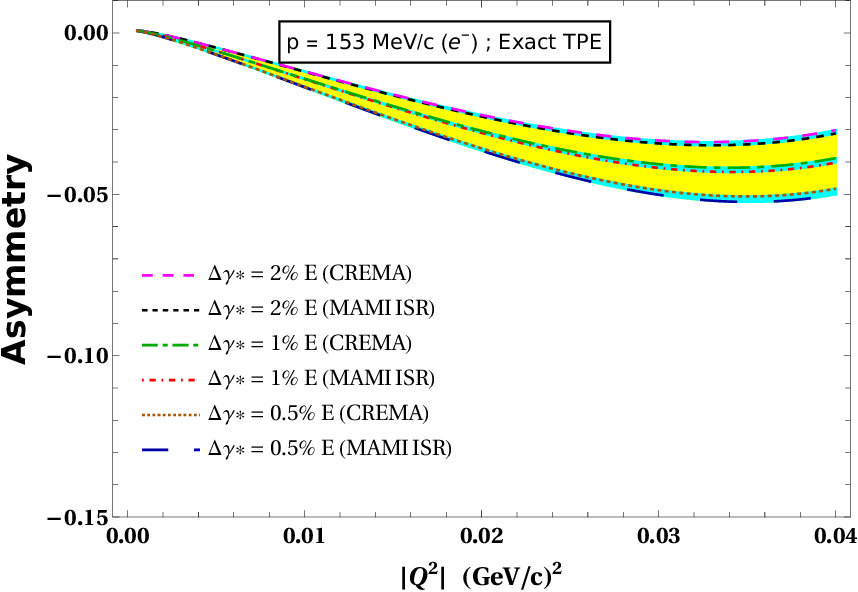}~\quad~\includegraphics[width=0.48\linewidth]{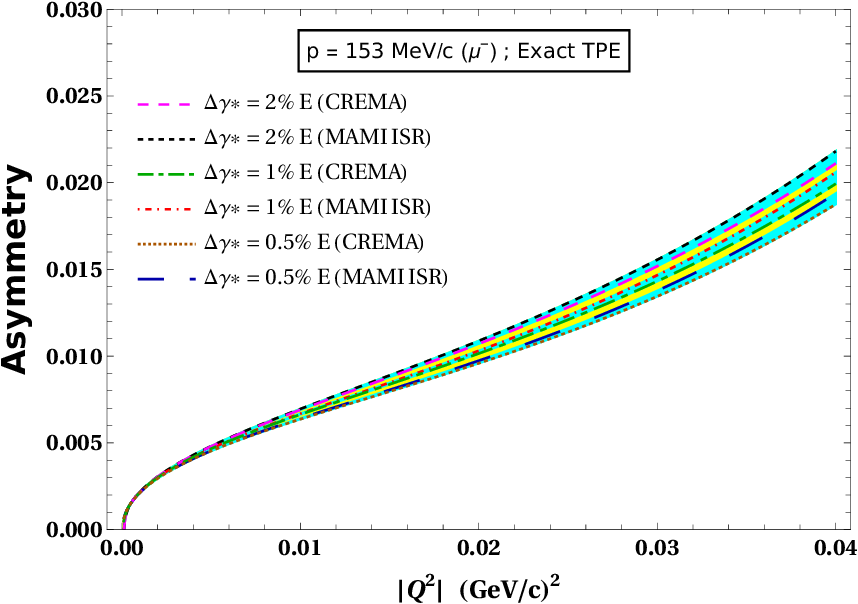}
     
\vspace{0.6cm}
    
\includegraphics[width=0.48\linewidth]{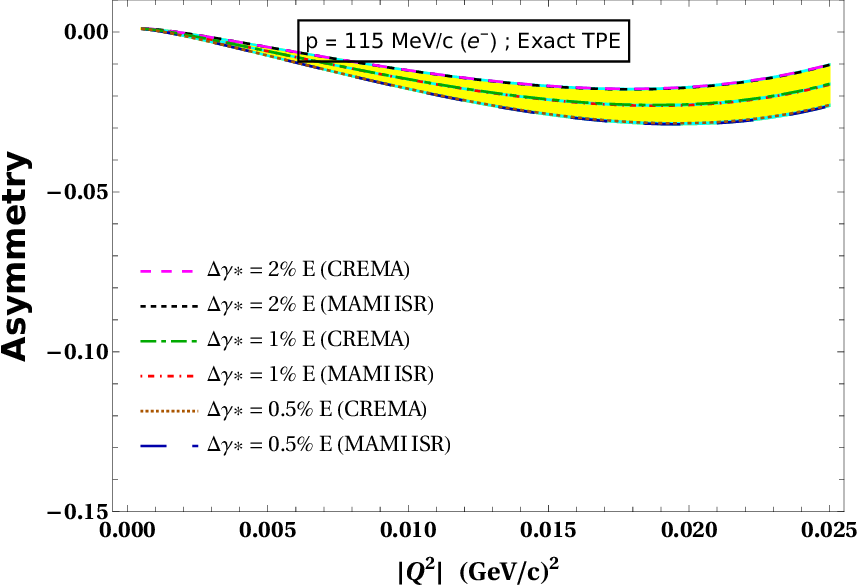}~\quad~\includegraphics[width=0.48\linewidth]{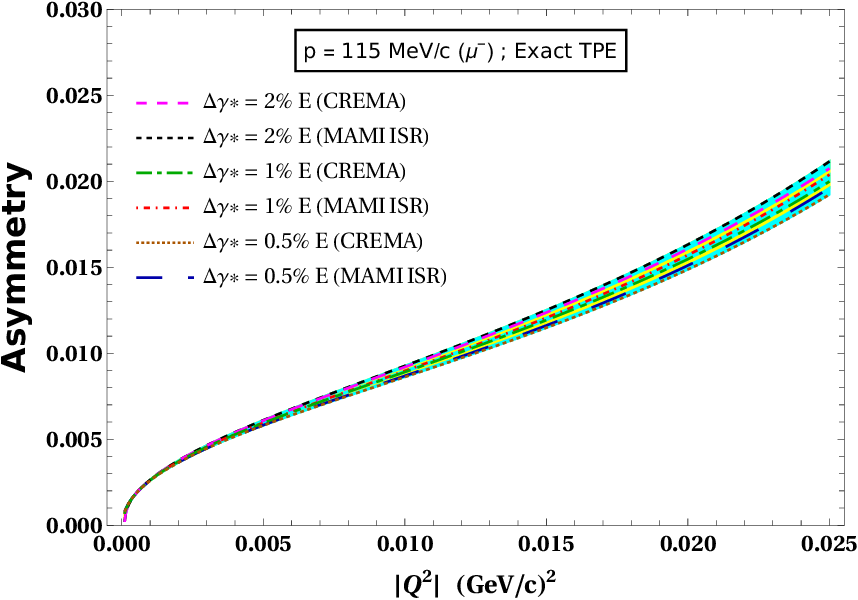}
\caption{Our results for the charge asymmetry (${\mathcal A}^{(1)}_{\ell^\pm}$) between lepton-proton 
        ($\ell^-$-p) and anti-lepton-proton ($\ell^+$-p) scattering cross-sections up-to-and-including NLO
        accuracy in HB$\chi$PT, using the {\it exact} TPE results of Choudhary 
        {\it et al}.~\cite{Choudhary:2023rsz}. The e-p ($\mu$-p) elastic scattering results are displayed on
        the left (right) panel as a function of the squared four-momentum transfer $|Q^2|$ at the three MUSE 
        proposed incoming lepton beam three-momenta: 210~MeV/c, 153~MeV/c, and 115~MeV/c. The yellow bands 
        depict the uncertainty in our predictions owing to the variation of the detector threshold, namely, 
        $0.5\%<\Delta_{\gamma^*}<2\%$ of the incoming lepton energy $E$ stemming from the soft-photon 
        bremsstrahlung contributions. Likewise, the cyan bands depict the uncertainty due to the variation 
        of the proton's rms charge radius within the range of the extracted values from the recent precision
        e-p scattering measurements by the A1 Collaboration~\cite{Mihovilovic:2019jiz}, and that from the 
        erstwhile high-precision muonic hydrogen atomic-spectroscopy measurements by the CREMA 
        Collaboration~\cite{{Antognini:2013txn,Pohl:2013yb}}.}
\label{fig:asymmetry2}
\end{center}
\end{figure*}
%%%%%%%%%%%%%%%%%%%%%%%%%%%%%%%%%%%%%%%%%%%%%%%%%%%%%%%%%%%%%%%%%%%%%%%%%%%%%%%%%%%%%%%%%%%%%%%%%%%%%%%%%%%%%%%%%

%
%Before turning to our asymmetry results, we 
%
We briefly comment on the theoretical error estimate for our SPA-based TPE prediction. As evident from the LO, NLO, 
and NNLO contributions, Eqs.~\eqref{eq:LO}, \eqref{eq:NLO_sum} and \eqref{eq:NNLO_sum}, respectively, our analytical
expressions do not contain arbitrary free parameters. Therefore, the only uncertainty stems from the perturbative
truncation at NNLO in the HB$\chi$PT power counting. In this work, we considered only a seemingly dominant class of 
NNLO contributions to the TPE at "one-loop" approximation. The exclusion of the non-factorizable two-loop NNLO 
contributions, Fig.~\ref{fig:pi-loops}, leads to uncertainties that scale as the same order as the NNLO 
contributions themselves. We, therefore, envisage a maximum uncertainty in our TPE results of about $\sim 0.6\%$ of
the LO Born differential cross-section, Eq.~\eqref{eq:diff_LO}, for both lepton flavors, as revealed easily by 
glancing at the results of Fig.~\ref{fig:delta_TPE}.  

Finally, we present in Figs.~\ref{fig:asymmetry1} and ~\ref{fig:asymmetry2}, our numerical estimation of the charge
asymmetry between lepton- and anti-lepton-proton scattering, Eq.~\eqref{eq:A_asy}, at the level of NLO accuracy, with 
our TPE results evaluated using SPA, and with the exactly evaluated TPE by Choudhary 
{\it et al}.~\cite{Choudhary:2023rsz}. Both the TPE and the charge-odd bremsstrahlung corrections contribute to the 
total charge-odd radiative corrections $\delta_{\rm odd}$, whereas the charge-even radiative and hadronic corrections 
to the lepton-proton OPE scattering process contribute to the total charge-even corrections $\delta_{\rm even}$. In 
particular, we only consider the leading hadronic corrections to the OPE $\delta^{(2)}_\chi$, which are NNLO in the 
chiral power counting.\footnote{Notably, the NLO radiative corrections of $\mathcal{O}\left(\alpha/M\right)$ are by
power counting argument of the same order as the NNLO hadronic corrections to the OPE, given the standard 
correspondence in $\chi$PT $\alpha \sim {\mathcal Q}/M$~\cite{Talukdar:2020aui}, where ${\mathcal Q}$ is a typical 
momentum scale. However, numerically the hadronic corrections supersede the former.} 

As we previously discussed, the soft-photon bremsstrahlung corrections are an integral part of charge asymmetry results.
The estimation involves the arbitrary ``energy" parameter $\Delta_{\gamma^*}$, which serves to distinguish between 
the soft and hard photons~\cite{Mo:1968cg}. The numerical value of this parameter is identified with the resolution of
the detector, which is determined based on the specifics of the experimental setup. Apart from $\Delta_{\gamma^*}$, the
only other free parameter in our analysis is the proton's rms electric charge radius $r_p$ which contributes to the 
hadronic chiral correction Eq.~\eqref{eq:hadronic}. To assess the sensitivity of our asymmetry results to these two 
free parameters, in Figs.~\ref{fig:asymmetry1} and ~\ref{fig:asymmetry2}, we varied $\Delta_{\gamma^*}$ within the range
$0.5\% < \Delta_{\gamma^*} < 2\%$ of lepton energy $E$. The radius $r_p$ is likewise varied within a range 
corresponding to the extracted value, $0.87\pm (0.014)_{\rm stat.}\pm (0.024)_{\rm syst.}\pm (0.003)_{\rm mod.}$~fm, 
from the recent precision e-p scattering measurements by the A1 Collaboration (MAMI)~\cite{Mihovilovic:2019jiz}, and the 
value $0.84087(39)$~fm, from the erstwhile high-precision muonic hydrogen spectroscopic measurements by the CREMA 
Collaboration (PSI)~\cite{Antognini:2013txn,Pohl:2013yb}. The latter range serves as a rough indicator to gauge the 
impact of the current status of the proton's radius discrepancy in our asymmetry estimates. 

A comparison between Figs.~\ref{fig:asymmetry1} (TPE using SPA) and ~\ref{fig:asymmetry2} (exact TPE) reflects that the
asymmetry results are more sensitive to the above parameters for the electrons with the constituent TPE evaluated in SPA.   
Furthermore, we find that there is an increasing sensitivity due to the variation of the parameters with increasing 
$|Q^2|$ values and lepton beam momentum, for both electron and muon scattering cases. We find that the sensitivities due 
to the variations of $\Delta_{\gamma^*}$ and $r_p$ are quite contrasting for muon scattering when comparing the SPA 
versus exact TPE results. We also find that when comparing Figs.~\ref{fig:asymmetry1} and ~\ref{fig:asymmetry2} for 
electron scattering, the asymmetry is more negative in the former, although in both cases their magnitudes increase with 
rising incoming electron momentum (notice the same ordinate scale used to compare both figures in regard to the 
individual leptons). In contrast, for muon scattering, the asymmetry remains predominantly positive in both SPA and exact 
TPE input cases, with their magnitudes slowly decreasing (increasing) in the former (latter) with rising incoming muon 
momentum. There is, however, one exception, namely, with the SPA-based TPE input for the muon scattering corresponding to 
the largest beam momentum, where the asymmetry behavior is found to be quite unusual (upper right panel in 
Fig.~\ref{fig:asymmetry1}). In this case, the asymmetry is positive for small-$|Q^2|$ values, exhibiting a 
positive maximum with increasing $|Q^2|$, but then decreasing to become mostly negative in the large-$|Q^2|$ region for 
MUSE kinematics. In contrast, with the exact TPE input, the asymmetry increases dramatically with increasing incoming muon
beam momentum, as seen in Fig.~\ref{fig:asymmetry2}. It is worth mentioning that regarding the overall sensitivity due to
$r_p$, we find up to 4\% uncertainty with the exact TPE input results for both electron and muon scatterings at the largest
values of the MUSE beam momenta. 

In conclusion, we find that the SPA-based TPE evaluation yields the results (presented in this paper)  are significantly 
different from the exact TPE results of Choudhary {\it et al.}~\cite{Choudhary:2023rsz}. Invoking SPA, with one or the 
other photon propagator in the TPE box- and cross-box diagrams set on-shell, the assumption turns out to be rather 
restrictive. In effect, this means that for a SPA-based evaluation which, for instance, neglects the contribution from the
kinematic region with two exchanged hard-photons within the TPE loops, a significant systematic uncertainty can be 
introduced. Consequently, one should avoid the SPA methodology altogether to obtain reasonably precise estimates of the 
TPE or observables that rely on the TPE. Regarding the charge asymmetry, we have made only a preliminary estimate in this
work. In the near future, we hope to improve upon our presented asymmetry results for both the electron and muon 
scatterings, removing further sources of systematic theoretical uncertainties. To that end, two important future 
directions of the present work are conceivable: 
(i) an improved treatment of the exact TPE results of Choudhary {\it et al}.~\cite{Choudhary:2023rsz} by including NNLO 
corrections, and  
(ii) inclusion of hard bremsstrahlung photons to assess the charge asymmetry, thereby removing the dependence on the 
arbitrary $\Delta_{\gamma^*}$ parameter. Currently, work is in progress in both these directions~\cite{Bhoomika24}.

%%%%%%%%%%%%%%%%%%%%%%%%%%%%ACKNOWLEDGEMENT%%%%%%%%%%%%%%%%%%%%%%%%%%%%%
\section*{Acknowledgments}
%%%%%%%%%%%%%%%%%%%%%%%%%%%%%%%%%%%%%%%%%%%%%%%%%%%%%%%%%%%%%%%%%%%%%%%%
RG acknowledges the organizers of the HADRON2025 International Conference at Osaka University for their financial support
and hospitality during the conference. UR acknowledges financial support from the Science and Engineering Research Board 
under grant number CRG/2022/000027. 
%%%%%%%%%%%%%%%%%%%%%%%%%%%%%%%%%%%%%%%%%%%%%%%%%%%%%%%%%%%%%%%%%%%%%%%%

\appendix
%%%%%%%%%%%%%%%%%%%%%%%%%%%%%%%%%%%%%%%%%%%%%%%%%%%%%%%%%%%%%%%%%%%%%%%%%%%%%%%%%%%%%%%%%%%%%%%%%%%%%%%
\section{Loop-integrations} 
\label{Integrations}
%%%%%%%%%%%%%%%%%%%%%%%%%%%%%%%%%%%%%%%%%%%%%%%%%%%%%%%%%%%%%%%%%%%%%%%%%%%%%%%%%%%%%%%%%%%%%%%%%%%%%%%
In this appendix, we collect the Feynman loop-integrations employed in our analytical evaluations of the TPE diagrams, 
where we denote the external lepton four-momenta as $p$ and $p^\prime$, such that the four-momentum transfer is 
$Q=p-p^\prime$, while the loop four-momentum is denoted by $k$. These are given by the following generic scaler and 
tensor integrals with indices $n_{1,2,3,4}\in {\mathbb Z}$ and a real-valued parameter $\omega$:
\begin{widetext}
%%%%%%%%%%%%%%%%%%%%%%%%%%%%%%%%%%%%%%%%%%%%%%%%%%%%%%%%%%  
\begin{eqnarray}
I^-(p,\omega|n_1,n_2,n_3,n_4) \!\!&=&\!\!\frac{1}{i} \int \frac{{\rm d}^4k}{(2\pi)^4}
\frac{1}{(k^2+i0)^{n_1} [(k-Q)^2+i0]^{n_2}(k^2-2k\cdot p+i0)^{n_3}(v\cdot k+\omega +i0)^{n_4}}\,,
\\
I^+(p^\prime ,\omega|n_1,n_2,n_3,n_4) \!\!&=&\!\!\frac{1}{i} 
\int \frac{{\rm d}^4k}{(2\pi)^4}
\frac{1}{(k^2+i0)^{n_1} [(k-Q)^2+i0]^{n_2}(k^2+2k\cdot p^\prime+i0)^{n_3}(v\cdot k+\omega +i0)^{n_4}}\,,
\end{eqnarray}
and 
\begin{eqnarray}
I^{-\mu}(p,\omega|n_1,n_2,n_3,n_4) \!\!&=&\!\!\frac{1}{i} 
\int \frac{{\rm d}^4k}{(2\pi)^4}
\frac{k^\mu}{(k^2+i0)^{n_1} [(k-Q)^2+i0]^{n_2}(k^2-2k\cdot p+i0)^{n_3}(v\cdot k+\omega +i0)^{n_4}}\,,
\end{eqnarray}
%%%%%%%%%%%%%%%%%%%%%%%%%%%%%%%%%%%%%%%%%%%%%%%%%%%%%%%%%%  
respectively. The specific integrals that were used for our calculations have $\omega=0$. First, we consider the two 
IR-divergent loop-integrals, namely, 
\begin{eqnarray}
I^-(p,0|1,0,1,1) \!\!&\equiv&\!\! I^{(0)}(p,0|1,0,1,1)=\frac{1}{i} 
\int \frac{{\rm d}^4k}{(2\pi)^4}\frac{1}{(k^2+i0)\,(k^2-2k\cdot p+i0)\,(v\cdot k+i0)}
\nonumber\\
\!\!&=&\!\!-\,\frac{1}{(4\pi)^2\beta E}\Bigg[\left\{\frac{1}{\epsilon}-\gamma_E
+\ln{\left(\frac{4\pi \mu^2}{m_l^2}\right)}\right\}\ln{\sqrt{\frac{1+\beta}{1-\beta}}}-{\rm Li}_2\left(\frac{2\beta}{1+\beta}\right)
-\ln^2{\sqrt{\frac{1+\beta}{1-\beta}}}
\nonumber\\
&&\hspace{1.75cm} -\,i\pi\left\{\frac{1}{\epsilon}-\gamma_E+\ln{\left(\frac{4\pi \mu^2}{m_l^2}\right)}\right\}\Bigg]\,, \quad \text{and}
\\
I^+(p^\prime,0|1,0,1,1) \!\!&=&\!\! \frac{1}{i} 
\int \frac{{\rm d}^4k}{(2\pi)^4}\frac{1}{(k^2+i0)\,(k^2+2k\cdot p^\prime+i0)\,(v\cdot k+i0)}
\nonumber\\
&=&\!\! \frac{1}{(4\pi)^2\beta^\prime E^\prime}\Bigg[\left\{\frac{1}{\epsilon}-\gamma_E
+\ln{\left(\frac{4\pi\mu^2}{m_l^2}\right)}\right\}\ln{\sqrt{\frac{1+\beta^\prime}{1-\beta^\prime}}}
-{\rm Li}_2\left(\frac{2\beta^\prime}{1+\beta^\prime}\right)-\ln^2{\sqrt{\frac{1+\beta^\prime}{1-\beta^\prime}}}
\nonumber\\
&&\hspace{1.75cm} -\,i\pi\left\{\frac{1}{\epsilon}-\gamma_E+\ln{\left(\frac{4\pi \mu^2}{m_l^2}\right)}\right\}\Bigg]
\nonumber\\
\!\!&\equiv&\!\! -\,I^{(0)}(p,0|1,0,1,1) +\delta^{(1/M)}I^+(p^\prime,0|1,0,1,1)+\delta^{(1/M^2)}I^+(p^\prime,0|1,0,1,1) 
+ \mathcal{O}\left(\frac{1}{M^3}\right)\,,
\end{eqnarray}
where in isolating the IR-divergent parts of integrals we utilize dimensional regularization by analytically continuing 
the integrals to $D$-dimensional ($D = 4 -2\epsilon$) space-time, where the pole $\epsilon < 0$ yields the IR-divergence. 
For convenience, we split $I^{+}(p^\prime,0|1,0,1,1)$ into the LO [i.e., $\mathcal{O}\left(1/{M^0} \right)$], NLO [i.e., 
$\mathcal{O}\left(1/M\right)$], and NNLO [i.e., $\mathcal{O}\left(1/M^2\right)$] parts, thereby yielding the following 
expressions:
\begin{eqnarray}
\delta^{(1/M)}I^+(p^\prime,0|1,0,1,1)\!\!&=&\!\! -\,\frac{Q^2}{2(4\pi)^2 M E^2 \beta^3}\Bigg[\left\{\frac{1}{\epsilon}-\gamma_E
+\ln{\left(\frac{4\pi \mu^2}{m_l^2}\right)}\right\}\left(\ln{\sqrt{\frac{1+\beta}{1-\beta}}}-\beta\right)
-{\rm Li}_2\left(\frac{2\beta}{1+\beta}\right)
\nonumber\\
&&\hspace{2.6cm} -\,\ln^2{\sqrt{\frac{1+\beta}{1-\beta}}}+2\ln{\sqrt{\frac{1+\beta}{1-\beta}}}- i\pi 
\left\{\frac{1}{\epsilon}-\gamma_E+\ln{\left(\frac{4\pi \mu^2}{m_l^2}\right)}\right\} \,\,\Bigg]\,,
\end{eqnarray}
and 
\begin{eqnarray}
\delta^{(1/M^2)}I^+(p^\prime,0|1,0,1,1)\!\!&=&\!\! \frac{Q^4}{8(4\pi)^2 M^2 E^3 \beta^5}\Bigg[\left\{\frac{1}{\epsilon}-\gamma_E
+\ln{\left(\frac{4\pi \mu^2}{m_l^2}\right)}\right\}\left((3-\beta^2)\ln{\sqrt{\frac{1+\beta}{1-\beta}}}-3\beta\right) -2\beta
\nonumber\\
&&\hspace{2.6cm} -\,(3-\beta^2){\rm Li}_2\left(\frac{2\beta}{1+\beta}\right)-(3-\beta^2)\ln^2{\sqrt{\frac{1+\beta}{1-\beta}}}
-2(\beta^2-4)
\nonumber\\
&&\hspace{2.6cm} \times\,\ln{\sqrt{\frac{1+\beta}{1-\beta}}}-i\pi (3-\beta^2)\left\{\frac{1}{\epsilon}-\gamma_E
+\ln{\left(\frac{4\pi \mu^2}{m_l^2}\right)}\right\} \Bigg]\,.
\end{eqnarray}
%%%%%%%%%%%%%%%%%%%%%%%%%%%%%%%%%%%%%%%%%%%%%%%%%%%%%%%%%%  
Here, $\beta=|{\bf p}\,|/E$ and $\beta^\prime=|{{\bf p}^\prime}\,|/E^\prime$ are the velocities of the incoming and 
outgoing lepton, respectively, and
\begin{eqnarray}
{\rm Li}_2(z) =- \int_0^z {\rm d}t\, \frac{\ln(1-t)}{t}\,,\quad  \forall z\in {\mathbb C}\,,
\label{eq:Li2}
\end{eqnarray}
is the standard di-logarithm (or Spence) function. The rest of the loop-integrals appearing in our TPE results are 
IR-finite. Splitting them into LO, NLO, and NNLO parts, they are respectively given by 
\begin{eqnarray}
I^-(p,0|0,1,1,1) \!\!&=&\!\! \frac{1}{i} 
\int \frac{{\rm d}^4k}{(2\pi)^4}\frac{1}{[(k-Q)^2+i0]\,(k^2-2k\cdot p+i0)\,(v\cdot k+i0)}
\nonumber\\
\!\!&\equiv&\!\! I^{(0)}(p,0|0,1,1,1) +\delta^{(1/M)}I^-(p,0|0,1,1,1)+\delta^{(1/M^2)}I^-(p,0|0,1,1,1) 
+ \mathcal{O}\left(\frac{1}{M^3}\right)\,,
\end{eqnarray}
where,
\begin{eqnarray}
I^{(0)}(p,0|0,1,1,1)\!\!&=&\!\! -\frac{1}{(4\pi)^2 \beta E} \Bigg[\frac{\pi^2}{6}-{\rm Li}_2\left(\frac{2\beta}{1+\beta}\right)
-{\rm Li}_2\left(\frac{1+\beta}{1-\beta}\right)-2\ln^2{\sqrt{\frac{1+\beta}{1-\beta}}}
\nonumber\\
&&\hspace{1.8cm}+\,2\ln\left({ -\frac{Q^2}{2 M E\beta}}\right)
\ln{\sqrt{\frac{1+\beta}{1-\beta}}}\,\,\Bigg]\,,
\\
\delta^{(1/M)}I^-(p,0|0,1,1,1)\!\!&=&\!\!-\frac{Q^2}{(4\pi)^2 M \beta^3 E^2} \Bigg[-\frac{\pi^2}{12}-\beta 
+\frac{1}{2}{\rm Li}_2\left(\frac{2\beta}{1+\beta}\right)+\frac{1}{2}{\rm Li}_2\left(\frac{1+\beta}{1-\beta}\right)
-(1+\beta)\ln{\sqrt{\frac{1+\beta}{1-\beta}}}
\nonumber\\
&&\hspace{2.55cm} +\,2\beta \ln{\sqrt{\frac{2\beta}{1-\beta}}}+\ln^2{\sqrt{\frac{1+\beta}{1-\beta}}}
-\ln{\left(-\frac{Q^2}{2 M E\beta}\right)}\ln{\sqrt{\frac{1+\beta}{1-\beta}}}
\nonumber\\
&&\hspace{2.55cm} +\,\beta \ln{\left( -\frac{Q^2}{2 M E\beta}\right)}+i\pi\beta \Bigg]\,, \quad \text{and}
\\
\delta^{(1/M^2)}I^-(p,0|0,1,1,1)\!\!&=&\!\! \frac{Q^4}{8 (4\pi)^2 M^2 E^3 \beta^5}\Bigg[5\beta 
+3(1+\beta)^2\ln{\sqrt{\frac{1+\beta}{1-\beta}}-12\beta \ln{\sqrt{\frac{2\beta}{1-\beta}}}}
-12\beta \ln{\sqrt{-\frac{Q^2}{2M E\beta}}}
\nonumber\\
&&\hspace{2.55cm}-\,(3-\beta^2)\Bigg\{\frac{\pi^2}{6}-{\rm Li}_2\left(\frac{2\beta}{1+\beta}\right)
-{\rm Li}_2\left(\frac{1+\beta}{1-\beta}\right)-2\ln^2{\sqrt{\frac{1+\beta}{1-\beta}}}
\nonumber\\
&&\hspace{2.55cm}+\,2\ln{\left(-\frac{Q^2}{2ME\beta}\right)}\ln{\sqrt{\frac{1+\beta}{1-\beta}}}\,\Bigg\}-6i \pi \beta\Bigg]\,.
\end{eqnarray}
%%%%%%%%%%%%%%%%%%%%%%%%%%%%%%%%%%%%%%%%%%%%%%%%%%%%%%%%%%  
Similarly, we have
\begin{eqnarray}
I^+(p^\prime,0|0,1,1,1) \!\!&=&\!\! \frac{1}{i} 
\int \frac{{\rm d}^4k}{(2\pi)^4}\frac{1}{[(k-Q)^2+i0]\,(k^2+2k\cdot p^\prime+i0)\,(v\cdot k+i0)}
\nonumber\\
&\equiv&\!\! -\,I^{(0)}(p,0|0,1,1,1) +\delta^{(1/M)}I^+(p^\prime,0|0,1,1,1)+\delta^{(1/M^2)}I^+(p^\prime,0|0,1,1,1)
+ \mathcal{O}\left(\frac{1}{M^3}\right)\,,\quad\,
\end{eqnarray}
where,
\begin{eqnarray}
\delta^{(1/M)}I^+(p^\prime,0|0,1,1,1) \!\!&=&\!\! -\frac{Q^2}{(4\pi)^2 M \beta^3 E^2}
\left[\ln{\sqrt{\frac{1+\beta}{1-\beta}}}-\beta \right]\,,
\\
\delta^{(1/M^2)}I^+(p^\prime,0|0,1,1,1) \!\!&=&\!\!  \frac{Q^4 }{8(4\pi)^2 M^2 E^3 \beta^5}
\left[(3-\beta^2)\ln\sqrt{{\frac{1+\beta}{1-\beta}}}-3\beta\right]\,.
\end{eqnarray}
Next, we consider
\begin{eqnarray}
I^-(p,0|0,1,1,2) \!\!&=&\!\! \frac{1}{i} 
\int \frac{{\rm d}^4k}{(2\pi)^4}\frac{1}{[(k-Q)^2+i0]\,(k^2-2k\cdot p+i0)\,(v\cdot k+i0)^2}
\nonumber\\
&\equiv&\!\!\, M I^{(0)}(p,0|0,1,1,2) +\delta^{(1/M)}I^+(p^\prime,0|0,1,1,2)
+ \mathcal{O}\left(\frac{1}{M^2}\right)\,,
\end{eqnarray}
where,
\begin{eqnarray}
I^{(0)}(p,0|0,1,1,2)\!\!&=&\!\! -\frac{4}{(4 \pi)^2 Q^2 E \beta}\ln{\sqrt{\frac{1+\beta}{1-\beta}}}\,,
\\
\delta^{(1/M)}I^-(p,0|0,1,1,2)\!\!&=&\!\!- \frac{3 Q^2}{(4\pi)^2 M E^3 \beta^4 }\,.
\end{eqnarray}
%%%%%%%%%%%%%%%%%%%%%%%%%%%%%%%%%%%%%%%%%%%%%%%%%%%%%%%%%%  
Similarly, we have
\begin{eqnarray}
I^+(p^\prime,0|0,1,1,2) \!\!&=&\!\! \frac{1}{i} 
\int \frac{{\rm d}^4k}{(2\pi)^4}\frac{1}{[(k-Q)^2+i0]\,(k^2+2k\cdot p^\prime+i0)\,(v\cdot k+i0)^2}
\nonumber\\
&\equiv&\!\!-MI^{(0)}(p,0|0,1,1,2) +\delta^{(M^0)}I^+(p^\prime,0|0,1,1,2)+\delta^{(1/M)}I^+(p^\prime,0|0,1,1,2) 
+ \mathcal{O}\left(\frac{1}{M^2}\right)\,,\quad \,
\end{eqnarray}
where,
\begin{eqnarray}
\delta^{(M^0)}I^+(p^\prime,0|0,1,1,2)\!\!&=&\!\! -\frac{2}{(4\pi)^2 E^2 \beta^3}\bigg[\ln{\sqrt{\frac{1+\beta}{1-\beta}}-\beta}\bigg]\,,
\\
\delta^{(1/M)}I^+(p^\prime,0|0,1,1,2)\!\!&=&\!\! \frac{ Q^2 }{2M (4\pi)^2 E^3 \beta^5}
\left(-3\beta +(3-\beta^2)\ln\sqrt{{\frac{1+\beta}{1-\beta}}}\,\,\right)\,.
\end{eqnarray}
%%%%%%%%%%%%%%%%%%%%%%%%%%%%%%%%%%%%%%%%%%%%%%%%%%%%%%%%%%  
It is notable that the last two integrals, namely, $I^-(p,0|0,1,1,2)$ and $I^+(p^\prime,0|0,1,1,2)$, have the leading term of 
${\mathcal O}(M)$, and consequently we need to consider their expansions no more than ${\mathcal O}(M^{-1})$, as clearly 
suggested by the NLO and NNLO amplitudes, Eqs.~\eqref{eq:ffNLO} and \eqref{eq:fNNLO}. Finally, we consider the tensor 
loop-integration
\begin{eqnarray}
I^{-\mu}_1(p,0|1,1,1,0) \!\!&=&\!\! \frac{1}{i} 
\int \frac{{\rm d}^4k}{(2\pi)^4}\frac{k^\mu}{(k^2+i0)[(k-Q)^2+i0]\,(k^2-2k\cdot p+i0)}
\nonumber\\
&&=-\,\frac{1}{4 \pi^2 Q^2 \nu_l^2}\left[\left(p^{\mu}-\frac{1}{2}Q^{\mu}\right)\ln\sqrt{-\frac{Q^2}{m_l^2}}
-4\pi^2(Q^2 p^{\mu}-2m_l^2 Q^{\mu})I(Q|1,1,1,0)\right]\,,
\end{eqnarray}
where
\begin{eqnarray}
%
%I(Q,0|1,1,1,0) 
%
I(Q|1,1,1,0) \!\!&\equiv&\!\! I^-(p,0|1,1,1,0) = \frac{1}{i} \int \frac{{\rm d}^4k}{(2\pi)^4}\frac{1}{(k^2+i0)\,[(k-Q)^2+i0]\,(k^2-2k\cdot p+i0)}
\nonumber\\
&\equiv&\!\! I^+(p^\prime,0|1,1,1,0) = \frac{1}{i} \int \frac{{\rm d}^4k}{(2\pi)^4}\frac{1}{(k^2+i0)\,[(k-Q)^2+i0]\,(k^2+2k\cdot p^\prime+i0)}
\nonumber\\
&=&\!\! \frac{1}{8 \pi^2 Q^2 \nu_l} \left[\frac{\pi^2}{3}+\ln^2{\sqrt{\frac{\nu_l+1}{\nu_l-1}}
+ {\rm Li}_2\left(\frac{\nu_l-1}{\nu_l+1}\right)}\right]\,.
\end{eqnarray}
%%%%%%%%%%%%%%%%%%%%%%%%%%%%%%%%%%%%%%%%%%%%%%%%%%%%%%%%%%  
\end{widetext}
%%%%%%%%%%%%%%%%%%%%%%%%%%%%%%%%%%%%%%%%%%%%%%%%%%%%%% 
\bibliographystyle{apsrev}

\end{document}